\documentclass[aps, prd, preprint, floats, epsf, superscriptaddress, longbibliography, nofootinbib]{revtex4-1}

\usepackage{cancel}
\usepackage{amsmath,amssymb,amsfonts}
\usepackage{dcolumn}
\usepackage{bm}
\usepackage[mathlines]{lineno}
\DeclareMathOperator{\sech}{sech}

\usepackage{graphicx}
\usepackage{verbatim}
\usepackage{color}
\usepackage{subcaption}
\usepackage[hyperfootnotes=true]{hyperref}
\usepackage{float}
\usepackage{tikz-feynman}
\usepackage{mathtools}
\usepackage{mathrsfs}
\usepackage{enumerate}
\usepackage{multirow}
\usepackage{tablefootnote}
\usepackage[normalem]{ulem}

\graphicspath{{Plots/}}

\begin{document}

\title{Distinguishing $W'$ Signals at Hadron Colliders Using Neural Networks}

\author{Spencer Chang}
\affiliation{Department of Physics and Institute for Fundamental Science\\ University of Oregon, Eugene, Oregon 97403, U.S.A.}

\author{Ting-Kuo Chen}
\affiliation{Department of Physics, National Taiwan University, Taipei 10617, Taiwan}

\author{Cheng-Wei Chiang}
\affiliation{Department of Physics, National Taiwan University, Taipei 10617, Taiwan}
\affiliation{Physics Division, National Center for Theoretical Sciences,
Taipei, Taiwan 10617, R.O.C.}

\begin{abstract}
We investigate a neural network-based hypothesis test to distinguish different $W'$ and charged scalar resonances through the $\ell+\cancel{E}_T$ channel at hadron colliders.  This is traditionally challenging due to a four-fold ambiguity at proton-proton colliders, such as the Large Hadron Collider.  Of the neural network approaches we studied, we find a multi-class classifier based on a fully-connected neural network trained upon 2D histograms made from kinematic variables of the final state $\ell$ to be the most powerful. Furthermore, by considering the 1-jet processes, we demonstrate that one can generalize to multiple $2D$ histograms to represent different variable pairs. Finally, as a comparison to traditional approaches, we compare our method with Bayesian hypothesis testing and discuss the pros and cons of each approach. The neural network scheme presented in this paper is a powerful tool that can help probe the properties of charged resonances.
\end{abstract}

\maketitle


\section{Introduction}\label{sec:1}

Ever since the discovery of the $W$ boson through the $e\nu$ decay channel in 1983 at the SPS collider~\cite{Arnison:1983rp, Banner:1983jy}, the search for $W'$  and other charged boson resonances has continued. The latest analyses include the $13$ TeV search in the di-jet channel~\cite{Aad:2019hjw,Sirunyan:2019vgj,Aad:2020cws}, the di-jet + lepton channel~\cite{Aad:2020kep}, the $\ell+\cancel{E}_T$~\cite{Sirunyan:2018mpc,Aad:2019wvl} channel, the $\tau\nu$ channel~\cite{Sirunyan:2018lbg}, and di-boson channels~\cite{Sirunyan:2019jbg,Aad:2019fbh} conducted by ATLAS and CMS.  So far, the mass limit for sequential $W'$ has been pushed above the TeV level (see Ref.~\cite{Zyla:2020zbs}), and thus future $W'$ signals are expected to occur at higher masses in high-energy hadron colliders. One such example is the CERN Large Hadron Collider (LHC), which is the main focus of our study. In this case, the leptonic search turns out to be a favorable choice, as it avoids the large QCD background. Some of the most important properties to be identified of a $W'$ would be the mass, decay width, and couplings to the Standard Model (SM) fermions; if we further include the study of charged scalar bosons, spin would also be important. However, determining the boson's couplings and spin in its center-of-mass (COM) frame at the LHC suffers from two ambiguities:
\begin{itemize}
 \item \textit{Unknown initial state}: To study the Lorentz structure of a charged current interaction, the incident partons must be identified so as to define the forward direction (e.g.~in the quark direction, not the anti-quark direction.).  Due to the parton distribution functions (PDFs), the best one can do is to make a reasonable guess for this from the PDF properties~\cite{Langacker:1984}. 
 \item \textit{Missing longitudinal momentum}: Since the COM frame of the colliding partons is typically boosted, we need to identify the missing longitudinal momentum associated with the neutrino to correctly determine the COM angular distribution in $\cos \theta_{\rm COM}$.  From kinematics, the longitudinal momentum can be solved from a quadratic equation assuming the mediating boson to be on-shell, but there is no event-by-event information that can be used to determine which of the two quadratic solutions is correct. This ambiguity has already been pointed out in several studies involving $\cancel{E}_T$, such as the reconstruction of $W\rightarrow e\nu$ at the SPS $p\overline{p}$ Collider~\cite{Arnison:1983rp} and top pair production at the Tevatron~\cite{Abe:1995hr}.
 
\end{itemize}
Even though the mentioned ambiguities have imposed an obstacle to such studies, several studies based on traditional approaches have still been conducted to reconstruct the information of the $W'$, such as Refs.~\cite{Rizzo:2007xs,Wang:2008sw,Gopalakrishna:2010xm,Eboli:2011bq,Chiang:2011kq}.

In this paper, we investigate deep-learning-based approaches to tackle the problem of determining the spin and interaction type of a heavy charged boson resonance through its leptonic decay channels.  In particular, we will consider $W'$ and $H$, generic spin-1 and spin-0 charged resonances  respectively.  Over the past few years, neural networks have made enormous strides on a variety of challenging problems in different fields.  Some recent high energy physics applications include Refs.~\cite{Kasieczka:2017nvn,Brehmer:2018kdj,Brehmer:2018eca,Simola:2018ntn,Kasieczka:2018lwf,Freitas:2019hbk,Kasieczka:2019dbj,Khosa:2019kxd,Chen:2019uar}.

The above ambiguities make event-by-event reconstruction by a neural network challenging, but classification based on a collection of events can still have significant distinguishing power.  Bosons with different leptonic couplings and spins will manifest distinctive kinematic features which become apparent as one accumulates events. Thus, instead of trying to reconstruct the spins and couplings directly, we can use a multi-class neural network classifier that takes measured lab quantities of a set of events as input.    There are two straightforward ways to input this collection of events: either simply feed them in event by event as an array, or combine a number of events and form a 2D histogram of a selected pair of variables. The latter would be similar to feeding in part of the probability density function on the chosen 2D kinematic plane. Based upon these possibilities, we have considered the following three neural network (NN) models for this problem:
\begin{itemize}
\item \textit{Fully Connected Neural Network upon Individual Events} (FNNi): We constructed a fully connected neural network (FNN)  trained upon the kinematic information of individual events. To utilize the scores of this network for hypothesis testing on a group of accumulated events, we use the normalized class score product of the group.  
\item \textit{Fully Connected Neural Network upon Histograms} (FNNh): We constructed an FNN trained upon flattened 2D histograms made from pairs of kinematic observables of a certain number of events.
\item \textit{Convolutional Neural Network} (CNN): We also constructed a CNN trained upon the 2D histograms mentioned above.
\end{itemize}

These methods have already been proposed and used in Ref.~\cite{Khosa:2019kxd}\footnote{We note that there are several other studies that also use ensembles of events and/or multi-dimensional histograms to perform machine learning. See Refs.~\cite{Lai:2018ixk,Du:2019civ,Mullin:2019mmh,Flesher:2020kuy,Lazzarin:2020uvv,Lai:2020byl}.} to distinguish the mono-jet and di-jet signatures of weakly interacting massive particles (WIMPs) from those of the SM and other dark matter models.  In our study, we investigate the application of these methods to the classification of samples into the following three coupling classes\footnote{These are the interactions familiar to us in the SM.  The proposed method can be generalized to include other interactions, such as other linear combinations of $\sim W'_\mu\overline{\psi}\gamma^{\mu}(a + b \gamma_5)\chi$.  The discriminating power, of course, will depend upon how close the different coupling classes are.
}:
\begin{itemize}
 \item \textit{Vector/Axial} (VA): This class corresponds to a $W'$ with vector-like (V) fermionic couplings, $W'_\mu\overline{\psi}\gamma^{\mu}\chi$, or axial-vector-like (A) fermionic couplings, $W'_\mu\overline{\psi}\gamma^{\mu}\gamma_5\chi$.
 \item \textit{Chiral} (CH): This class corresponds to a $W'$ with left-handed (LH) fermionic couplings, $ W'_\mu\overline{\psi}\gamma^{\mu}(1-\gamma_5)\chi$, or right-handed (RH) fermionic couplings, $ W'_\mu\overline{\psi}\gamma^{\mu}(1+\gamma_5)\chi$.
 \item \textit{Scalar} (SC): This class corresponds to an $H^\pm$ with Yukawa fermionic couplings, $ H\overline{\psi}\chi$ and $H\overline{\psi}\gamma_5\chi$.
\end{itemize}
For a $pp$ collider, we will show that for signal alone the $p_T$ and $\eta$ variables of the lepton cannot distinguish between the V and A hypotheses or between the LH and RH hypotheses.  Interference between a $W'$ and the SM $W$ background could in principle break this degeneracy, yet such effects are found to be negligible for the TeV-mass bosons considered in this study.  Thus, under our approximations the VA, CH and SC hypotheses comprise three distinct signals.

We prepare the samples assuming $14$-TeV $pp$ collisions, which is the expected COM energy of HL-LHC.  Going beyond the signal-only hypothesis testing of Ref.~\cite{Khosa:2019kxd}, we will also include the SM background from the $W$ boson.  We will investigate scenarios of different $S/B$ ratios, assuming the HL-LHC integrated luminosity of $\mathcal{L}=3\ \text{ab}^{-1}$.

To choose the masses for our study, we use Ref.~\cite{Aad:2019wvl} to determine the 95\% C.L.~cross section upper limits of different charged resonance masses.  Since we anticipate that our technique requires $S/B \agt 1$ to be effective, we consider masses where the current cross section limits allow $S/B \sim 1$ and where we can still expect to get a 5$\sigma$ discovery in the HL-LHC era.    
These conditions force the mass to be $\geq4.5$~TeV, so we will focus on the mass of $4.5$-TeV.  As a comparison, we will also explore $6$-TeV resonances, where the signal purity can be higher but the hypothesis testing is more challenging due to low statistics.

We only study the $e\nu$ decay channel, though this method can be readily applied to the $\mu \nu$ channel and improve its efficiency.  Also, we assume that the coupling strength and structure are universal to all generations in both quark and lepton sectors (even for $H^\pm$).

We also take into consideration the effects of different boson resonance widths, varying from $500$, $200$ to $50$~GeV for the $4.5$-TeV resonances.  It is observed that the training outcomes upon different widths are quite similar.  We will focus on the samples of $200$~GeV width, a choice to mimic the SM $W$ width-to-mass ratio $\Gamma_W/m_W\approx1/40$, in most of our presentation below. As for the $6$-TeV resonances, we only study the case of $300$~GeV width.

Beside the 0-jet process, we have also studied the 1-jet process in which an extra jet is included in the final state. Since in real experiments jets can be copiously produced through either soft radiation or hard interactions, we consider all processes of jet multiplicities up to 2, and extract from them the 0-jet and 1-jet samples with criteria to be mentioned in Sec.~\ref{sec:3}. To make use of the extra information provided by the jet, we will further extend the 2D histogram inputs to include more variable pairs by using ``RGB'' colors to demonstrate that the histogram approach of Ref.~\cite{Khosa:2019kxd} can be generalized to higher dimensions.  We will formulate a few different input schemes for these 1-jet histograms, although there is no major performance difference among them.  To understand the results, we will also study the importance and contributions of the different variable pairs in these schemes.  It is worth noting here that for situations involving more kinematic variables like the current study, our results show that the NN approach is more convenient than and superior to conventional methods, such as Bayesian hypothesis or $\chi^2$ tests.

In the Appendix, we further provide detailed technical studies of the NN performance when the bin resolution and kinematic window are varied.  In addition, we compare the performances of binary classifiers to those of the original ternary classifiers by performing a \textit{projection} on the testing scores of the latter, which demonstrates that our ternary classifier is as capable as individual binary classifiers.  Finally, we investigate the results of applying to the testing samples models trained for incorrect assumption of significance or decay width, testing the flexibility of our methods.

This paper is organized as the following. In Sec.~\ref{sec:2}, we briefly review the kinematic properties of bosons of different coupling classes.  In Sec.~\ref{sec:3}, we discuss the 0-jet and 1-jet samples and analyze their kinematic features.  In Sec.~\ref{sec:4}, we describe the details of our NN models as well as the training specifications.  In Sec.~\ref{sec:5}, we present and discuss the 0-jet and 1-jet training results.  In Sec.~\ref{sec:6}, we compare our NN method with the Bayesian hypothesis test and discuss the pros and cons.  In Sec.~\ref{sec:7}, we draw conclusions and propose possible further studies.  More technical details of our investigations are provided in Appendix~\ref{sec:a}.

\section{Parton-level analysis of general singly-charged bosons}\label{sec:2}

Consider the following processes:
\begin{equation}
	pp\rightarrow W/{W'}/H\rightarrow e\nu_e
	~.
\end{equation}
The corresponding $p_T$ and $\eta$ differential cross sections of $e$ are given by
\begin{equation}
\label{Diff:Proton}
\begin{aligned}
	\frac{d\sigma}{d\chi} = \sum_{q,q'}\int &dxdy\frac{d\hat{\sigma}(x,y)}{d\chi}
	\cdot q(x,Q^2)\bar{q}'(y,Q^2) ~, 
	\qquad (\chi=p_T,\eta)
\end{aligned}
\end{equation}
where $q(x,Q^2)$, $\bar{q}'(y,Q^2)$ are the parton distribution functions (PDFs).

The parton-level $p_T$ and $\eta$ differential cross sections for $H$ and $W'$ are given respectively by  
\begin{subequations}
\begin{align}
	\frac{d\hat{\sigma}_{H}}{dp_T} &= \frac{1}{2\pi}\frac{y_H^4}{(p^2-m_H^2)^2+m_H^2\Gamma_H^2}\frac{p_T}{\sqrt{1-\frac{4p_T^2}{p^2}}}
	~,
	\label{Diff:PT:Parton:H} \\
	\frac{d\hat{\sigma}_{W'}}{dp_T} &= \frac{1}{2\pi}\frac{2 \left( c_V^2+c_A^2 \right)^2 \left( 1-\frac{2p_T^2}{p^2} \right)}{(p^2-m_{W'}^2)^2+m_{W'}^2\Gamma_{W'}^2}\frac{p_T}{\sqrt{1-\frac{4p_T^2}{p^2}}}
	~,
	\label{Diff:PT:Parton:W}
\end{align}
\end{subequations}
and
\begin{subequations}
	\begin{align}
\frac{d\hat{\sigma}_{H}}{d\eta} &= \frac{\sech^2\eta}{32\pi}\frac{128E_1^2E_2^2}{(p^2-m_H^2)^2+m_H^2\Gamma_H^2}\cdot y_H^4\frac{F(E_1,E_2,\eta)}{G^2(E_1,E_2,\eta)}
~,
\label{Diff:Parton:General:H} 
\\
\frac{d\hat{\sigma}_{W'}}{d\eta} &= \frac{\sech^2\eta}{32\pi}\frac{128E_1^2E_2^2}{(p^2-m_{W'}^2)^2+m_{W'}^2\Gamma_{W'}^2} 
\left\{2(c_V^2+c_A^2)^2\left[\frac{I(E_1,E_2,\eta)}{H(E_1,E_2,\eta)}+\frac{I(E_2,E_1,\eta)}{H(E_2,E_1,\eta)}\right]
\right.
\nonumber \\
&\qquad\qquad\qquad \left.
+ 4c_V^2c_A^2\left[\frac{J(E_1,E_2,\eta)}{H(E_1,E_2,\eta)}+\frac{J(E_2,E_1,\eta)}{H(E_2,E_1,\eta)}\right]\right\} 
~,
\label{Diff:Parton:General:W}
	\end{align}
\end{subequations}
where $p^2=xys$, $E_1=\frac{x\sqrt{s}}{2}$, $E_2=\frac{y\sqrt{s}}{2}$, $\sqrt{s} = 14$~TeV and $F,G,H,I,J$ are given by
\begin{align}
\begin{split}
	F(A,B,\eta) &\equiv(A+B)^2+(A-B)^2\tanh^2\eta ~, \\
	G(A,B,\eta) &\equiv(A+B)^2-(A-B)^2\tanh^2\eta ~, \\
	H(A,B,\eta) &\equiv[(A+B)-(A-B)\tanh\eta]^4 ~, \\
	I(A,B,\eta) &\equiv A^2(1-\tanh\eta)^2+B^2(1+\tanh\eta)^2 ~, \\
	J(A,B,\eta) &\equiv A^2(1-\tanh\eta)^2-B^2(1+\tanh\eta)^2 ~.
\end{split}
\end{align}

From these parton-level differential cross sections, one can tell $H$ and $W'$ apart from the $p_T$ distributions alone.  However, the $W'$ bosons of different coupling structures would give identical $p_T$ distributions up to the normalization $(c_V^2+c_A^2)^2$ factor in Eq.~(\ref{Diff:PT:Parton:W}).  On the other hand, the second term in the curly brackets of Eq.~(\ref{Diff:Parton:General:W}) is proportional to $c_V^2c_A^2$ and would lead to distinct $\eta$ distributions for different $W'$ coupling scenarios.  Thus, combining the {\it parton-level} $p_T$ and $\eta$ distributions, one should be able to readily distinguish among the three classes but cannot distinguish between V and A nor between LH and RH from the shape of the distributions alone.  After convoluting with the PDF's, the distribution differences among the classes become less obvious, but will still be detectable through our technique.

\section{Sample generation and analysis}\label{sec:3}

We prepare our parton-level samples using \texttt{MadGraph5\_aMC@NLO v2.7.3}~\cite{MG5}, followed by parton shower and hadronization performed with \texttt{Pythia 8.2.44}~\cite{PY8,PY8_2}. To properly interface these two softwares as we include processes of jet multiplicities of 0 - 2, we utilize MLM matching with a jet merging scale of $30$~GeV. The cuts imposed at the generator level are summarized in TABLE \ref{Sample:cut}. The selection cut is imposed to suppress the SM $W$ background while retaining a sufficient amount of the new-physics (NP) signals below the Jacobian peak at $p_T^\ell = m_{W'} / 2 = m_{H} / 2$.  This $p_T$ cut is a practical one so that the NN training samples are not background dominated at the low end of this cut, which assists in training while allowing our $p_T$ binning to be sufficiently high in resolution.  In the Appendix, we will explore how the NN performance depends on the $p_T$ cut and show that there can be a trade-off between information loss (too high of a cut) and $p_T$ resolution (too low of a cut).

\begin{table}[!ht]
\centering
\begin{tabular}{c|c}
\toprule
Basic cuts & ~~$p_T^j >30$~GeV ~;~ $\vert\eta^j\vert<5.0$ ~;~ $\left| \eta^\ell \right| <4.0$~~~ \\
\colrule
~~Selection cuts~~ & $p_T^\ell,\cancel{E}_T > 0.3m_{W',H}$~~~ \\
\botrule
\end{tabular}
\caption{Summary of cuts imposed on the samples at the generator level.}
\label{Sample:cut}
\end{table}

The samples are then passed to \texttt{Delphes 3.4.2}~\cite{Delphes,Delphes_2,Delphes_3} for detector simulation using the Phase-II CMS card. The events are reconstructed with \texttt{FastJet 3.3.2}~\cite{FastJet}. In particular, the final-state jets are reconstructed using the anti-$k_T$ clustering algorithm~\cite{anti-kt} with the cone radius $R=0.4$.

The processes are simulated for $14$-TeV LHC collisions with the \texttt{NNPDF23\_nlo\_as\_0119}~\cite{NNPDF} PDF set.  The $W'$- and $H$-mediated processes are generated respectively with the Wprime model and General 2HDM from the \texttt{FeynRules}~\cite{FeynRules} model database.  In what follows, we describe the details of the 0- and 1-jet samples.

\subsection{0-jet samples}\label{subsec:3:1}

The 0-jet samples simply include all events with an observable electron and have $\geq 0$ jets. For these samples, we only make use of the electron observables and ignore all the jet information. We denote the new boson width by $\Gamma_{\rm NP}$ and consider three different values: $500$, $200$, and $50$~GeV for $4.5$-TeV resonances. We will show in Sec.~\ref{sec:5} that the width varying in this range does not affect the training outcomes much, and thus we consider only $\Gamma_{\rm NP}\approx300$~GeV for the heavier $6$-TeV resonances.  At the generator level, we generate $0.5$M events for each of the VA, CH, SC, and SM classes.  After detector simulation, the successfully tagged event numbers of all three NP classes, including different widths and masses, and the SM class are all roughly around $300$K.

\begin{figure}[H]
\centering
\begin{subfigure}[t!]{0.4\textwidth}
	\includegraphics[width=\textwidth]{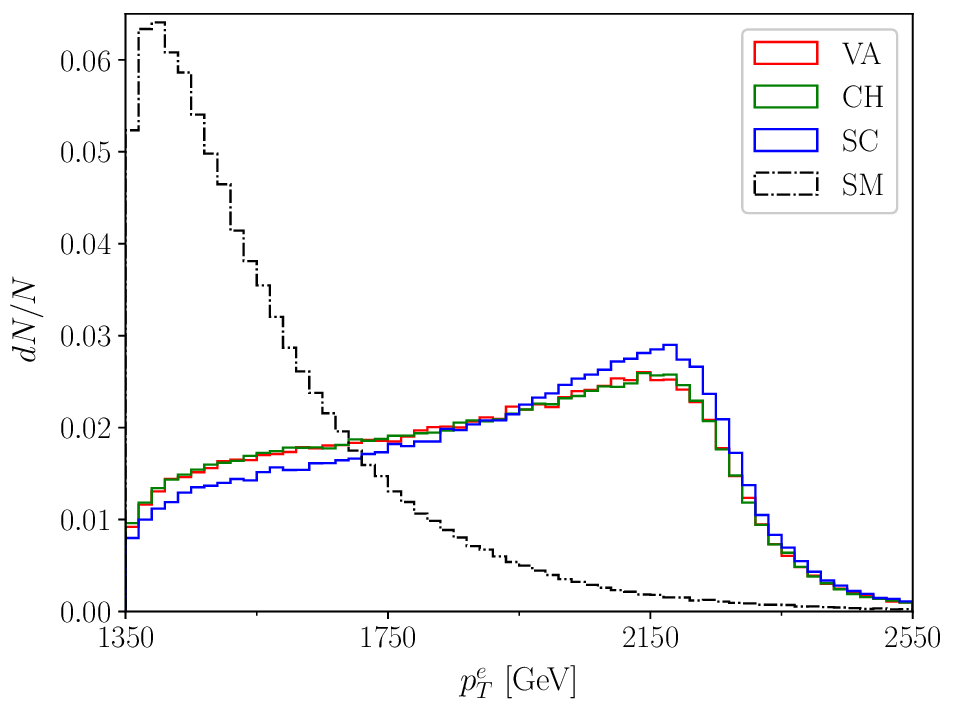}
	\vspace{-1.2cm}
	\caption{}
\end{subfigure}
\begin{subfigure}[t!]{0.4\textwidth}
	\includegraphics[width=\textwidth]{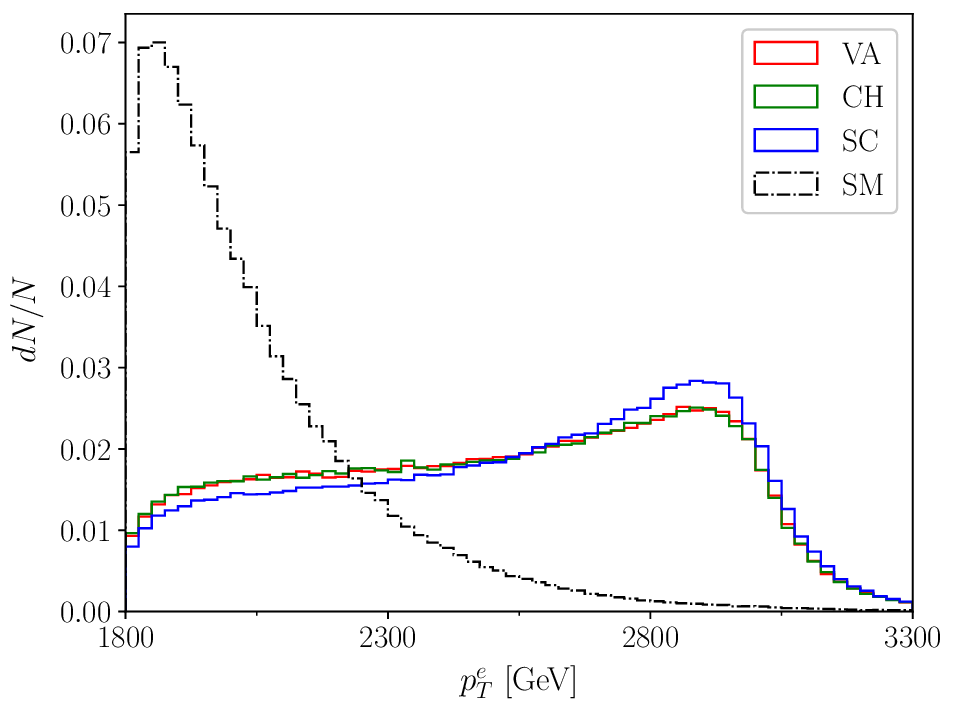}
	\vspace{-1.2cm}
	\caption{}
\end{subfigure}
\begin{subfigure}[t!]{0.4\textwidth}
	\includegraphics[width=\textwidth]{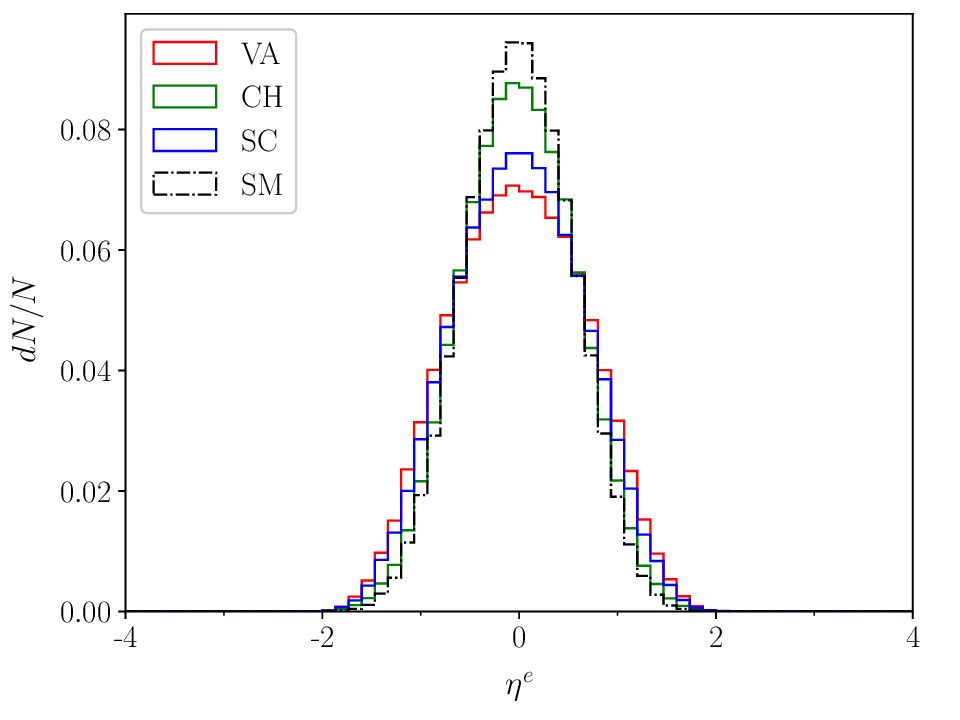}
	\vspace{-1.2cm}
	\caption{}
\end{subfigure}
\begin{subfigure}[t!]{0.4\textwidth}
	\includegraphics[width=\textwidth]{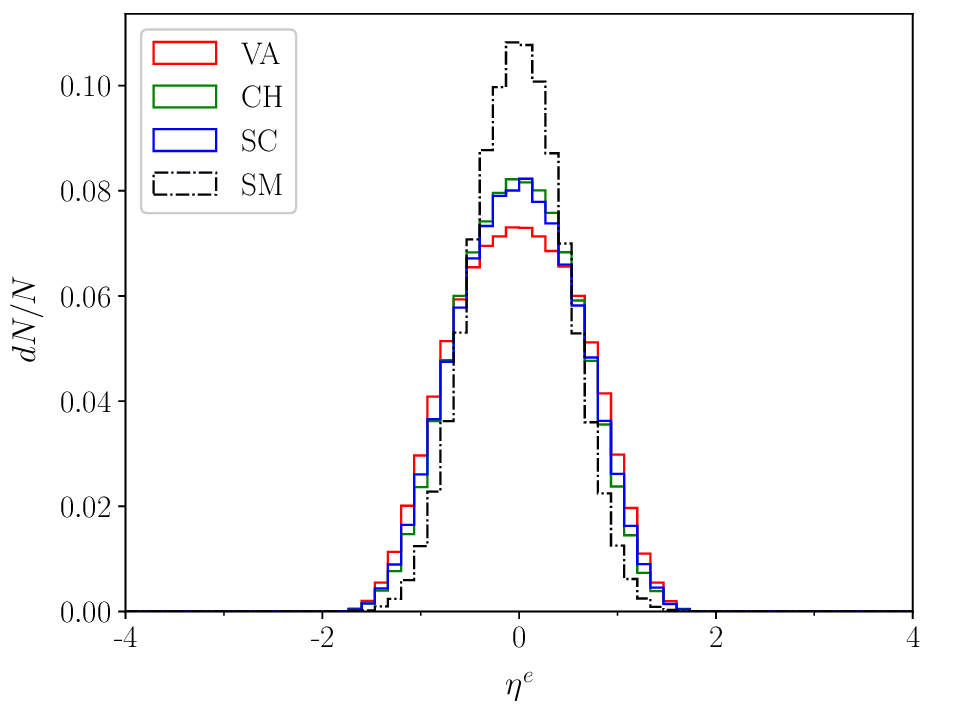}
	\vspace{-1.2cm}
	\caption{}
\end{subfigure}
\begin{subfigure}[t!]{0.4\textwidth}
	\includegraphics[width=\textwidth]{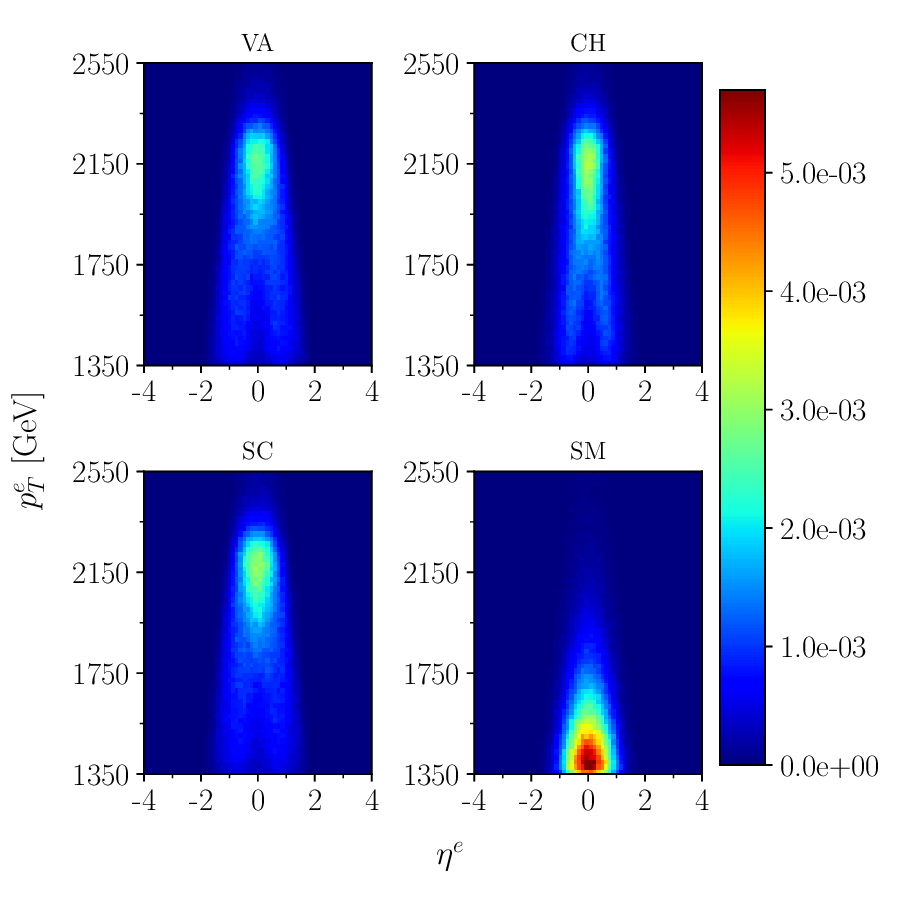}
	\vspace{-1.2cm}
	\caption{}
\end{subfigure}
\begin{subfigure}[t!]{0.4\textwidth}
	\includegraphics[width=\textwidth]{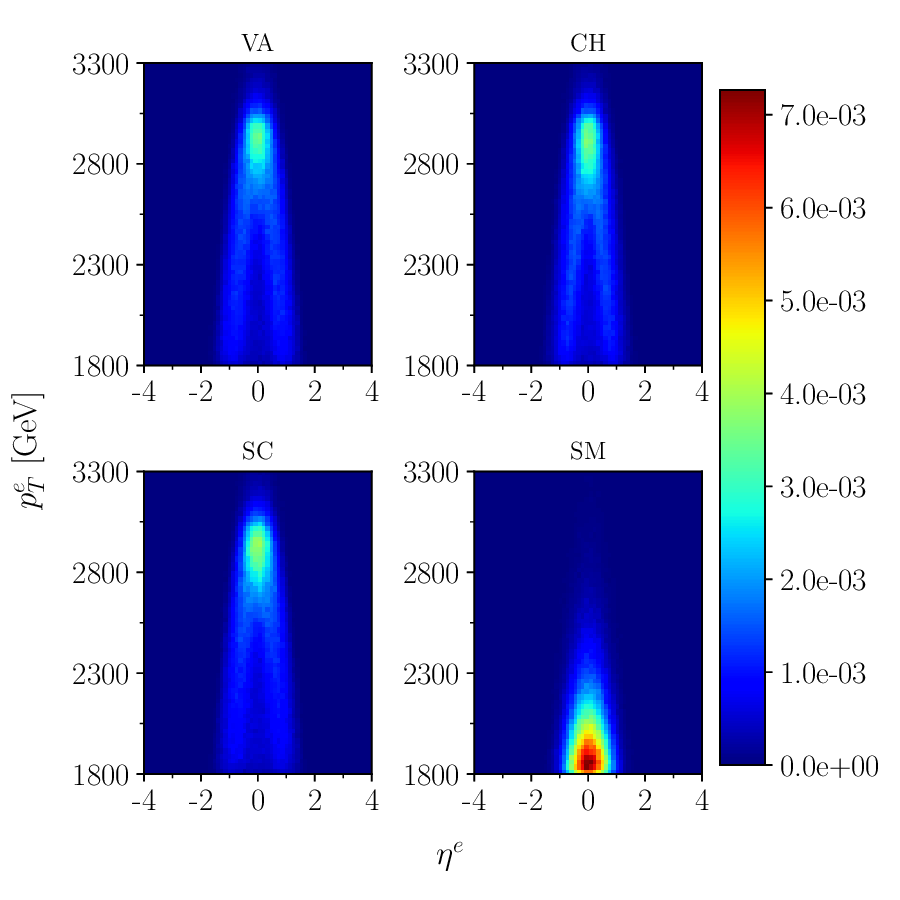}
	\vspace{-1.2cm}
	\caption{}
\end{subfigure}
\caption{(a) $p_T^e$, (c) $\eta^e$, (e) $p_T^e$ vs. $\eta^e$ distributions for the $4.5$-TeV 0-jet samples with $\Gamma_{\rm NP}\approx200$~GeV and (b) $p_T^e$, (d) $\eta^e$, (f) $p_T^e$ vs. $\eta^e$ distributions for the $6$-TeV 0-jet samples with $\Gamma_{\rm NP}\approx300$~GeV. In plots (a), (b), (c) and (d), VA is depicted in red, CH in green, SC in blue, and SM in black. In plot (e) and (f), the color scale range goes from 0 to the maximum entry among all four classes, with the warmer/colder regions denoting more/fewer entries. The same color scheme is applied to all the following figures. All the distributions are normalized to unity. }
\label{LO:Kinematics}
\end{figure}

We choose to divide both $p_T^e$ and $\eta^e$ into $60$ bins so that our NNs remain trainable. We only show the corresponding unit-normalized $p_T^e$, $\eta^e$, and $p_T^e$ vs. $\eta^e$ distributions for $\Gamma_{\rm NP} \approx 200$~GeV for $4.5$-TeV (left column) and for $6$-TeV (right column) resonances in FIG.~\ref{LO:Kinematics}.  As the boson width increases, the Jacobian peak in the $p_T^e$ distribution would become broader, while the $\eta^e$ distribution would remain identical.

As discussed in Sec.~\ref{sec:2}, na{\"i}vely the $p_T$ curves of the VA and CH classes should be identical in FIG.~\ref{LO:Kinematics}.  However, there is a slight difference between the two due to the $\eta^e$ cut mentioned in TABLE~\ref{Sample:cut}. Since there is a much larger difference in the $\eta^e$ distributions, we do not expect this difference to strongly affect the training or performance of our classifiers. The same issue will also occur in the 1-jet case.

The color scheme for FIG.~\ref{LO:Kinematics}(e) and \ref{LO:Kinematics}(f), and also for the remaining 2D histograms, are as follows: the coldest color (blue) denotes a $0$ entry, while the warmest color (red) denotes the maximum entry among all four classes. As shown in the plots, the Jacobian peaks are at around $p_T=m_{W',H}/2$ for all the NP classes, with the CH class possessing the longest tail toward low $p_T^e$, while the VA and SC class have similar tails but with different $\eta^e$ distribution. Such differences in the $p_T^e$ tail and the $\eta^e$ distribution show the kinematic information that can be used to distinguish among the three classes, even after including the background.

Within the selected phase space, the expected number of SM 0-jet events are
\begin{equation}
\label{num:B:LO}
	B_0=\sigma_{B_0}\times\mathcal{L}\approx\begin{cases}
	84 & \text{ for }4.5~\text{TeV} \\
	7 & \text{ for }6~\text{TeV}
	\end{cases}
	~.
\end{equation}
Thus, the total number of events we expect to observe is 
\begin{equation}
\label{num:total:LO}
	N_0 = B_0\times\left(1+\frac{S_0}{B_0}\right)
	~,
\end{equation}
where $S_0$ denotes the number of signal events.  We will vary the signal-to-background ratio $S_0 / B_0$ in our considerations.

We study scenarios of different $S_0/B_0$ ratios within the range specified as the following: given our selection criteria, the lower bound is set by the requirement that a $\geq 5\sigma$ excess by the end of the HL-LHC is to be expected and the upper bound is set by the current upper limit on cross section from ATLAS~\cite{Aad:2019wvl}. The corresponding range for $4.5$-TeV resonances is $S_0/B_0 \in [0.6,1.0]$, while that for $6$-TeV is $S_0/B_0 \in [2.5,5.5]$. We extend both ranges a little bit to better understand the trend of varying $S_0/B_0$, hence extending the ranges to $[0.4,1.2]$ and $[1.5,6.0]$, respectively.  For $4.5$-TeV resonances, we shuffle the samples repetitively until $15K$ histograms per class are generated.  As for $6$-TeV resonances, we generate $50K$ histograms per class to make up for the low event statistics in individual histograms. The same setting is also applied to 1-jet scenarios.

\begin{figure}[H]
\centering
\begin{subfigure}[h]{0.48\textwidth}
	\includegraphics[width=\textwidth]{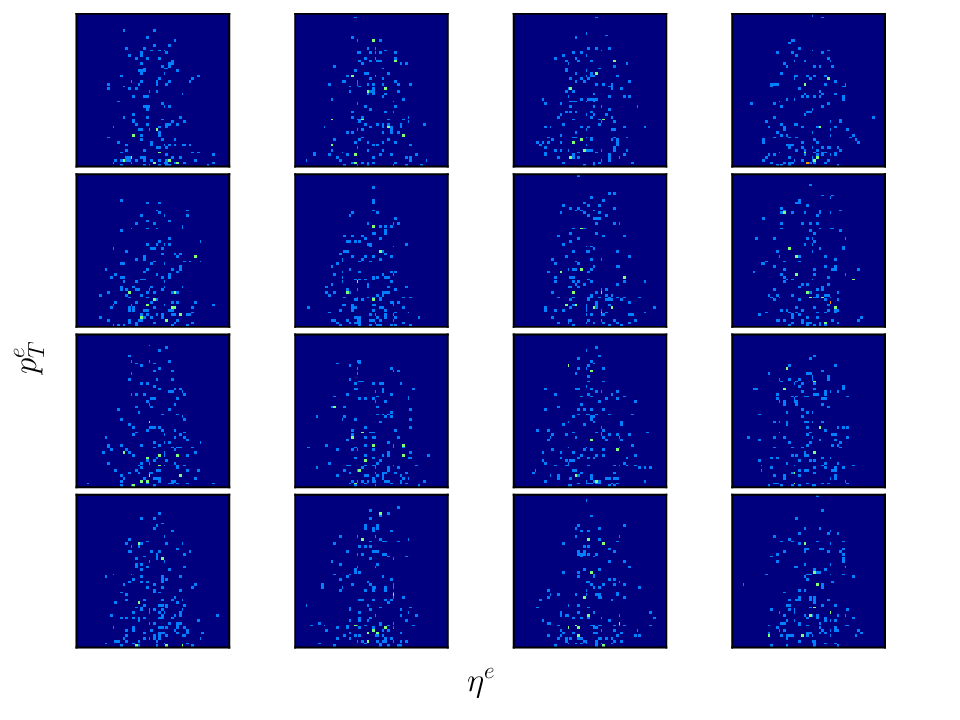}
	\vspace{-1.2cm}
	\caption{}
\end{subfigure}
\begin{subfigure}[h]{0.48\textwidth}
	\includegraphics[width=\textwidth]{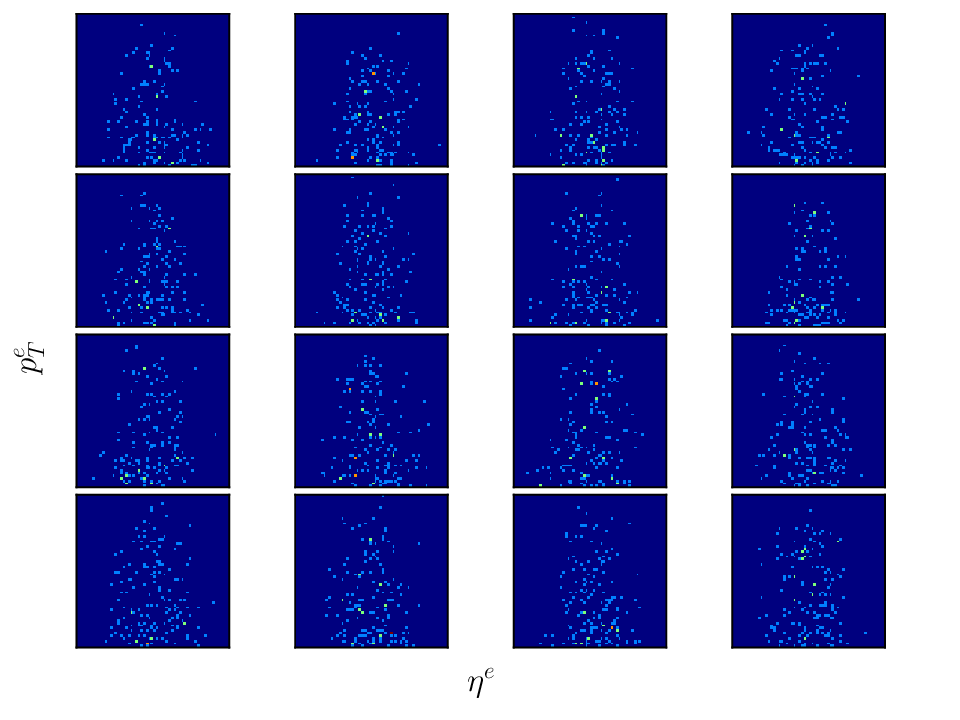}
	\vspace{-1.2cm}
	\caption{}
\end{subfigure}
\begin{subfigure}[h]{0.48\textwidth}
	\includegraphics[width=\textwidth]{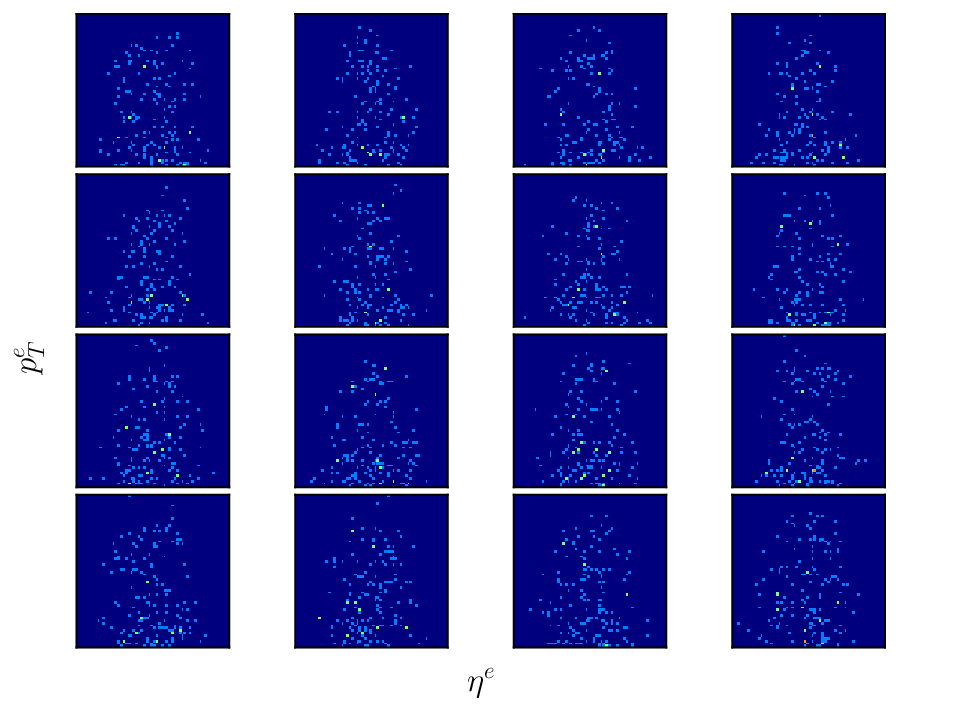}
	\vspace{-1.2cm}
	\caption{}
\end{subfigure}
\caption{Examples of 0-jet input histograms for (a) VA, (b) CH, and (c) SC samples for $4.5$-TeV resonances of $\Gamma_{NP}=200$~GeV with $S_0/B_0=1.0$.}
\label{Sample:LO}
\end{figure}

As seen in FIG.~\ref{LO:Kinematics}(c) and FIG.~\ref{LO:Kinematics}(d), $\eta^e$ is mostly confined within $[-2,2]$.  We therefore only bin the data within this range when making the histograms. The same procedure is also applied to the 1-jet samples. A few $p_T^e~\text{vs.}~\eta^e$ sample histograms for $4.5$-TeV resonances of $\Gamma_{NP}=200$~GeV with $S_0/B_0=1.0$ are shown in FIG.~\ref{Sample:LO}.  Note that it is quite challenging to distinguish them by eye at high accuracy but will be a manageable job for the NNs.

\subsection{1-jet samples}\label{subsec:3:2}

The 1-jet samples include those events that have a leading jet with $p_T^j>30$~GeV and are a subset of the 0-jet samples. Such events take up roughly 83\% of all NP samples and 69\% of the SM samples. For these samples, we ignore any sub-leading jet information. Therefore, the SM 1-jet event numbers within this phase space are given by
\begin{align}
\label{sigma:B:NLO}
	B_1=\sigma_{B_1}\times\mathcal{L}\approx\begin{cases}
	58 &\text{ for }4.5~\text{TeV} \\
	4 &\text{ for }6~\text{TeV}
	\end{cases}
	~.
\end{align}
Hence, the corresponding $S_1/B_1$ are scaled up from $S_0/B_0$ by a factor of $0.83/0.69 \simeq 1.2$. For the convenience of an easy comparison with the 0-jet analysis, we will still label the signal-to-background ratio of 1-jet samples by the 0-jet $S_0/B_0$ ratio, even though the true mixing ratio is $S_1/B_1$.

\begin{figure}[H]
\centering
\begin{subfigure}[!t]{0.4\textwidth}
	\includegraphics[width=\textwidth]{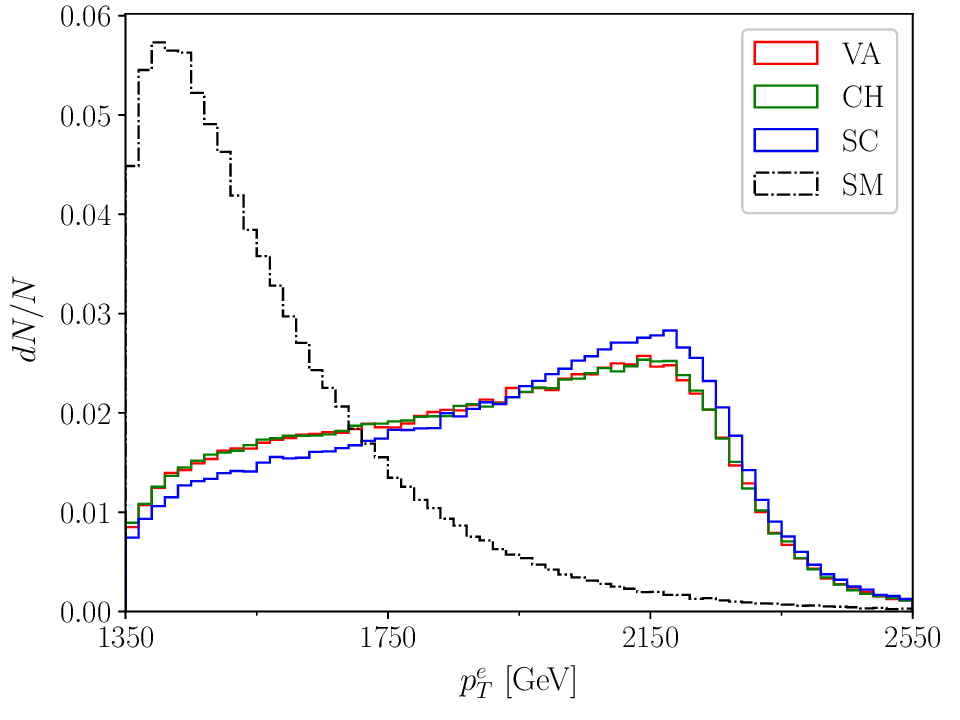}
	\vspace{-1.2cm}
	\caption{}
\end{subfigure}
\begin{subfigure}[!t]{0.4\textwidth}
	\includegraphics[width=\textwidth]{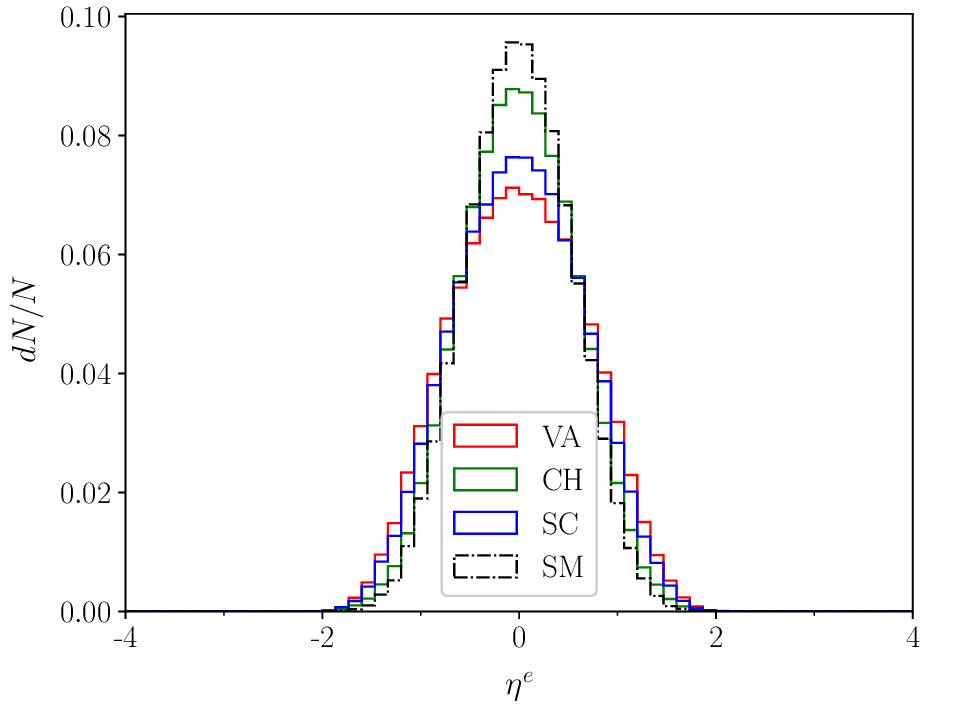}
	\vspace{-1.2cm}
	\caption{}
\end{subfigure}
\begin{subfigure}[!t]{0.4\textwidth}
	\includegraphics[width=\textwidth]{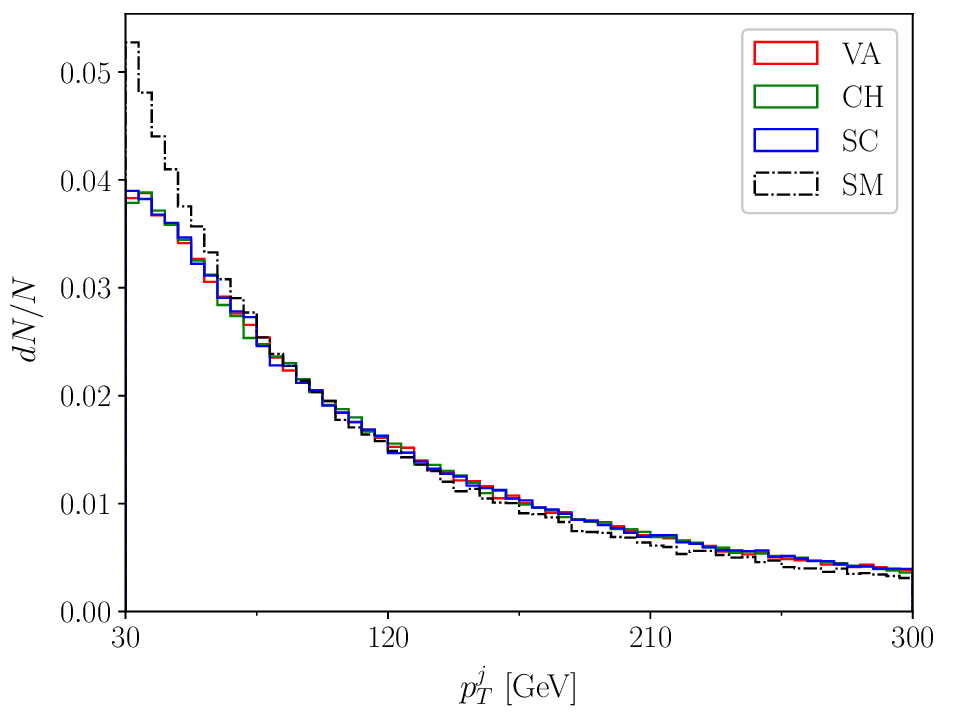}
	\vspace{-1.2cm}
	\caption{}
\end{subfigure}
\begin{subfigure}[!t]{0.4\textwidth}
	\includegraphics[width=\textwidth]{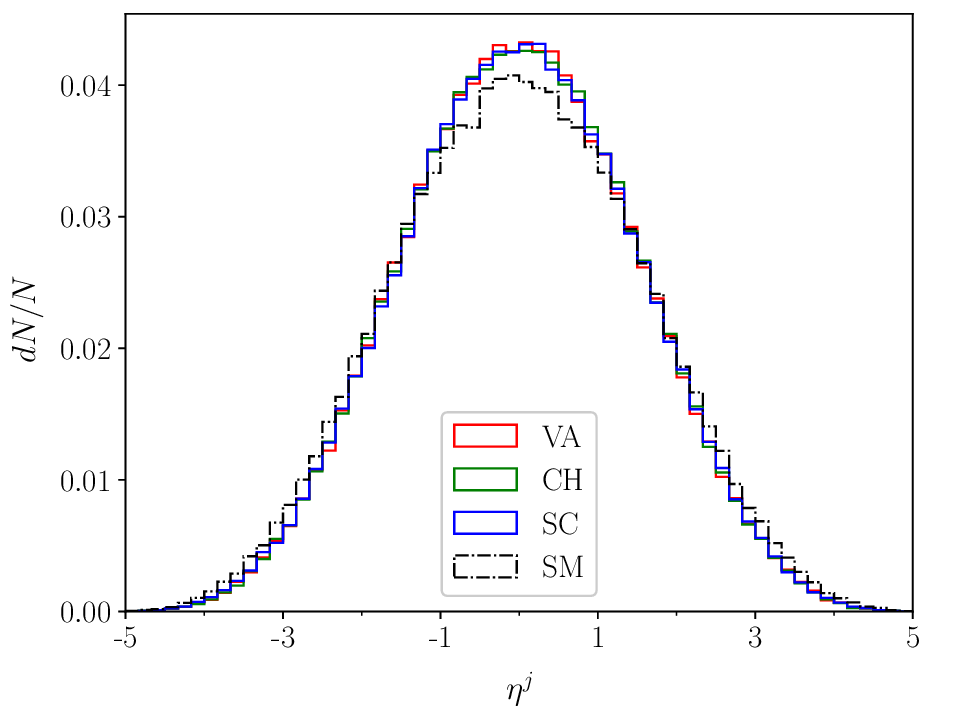}
	\vspace{-1.2cm}
	\caption{}
\end{subfigure}
\begin{subfigure}[!t]{0.4\textwidth}
	\includegraphics[width=\textwidth]{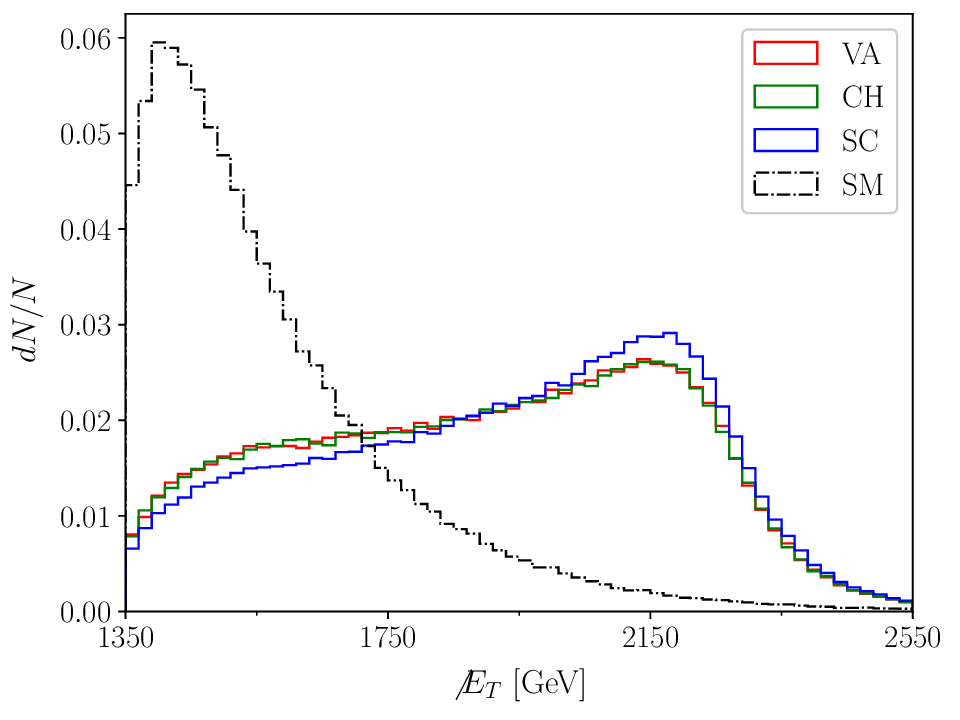}
	\vspace{-1.2cm}
	\caption{}
\end{subfigure}
\begin{subfigure}[!t]{0.4\textwidth}
	\includegraphics[width=\textwidth]{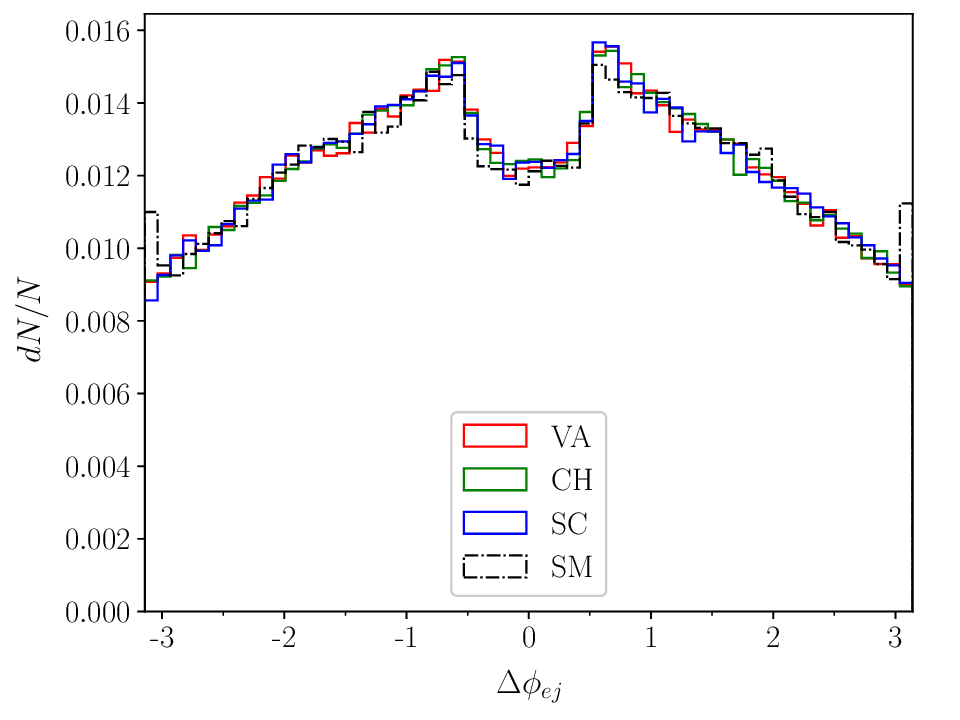}
	\vspace{-1.2cm}
	\caption{}
\end{subfigure}
\begin{subfigure}[!t]{0.4\textwidth}
	\includegraphics[width=\textwidth]{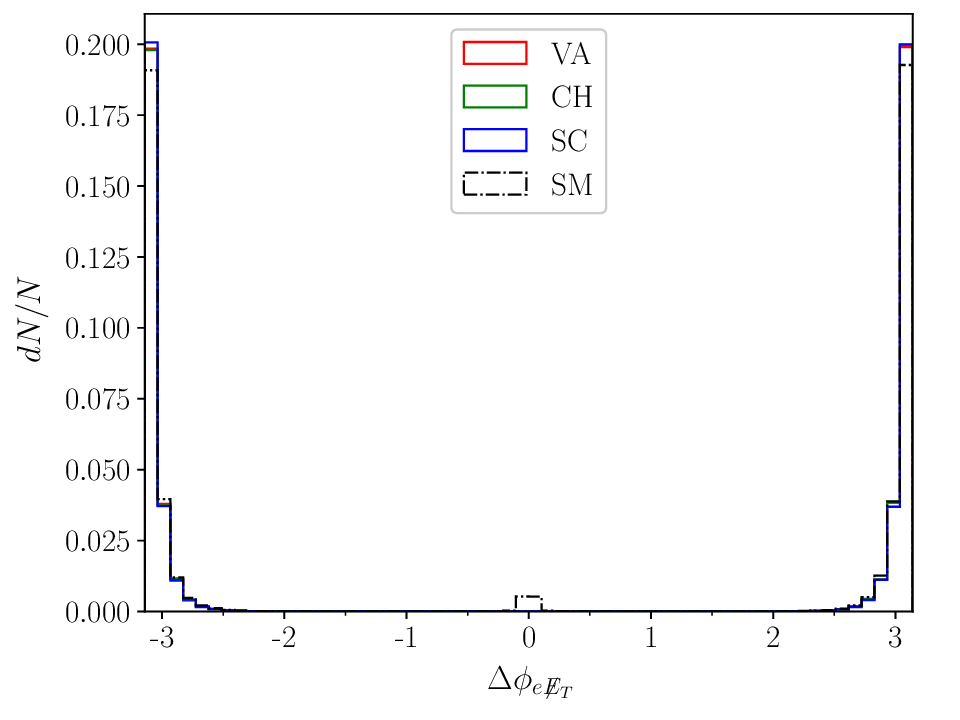}
	\vspace{-1.2cm}
	\caption{}
\end{subfigure}
\begin{subfigure}[!t]{0.4\textwidth}
	\includegraphics[width=\textwidth]{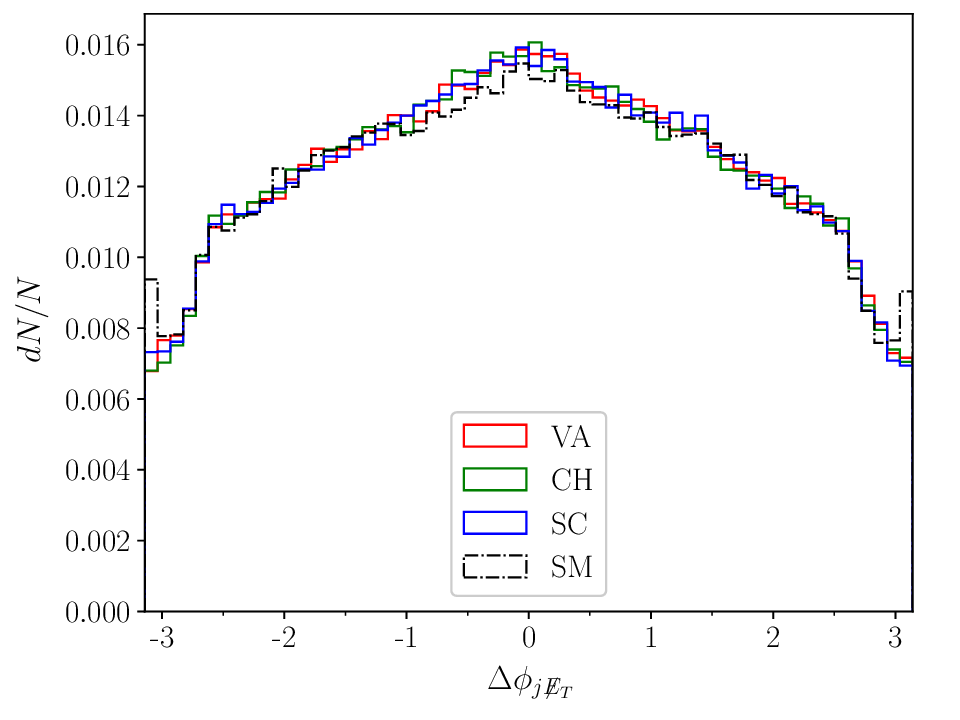}
	\vspace{-1.2cm}
	\caption{}
\end{subfigure}
\caption{Distributions of kinematic observables: (a) $p_T^e$, (b) $\eta^e$, (c) $p_T^j$, (d) $\eta^j$, (e) $\cancel{E}_T$, (f) $\Delta\phi_{ej}$, (g) $\Delta\phi_{e\cancel{E}_T}$, (h) $\Delta\phi_{j\cancel{E}_T}$ for 1-jet samples of mass $4.5$~TeV and $\Gamma_{\rm NP}\approx200$~GeV.}
\label{NLO:Kinematics}
\end{figure}

The kinematic observables of a 1-jet process are:
\begin{itemize}
	\item $p_T^{e}$ and $p_T^{j}$: transverse momenta of $e$ and leading jet $j$, respectively.
	\item $\eta^{e}$ and $\eta^{j}$: pseudorapidities of $e$ and $j$, respectively.
	\item $\Delta\phi_{ej}$: azimuthal separation between $e$ and $j$.
\end{itemize}
To form the required histograms and at the same time to involve as much information as possible, we further consider three derived observables:
\begin{itemize}
	\item $\cancel{E}_T$: missing transverse energy.
	\item $\Delta\phi_{e\cancel{E}_T}$ and $\Delta\phi_{j\cancel{E}_T}$: azimuthal separations between $e$ and $\cancel{E}_T$ and between $j$ and $\cancel{E}_T$, respectively.
\end{itemize}
We show the distributions of these kinematic observables for $4.5$-TeV resonances in FIG.~\ref{NLO:Kinematics}.

To utilize the additional information contained in these kinetic observables, we will make ``RGB'' histograms by choosing three pairs of variables.  We propose the following four schemes:
\begin{itemize}
	\item Scheme 1 -- \textit{Physical Relationship}: Intuitively, the kinematic information measured from a single object should manifest high correlation. Therefore, we first pair up $p_T^e$ and $\eta^e$ as well as $p_T^j$ and $\eta^j$. Then, guessing that observables of the same mass dimension could be correlated, we choose two out of the three azimuthal separation variables, $\Delta\phi_{e\cancel{E}_T}$ and $\Delta\phi_{j\cancel{E}_T}$, to form the third pair.
	
	\item Scheme 2 -- \textit{Principal Component Analysis}: Following Ref.~\cite{Khosa:2019kxd}, we also select another three pairs of variables by performing a principal component analysis (PCA). The results are shown in TABLE~\ref{PCA}. We start from the principal component (PC) with the highest variance. In each PC, we select the two variables with the highest (absolute) correlations to form a pair. Thus, from PC-1, we pair up $p_T^e$ and $\cancel{E}_T$; and from PC-2, we pair up $\Delta\phi_{ej}$ and $\Delta\phi_{e\cancel{E}_T}$.  Since $\Delta\phi_{ej}$ is already paired, we skip PC-3 and use PC-4 to pair up $\eta^e$ and $\eta^j$.
	
	\item Scheme 3 -- \textit{Common Axis}: In this scheme, we investigate whether spatial correlation among the RGB channels provides better discriminating power. If we set one of the two axes of the three channels to always be $p_T^e$, the NN can then possibly make use of the correlations of the other variables to $p_T^e$, as it now becomes physically meaningful to compare the corresponding pixels with a common $p_T^e$ coordinate.  In light of this, we choose the following three pairs for this scheme: $p_T^e$ and $\eta^e$, $p_T^e$ and $\cancel{E}_T$, and $p_T^e$ and $\Delta\phi_{ej}$.  
	
	\item Scheme 4 -- \textit{Best Individuals}: After obtaining the training results of all these individual pairs (also including the $\Delta\phi_{e\cancel{E}_T}$~vs.~$\Delta\phi_{j\cancel{E}_T}$ pair omitted in Scheme 2), to be shown in Sec.~\ref{sec:5}, we further combine the three most powerful pairs to formulate the scheme using $p_T^e$~vs.~$\eta^e$, $\eta^e$~vs.~$\eta^j$, and $p_T^e$~vs.~$\cancel{E}_T$.
\end{itemize}
\begin{table}[h!]
\centering
\begin{tabular}{c|c||c|c|c|c|c|c|c|c}
\toprule
\multirow{2}{*}{\hfill} & \multirow{2}{*}{~Variance~} & \multicolumn{8}{c}{Correlations} \\ \cline{3-10}
& & ~~~$p_T^{e}$~~~ & ~~~$\eta^{e}$~~~ & ~~~$p_T^j$~~~ & ~~~$\eta^j$~~~ & ~~~$\cancel{E}_T$~~~ 
& ~~$\Delta\phi_{ej}$~~ & ~~$\Delta\phi_{e\cancel{E}_T}$~~ & ~~$\Delta\phi_{j\cancel{E}_T}$~~ \\
\botrule
PC-1 & 1.78 & 0.707 & 0.001 & 0.040 & 0.003 & 0.706 & 0.009 & -0.020 & -0.015 \\
\colrule
PC-2 & 1.73 & -0.019 & 0.001 & -0.001 & -0.001 &-0.019 & 0.473 & -0.760 & -0.446 \\
\colrule
PC-3 & 1.27 & 0.003 & -0.000 & -0.001 & 0.000 & 0.004 & 0.695 & 0.011 & 0.719 \\
\colrule
PC-4 & 1.01 & -0.001 & -0.706 & -0.014 & 0.708 & 0.000 & 0.000 & -0.001 & -0.001 \\
\colrule
PC-5 & 0.999 & -0.011 & -0.110 & 0.989 & -0.089 & -0.044 & 0.001 & 0.000 & 0.001 \\
\colrule
PC-6 & 0.991 & -0.003 & 0.699 & 0.140 & 0.701 & -0.008 & 0.000 & 0.000 & 0.000 \\
\colrule
PC-7 & 0.221 & -0.707 & 0.000 & 0.024 & 0.000 & 0.706 & 0.000 & 0.000 & 0.000 \\
\colrule
PC-8 & 0.000 & 0.000 & 0.000 & 0.000 & 0.000 & 0.000 & -0.542 & -0.650 & 0.533 \\
\botrule
\end{tabular}
\captionsetup{singlelinecheck = false, justification=raggedright}
\caption{PCA result on 1-jet samples. The correlations indicate the linear components of the principal components (PCs). The higher the variance is in its absolute value, the more significant the component contributes to the diversity of the samples.}
\label{PCA}
\end{table}

\begin{figure}[H]
\centering
\begin{subfigure}[t]{0.4\textwidth}
	\includegraphics[width=\textwidth]{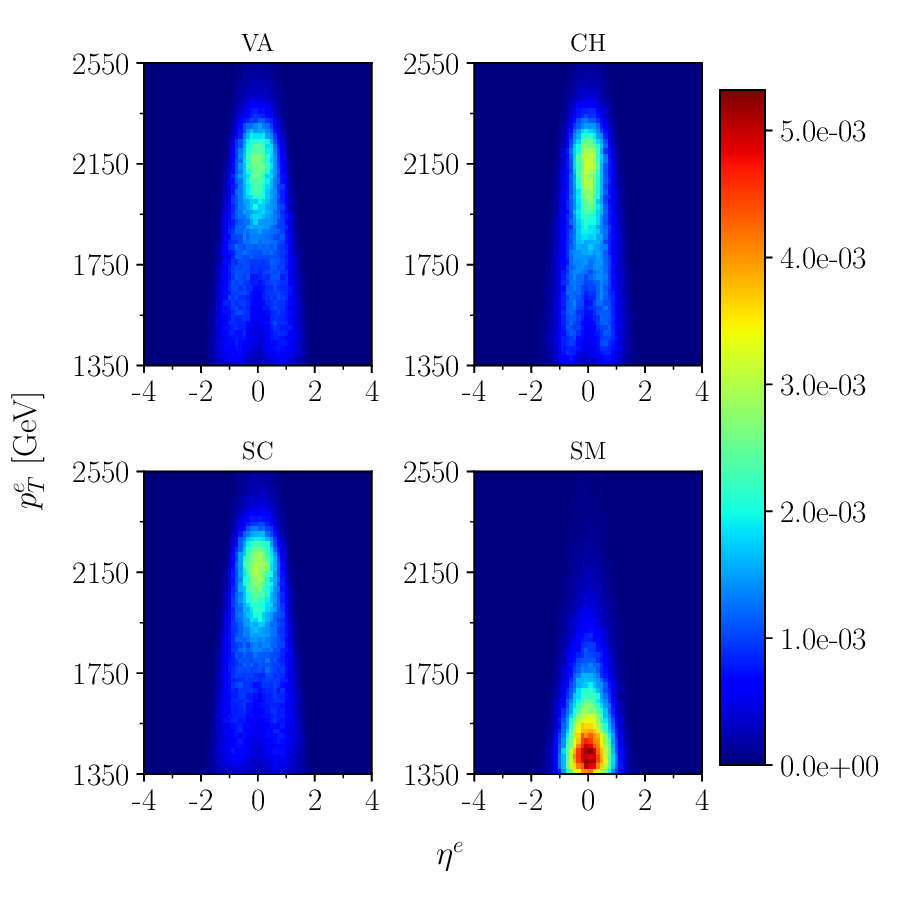}
	\vspace{-1.2cm}
	\caption{$p_T^e$ vs. $\eta^e$}
\end{subfigure}
\begin{subfigure}[t]{0.4\textwidth}
	\includegraphics[width=\textwidth]{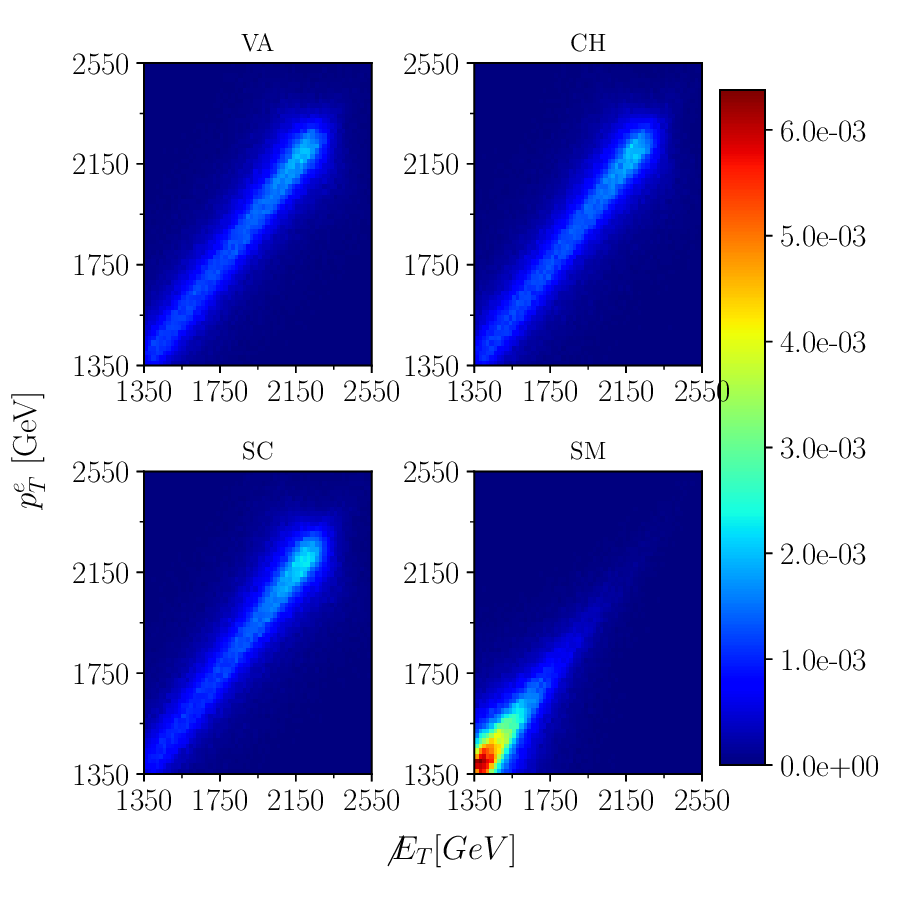}
	\vspace{-1.2cm}
	\caption{$p_T^e$ vs. $\cancel{E}_T$}
\end{subfigure}
\begin{subfigure}[t]{0.4\textwidth}
	\includegraphics[width=\textwidth]{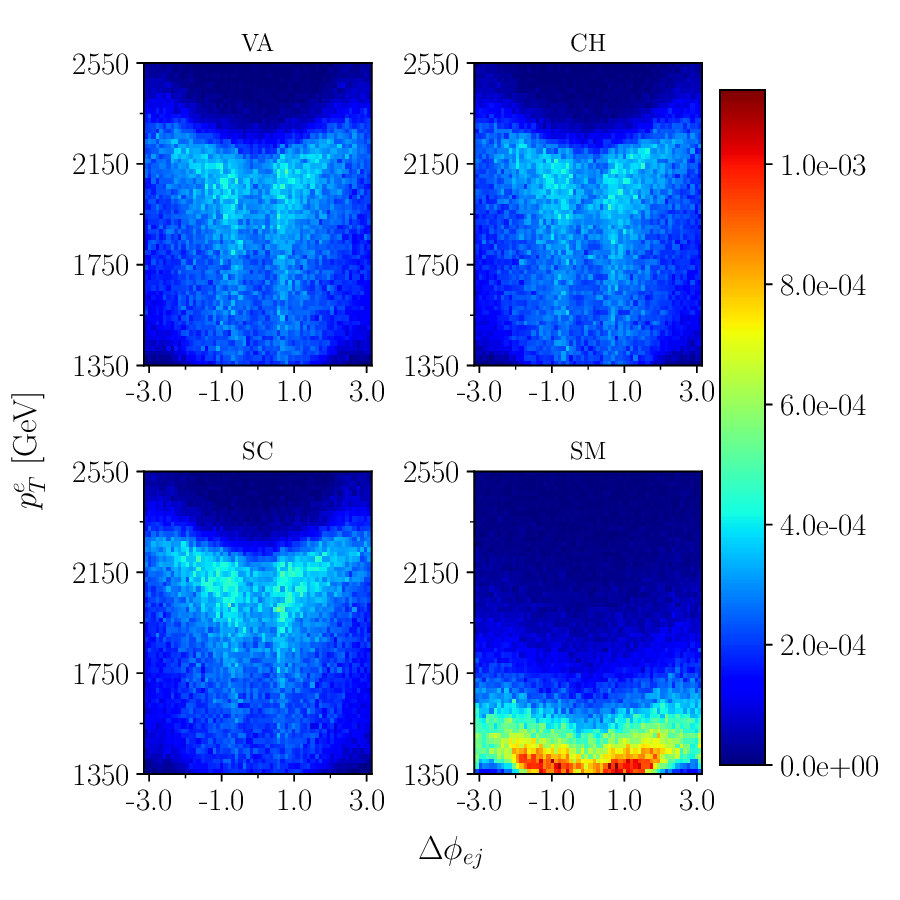}
	\vspace{-1.2cm}
	\caption{$p_T^e$ vs. $\Delta\phi_{ej}$}
\end{subfigure}
\caption{1-jet 2D histograms formed from variable pairs determined according to Scheme 3 for samples of mass $4.5$~TeV and $\Gamma_{\rm NP}\approx200$~GeV. (a) $p_T^e$ vs. $\eta^e$; (b) $p_T^e$ vs. $\cancel{E}_T$; (c) $p_T^e$ vs. $\Delta\phi_{ej}$. }
\label{hist:NLO}
\end{figure}

It turns out that all these four schemes give similar results.  Even Scheme 4, which one na{\"i}vely expects to have the best efficiency, does not show noticeable superiority to the others.  Since fixing one axis for all three color channels makes it easier to apply the Bayesian hypothesis test, to be discussed in Sec.~\ref{sec:6}, we will only focus on Scheme 3 in this paper. The corresponding 2D histograms for $4.5$-TeV resonances are shown in FIG.~\ref{hist:NLO}. 

\section{Model structure and training specifications}\label{sec:4}

In this section, we describe in detail the structure of our FNNi, FNNh, and CNN models, which are constructed with the \texttt{Keras}~\cite{Keras} library along with \texttt{TensorFlow}~\cite{tensorflow2015-whitepaper} for backend implementation. We will also describe our training specifications, including the training parameters and strategies.

\subsection{FNNi structure}\label{subsec:4:1}

Our FNNi is designed to read the 1D arrays of individual event observables as input, and to classify each histogram into one of the three signal classes.  For the 0-jet samples, we input two variables: $p_T^e$ and $\eta^e$; while for 1-jet, we input $p_T^e,\ p_T^j,\ \eta^e\,\ \eta^j,\ \cancel{E}_T,\ \Delta\phi_{ej},\ \Delta\phi_{e\cancel{E}_T}$, and $\Delta\phi_{j\cancel{E}_T}$.  The FNNi structure is specified in TABLE~\ref{Structure:FNNi}.

\begin{table}[h]
\centering
\begin{tabular}{c||c|c}
\toprule
\hfill & 0-jet & 1-jet \\
\colrule
\multirow{2}{*}{Input} & \multirow{2}{*}{$p_T^{e},\eta^{e},\phi^{e}$} & $p_T^{e},\eta^{e},p_T^j,\eta^j$ \\
& & $\cancel{E}_T,\Delta\phi_{ej},\Delta\phi_{e\cancel{E}_T},\Delta\phi_{j\cancel{E}_T}$ \\
\colrule
\multirow{3}{*}{Layers} & \multicolumn{2}{c}{batch normalization layer} \\
& \multicolumn{2}{c}{dense layer: 256$^a$} \\
& \multicolumn{2}{c}{dense layer: 256} \\
\colrule
\multirow{2}{*}{~~Layer settings~~} & \multicolumn{2}{c}{hidden layer activation = \texttt{relu}} \\
& \multicolumn{2}{c}{output layer activation = \texttt{softmax}} \\
\colrule
\multirow{3}{*}{Compilation} & \multicolumn{2}{c}{loss = \texttt{categorical\_crossentropy}} \\
& \multicolumn{2}{c}{optimizer = \texttt{adam}~\cite{2014arXiv1412.6980K}} \\
& \multicolumn{2}{c}{metric = \texttt{accuracy}} \\
\botrule
\end{tabular}
\\
\begin{flushleft}
{$^a$ This means that there are 256 nodes in the dense layer. \\}
\end{flushleft}
\caption{0-jet and 1-jet FNNi structure specifications.}
\label{Structure:FNNi}
\end{table}

\subsection{FNNh structure}\label{subsec:4:2}

Our FNNh is designed to read the flattened $60\times60$ 2D histograms of kinematic variable pairs as input, and to classify each histogram into one of the three signal classes.  For the 0-jet samples we only input one channel: $p_T^e$ vs. $\eta^e$, while for 1-jet we input three channels based on the four different schemes described above, though only the results of Scheme 3 are presented in this paper.  The FNNh structure is specified in TABLE~\ref{Structure:FNNh}. 

\begin{table}[h]
\centering
\begin{tabular}{c||c|c}
\toprule
\hfill & 0-jet & 1-jet \\
\colrule
\multirow{2}{*}{Input} & \multicolumn{2}{c}{Flattened $60\times60$ images} \\\cline{2-3}
 & ~~$p_T^e$ vs. $\eta^e$~~  & $p_T^e$ vs. $\eta^e$, $p_T^e$ vs. $\cancel{E}_T$, $p_T^e$ vs. $\Delta\phi_{ej}$ \\
\colrule
\multirow{3}{*}{Layers} & \multicolumn{2}{c}{batch normalization layer} \\
& \multicolumn{2}{c}{dense layer: 1024} \\
& \multicolumn{2}{c}{dense layer: 256} \\
\colrule
\multirow{2}{*}{~~Layer settings~~} & \multicolumn{2}{c}{hidden layer activation = \texttt{relu}} \\
& \multicolumn{2}{c}{output layer activation = \texttt{softmax}} \\
\colrule
\multirow{3}{*}{Compilation} & \multicolumn{2}{c}{loss = \texttt{categorical\_crossentropy}} \\
& \multicolumn{2}{c}{optimizer = \texttt{adam}} \\
& \multicolumn{2}{c}{metric = \texttt{accuracy}} \\
\botrule
\end{tabular}
\\
\caption{0-jet and 1-jet FNNh structure specifications.}
\label{Structure:FNNh}
\end{table}

\subsection{CNN structure}\label{subsec:4:3}

Our CNN is designed to read $60\times60$ 2D histograms of kinematic variable pairs as input with the RGB schemes mentioned previously, and to classify each histogram into one of the three signal classes. The CNN structure is specified in TABLE~\ref{Structure:CNN}.

\begin{table}[h]
\centering
\begin{tabular}{c||c|c}
\toprule
\hfill & 0-jet & 1-jet \\
\colrule
\multirow{2}{*}{Input} & \multicolumn{2}{c}{$60\times60$ images} \\\cline{2-3}
 & ~~$p_T^e$ vs. $\eta^e$~~  & RGB colors:  $p_T^e$ vs. $\eta^e$, $p_T^e$ vs. $\cancel{E}_T$, $p_T^e$ vs. $\Delta\phi_{ej}$ \\
\colrule
\multirow{7}{*}{Layers} & \multicolumn{2}{c}{batch normalization layer} \\
& \multicolumn{2}{c}{convolutional 2D layer: 3-32$^b$} \\
& \multicolumn{2}{c}{max pooling 2D layer: 2-2$^c$} \\
& \multicolumn{2}{c}{convolutional 2D layer: 3-32} \\
& \multicolumn{2}{c}{max pooling 2D layer: 2-2} \\
& \multicolumn{2}{c}{flatten layer} \\
& \multicolumn{2}{c}{dense layer: 128} \\
& \multicolumn{2}{c}{dense layer: 64} \\
\colrule
\multirow{2}{*}{~~Layer settings~~} & \multicolumn{2}{c}{hidden layer activation = \texttt{relu}} \\
& \multicolumn{2}{c}{output layer activation = \texttt{softmax}} \\
\colrule
\multirow{3}{*}{Compilation} & \multicolumn{2}{c}{loss = \texttt{categorical\_crossentropy}} \\
& \multicolumn{2}{c}{optimizer = \texttt{adam}} \\
& \multicolumn{2}{c}{metric = \texttt{accuracy}} \\
\botrule
\end{tabular}
\\
\begin{flushleft}
{$^b$ This means that the filter kernel dimension is $3\times3$, and that there are 32 nodes in the convolutional layer. \\} 
{$^c$ This means that the max pooling kernel dimension is $2\times2$, and that each stride is $2$ pixels. \\}
\end{flushleft}
\caption{0-jet and 1-jet CNN structure specifications.}
\label{Structure:CNN}
\end{table}

\subsection{Training specifications}\label{subsec:4:4}

In all trainings, we generate $15$k histograms per class for $4.5$-TeV and $50$k for $6$-TeV resonances. As for FNNi, we use $300$k SM samples for the 0-jet study and $200$k for the 1-jet study, while the numbers of the NP samples are determined by $S/B$. We then split the dataset into three subsets: training, validation, and testing sets, in the proportion of $0.64:0.16:0.20$. We set the batch size to $128$ and the maximum training epoch to $1000$. To avoid over-training, we call for an early stopping if the validation loss has not improved by more than $2\times10^{-4}$ for over $100$ epochs.

To evaluate the performance of our NNs, we determine the receiver operating characteristic (ROC) curve in terms of the \textit{one-against-all} strategy: we only consider the binary comparisons between class $i$ and a combination of the other two classes, where $i$ is the target class to be tested.  Then, we calculate the areas under the ROC curves (AUCs) as a measure of the NN performance.

\section{Training results}\label{sec:5}

In this section, we present the trained NN results of the 0- and 1-jet processes for various 0-jet $S/B\equiv S_0/B_0$ ratios.  We mainly focus on FNNh since it gives the best performance. We refer some more technical details of FNNh training to Appendix~\ref{sec:a}.  For the 1-jet samples, we further investigate the importance of individual kinematic observable pairs.

\subsection{0-jet results}\label{subsec:5:1}

Since we only make use of electron information of the 0-jet samples, ignoring the jet information, the analysis in this section will determine how useful the visible electron information can distinguish the signal hypotheses. We first present the CNN, FNNh, and FNNi training outcomes of $4.5$-TeV resonances with $200$~GeV width and of $6$-TeV resonances with 300~GeV width in FIG.~\ref{LO:Light}.  In the shaded regions on the figures, we denote two regions of $S/B$, one where HL-LHC will not achieve a $5\sigma$ excess and the other that violates the  current constraint from ATLAS~\cite{Aad:2019wvl}.  For $4.5$-TeV resonances, all the NNs can already start to distinguish the signal scenarios when $S/B \agt 0.4$ and steadily improve with higher signal purities.  At the 5 $\sigma$ discovery level, which corresponds to $S/B=0.6$ in this case, both CNN and FNNh can distinguish with AUCs over $0.7$ for all three classes, while FNNi just barely reaches this value for the SC class. Also, FNNh is always the best in terms of the identification of VA and SC classes. On the other hand for $6$-TeV resonances, the differences among the three neural networks are milder, with FNNh still performing the best. For the 6 TeV plots,  5$\sigma$ discovery level requires $S/B=2.5$, where the FNNh can reach AUCs $\geq0.65$ for all three classes; while at the current $95\%$ C.L.~limit $S/B=5.5$, it can reach AUCs of around $0.75$. Note that the CH class is always the easiest to be identified, while VA and SC are more difficult. Moreover, even though the valid $S/B$ values for $6$-TeV resonances are much higher than those for $4.5$-TeV resonances, the corresponding AUCs are significantly lower suggesting that event statistics can be more critical than signal purity for this method. Note however, as pointed out in Ref.~\cite{Nachman:2021yvi}, that from a statistical point of view there should be no general superiority of FNNh over FNNi. One major reason and benefit of using the FNNh approach is to enable the simplification of the model structure and training procedures.  Hence, even though we identify FNNh as the best approach in our study, this fact is based upon the specific simple designs of our NN models.  This argument also holds in the 1-jet study.

Since FNNh gives the best the results of the three or comparable results to the other two in all scenarios, we further present the results for $4.5$-TeV resonances with $\Gamma_{\rm NP}\approx500,~50$~GeV using FNNh in FIG.~\ref{LO:Width}. For all the three different $\Gamma_{\rm NP}$ samples, the AUCs are roughly consistent with one another, suggesting that the information of boson width does not affect the NN performance very much.  This is believed to be mainly due to the fact that only the $p_T$ distribution is changed by the width, only making it harder to distinguish between $W'$ and the $H'$ hypotheses.  Thus, we will focus exclusively on the samples of $\Gamma_{\rm NP}\approx200$~GeV for $4.5$-TeV resonances in what follows. 

\begin{figure}[H]
\hspace{1.7cm} 4.5~TeV \hspace{6cm} 6~TeV
\centering
\\
CNN
\begin{subfigure}[t!]{0.44\textwidth}
	\includegraphics[width=\textwidth]{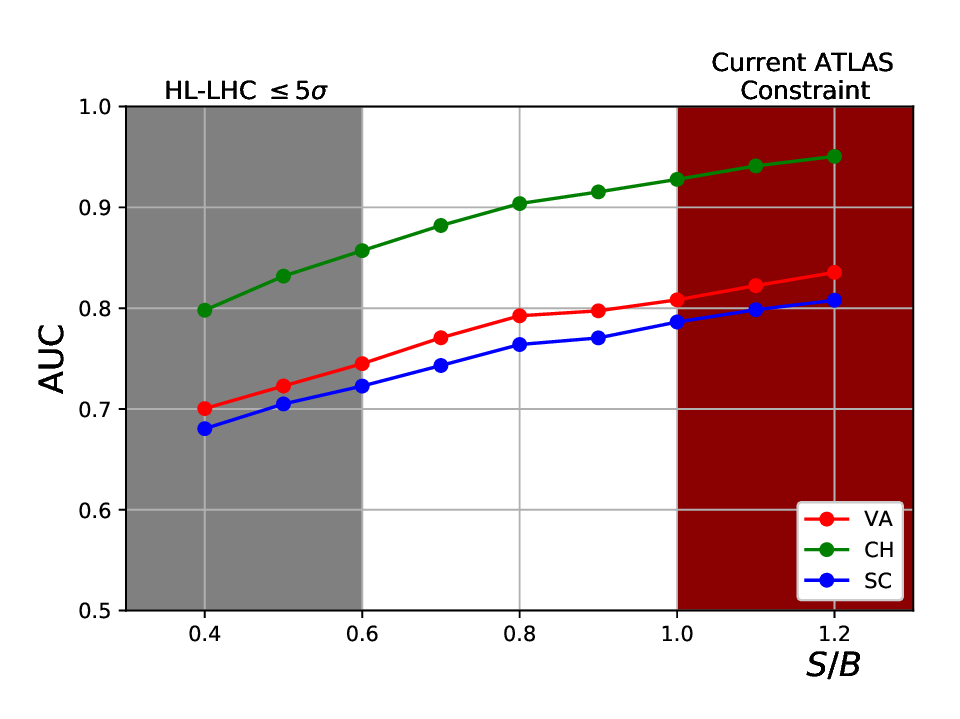}
	\vspace{-1.2cm}
	\caption{}
\end{subfigure}
\begin{subfigure}[t!]{0.44\textwidth}
	\includegraphics[width=\textwidth]{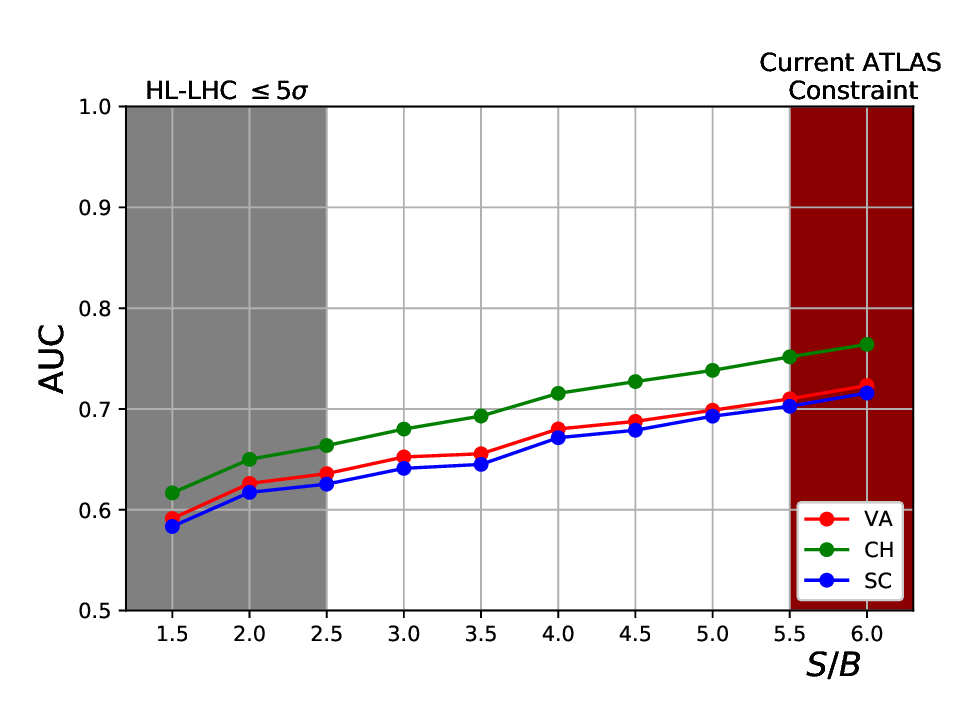}
	\vspace{-1.2cm}
	\caption{}
\end{subfigure}
FNNh
\begin{subfigure}[t!]{0.44\textwidth}
	\includegraphics[width=\textwidth]{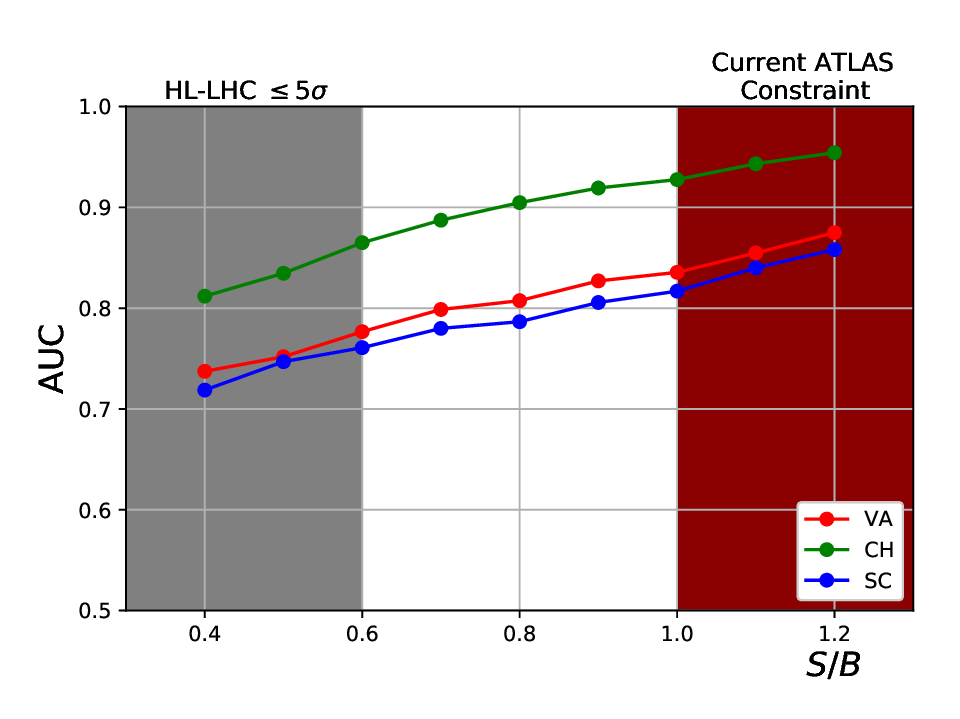}
	\vspace{-1.2cm}
	\caption{}
\end{subfigure}
\begin{subfigure}[t!]{0.44\textwidth}
	\includegraphics[width=\textwidth]{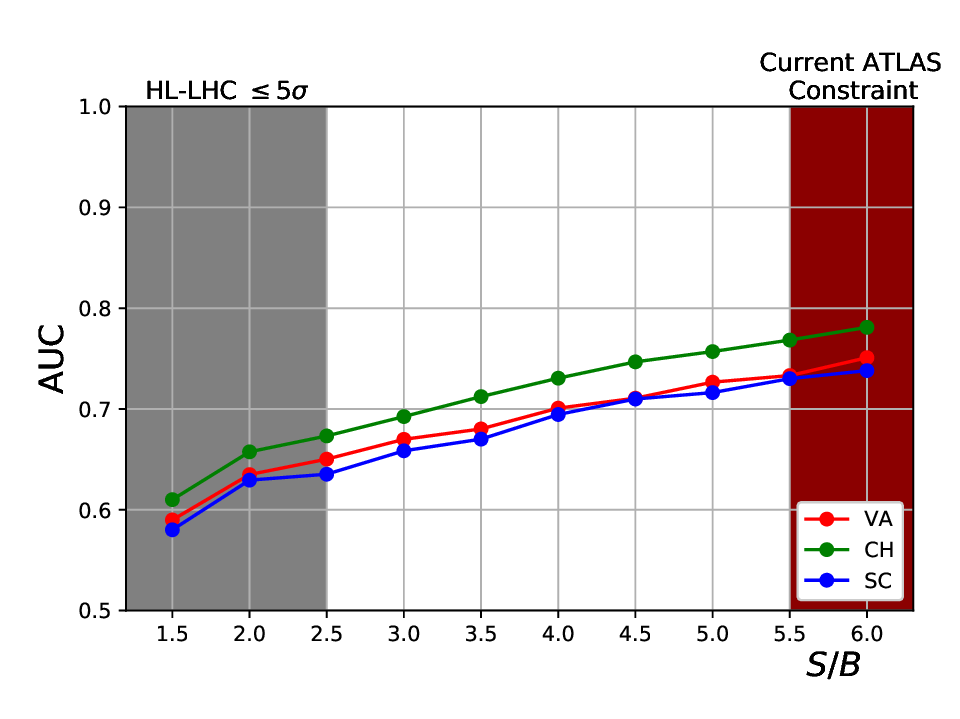}
	\vspace{-1.2cm}
	\caption{}
\end{subfigure}
FNNi
\begin{subfigure}[t!]{0.44\textwidth}
	\includegraphics[width=\textwidth]{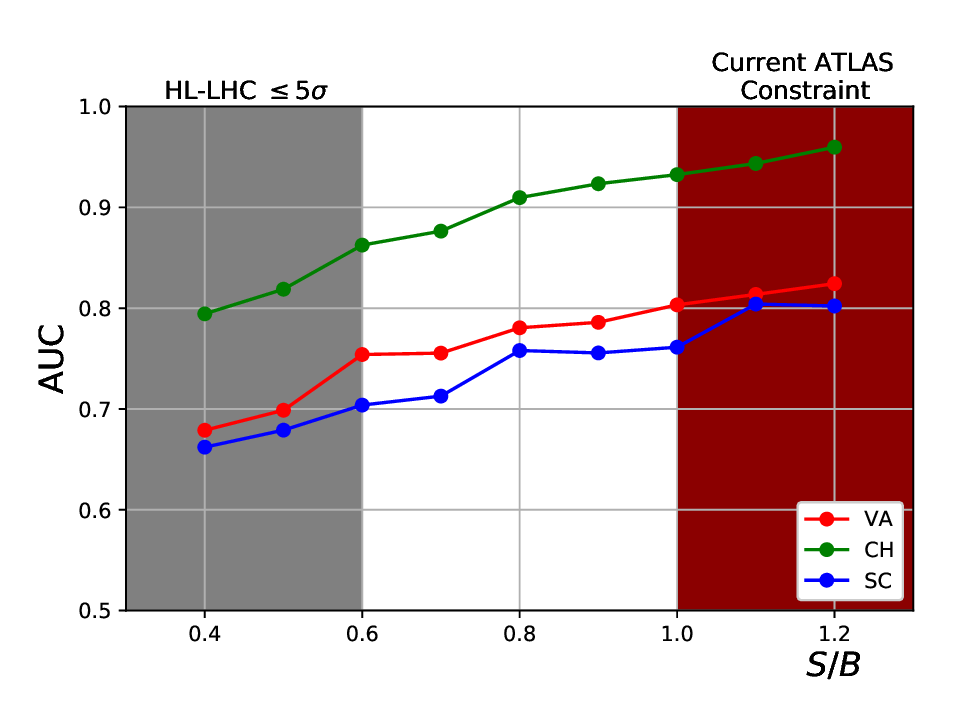}
	\vspace{-1.2cm}
	\caption{}
\end{subfigure}
\begin{subfigure}[t!]{0.44\textwidth}
	\includegraphics[width=\textwidth]{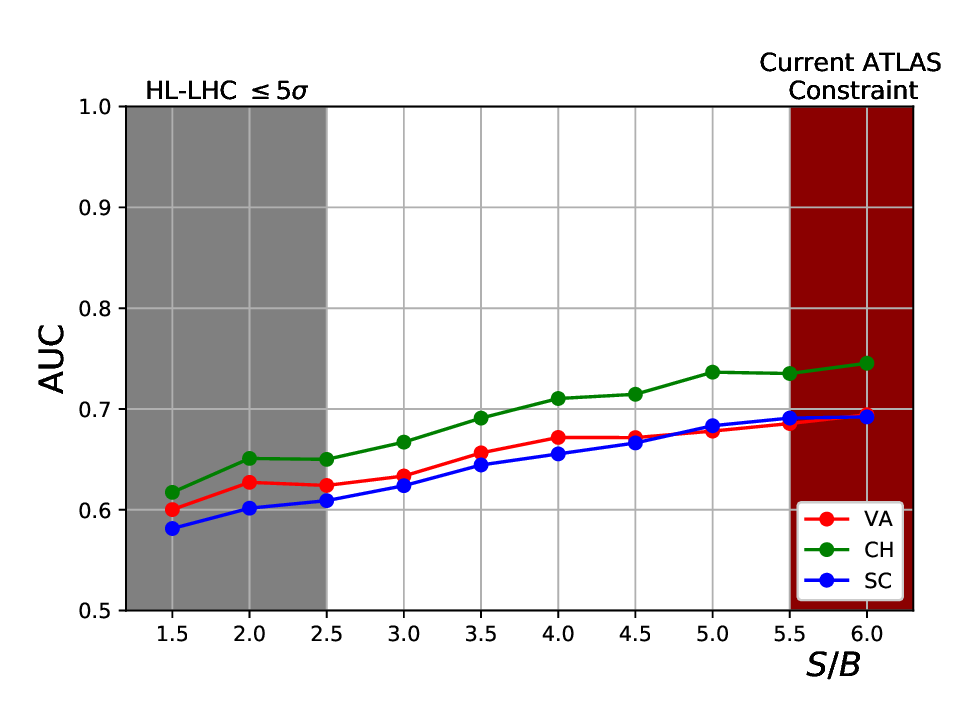}
	\vspace{-1.2cm}
	\caption{}
\end{subfigure}
\end{figure}
\begin{figure}[H]\ContinuedFloat
\centering

\caption{AUC as a function of the $S/B$ ratio for 0-jet samples.  The left column is for a $4.5$-TeV resonance with $\Gamma_{\rm NP}\approx200$~GeV, while the right column is for a 6-TeV resonance with $\Gamma_{\rm NP}\approx300$~GeV.  The first row uses CNN, the second row FNNh, and the third row FNNi.  The AUCs for the NNs to identify VA against non-VA are depicted red, CH against non-CH in green, and SC against non-SC in blue.  The shaded regions denote $S/B$ values where HL-LHC will not achieve a $5\sigma$ excess (gray) and the ATLAS constraint~\cite{Aad:2019wvl} is violated (red). The same color scheme applies to all the subsequent figures.}
\label{LO:Light}
\end{figure}
\begin{figure}[H]
\centering
\begin{subfigure}[t!]{0.49\textwidth}
	\includegraphics[width=\textwidth]{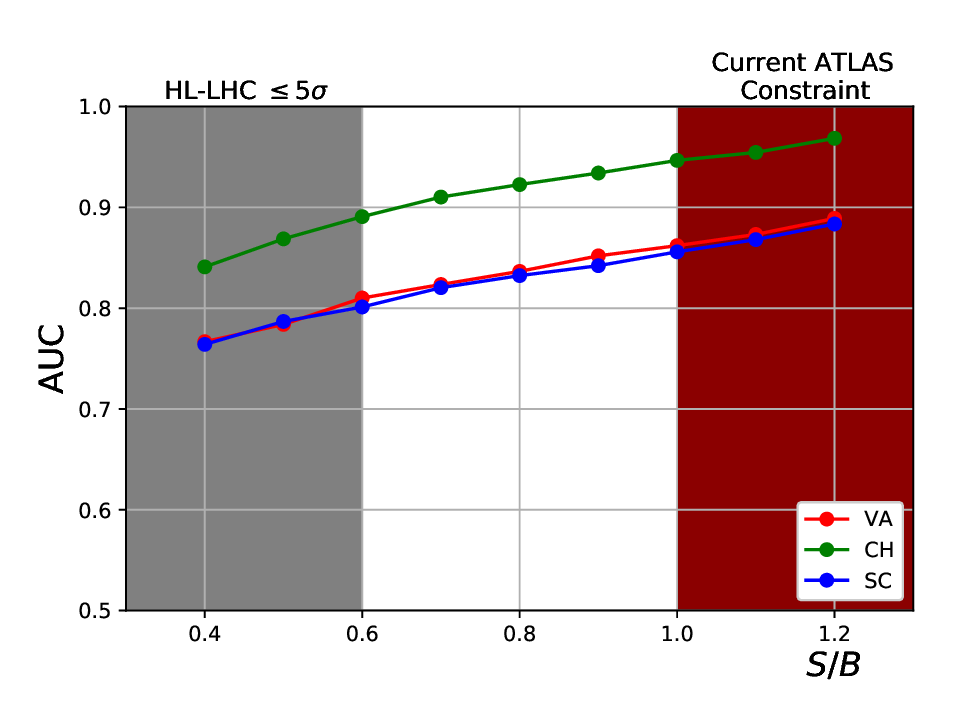}
	\vspace{-1.2cm}
	\caption{}
\end{subfigure}
\begin{subfigure}[t!]{0.49\textwidth}
	\includegraphics[width=\textwidth]{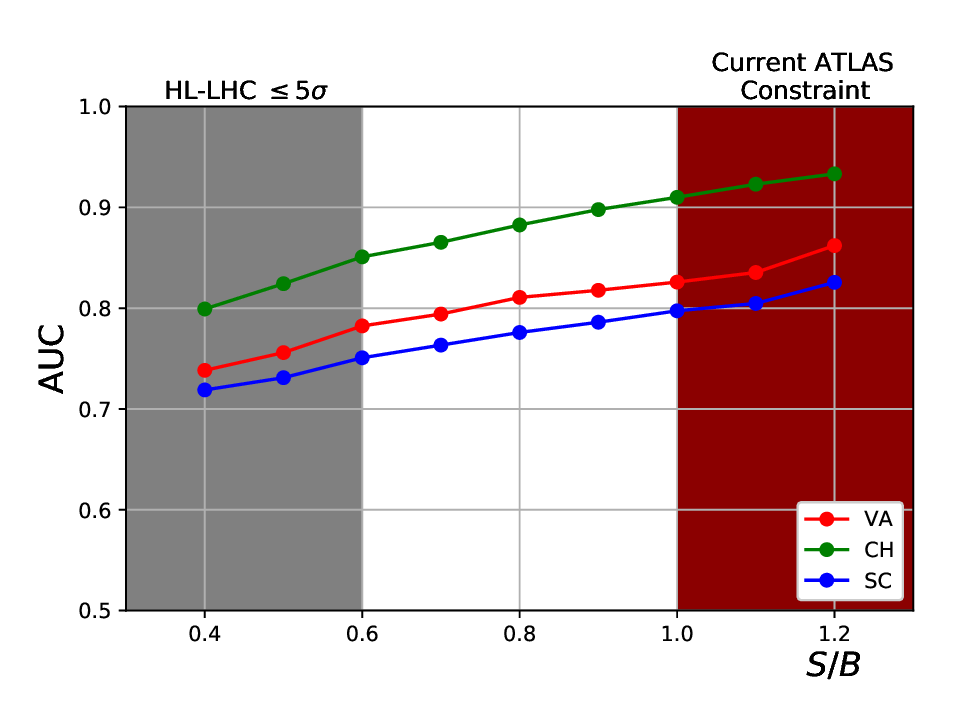}
	\vspace{-1.2cm}
	\caption{}
\end{subfigure}

\caption{FNNh training outcomes for 0-jet samples of $4.5$-TeV resonance with $\Gamma_{\rm NP}\approx$ (a) $500$ and (b) $50$~GeV.  Same color scheme as Fig.~\ref{LO:Light}.}
\label{LO:Width}
\end{figure}

\begin{figure}[H]
\hspace{3.3cm} w/o CV \hspace{6.1cm} with CV
\\
\centering
4.5~TeV
\begin{subfigure}[t!]{0.45\textwidth}
	\includegraphics[width=\textwidth]{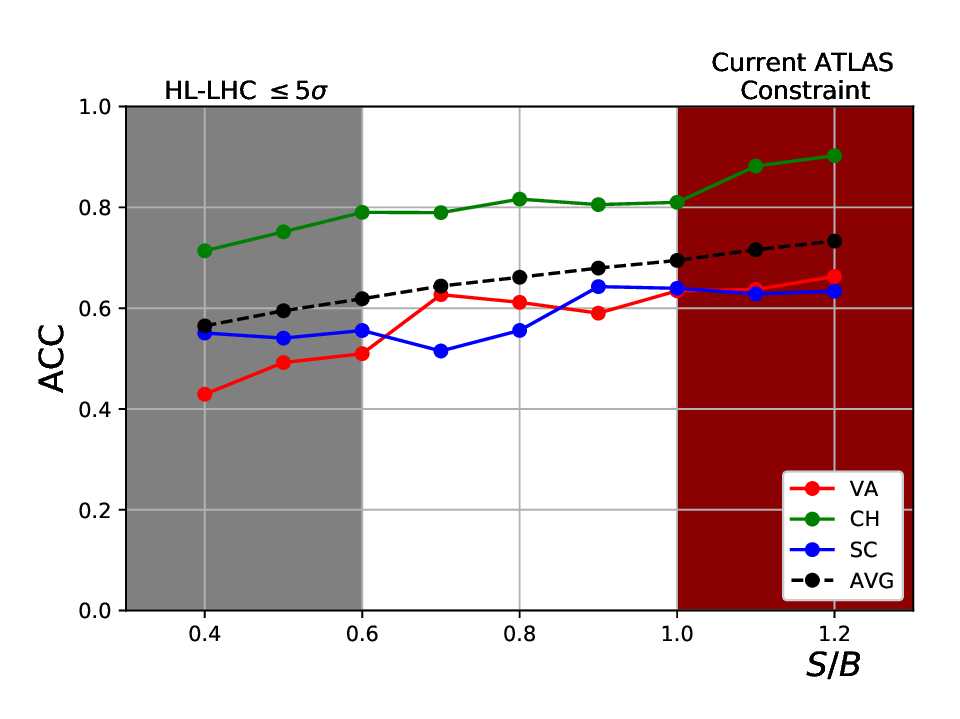}
	\vspace{-1.2cm}
	\caption{}
\end{subfigure}
\begin{subfigure}[t!]{0.45\textwidth}
	\includegraphics[width=\textwidth]{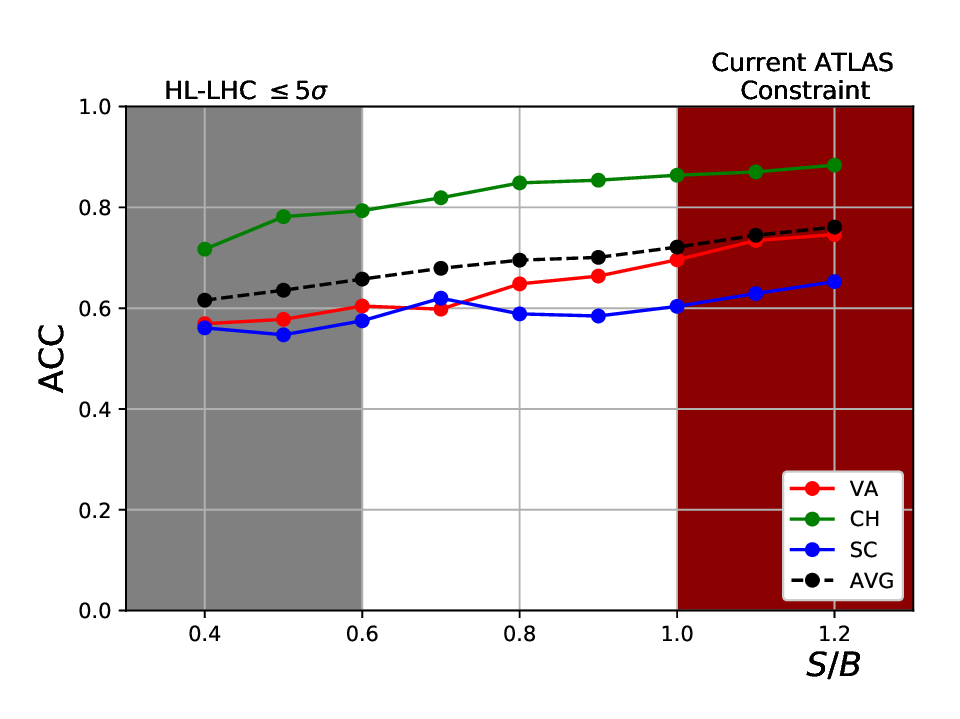}
	\vspace{-1.2cm}
	\caption{}
\end{subfigure}
\\
6~TeV
\begin{subfigure}[t!]{0.45\textwidth}
	\includegraphics[width=\textwidth]{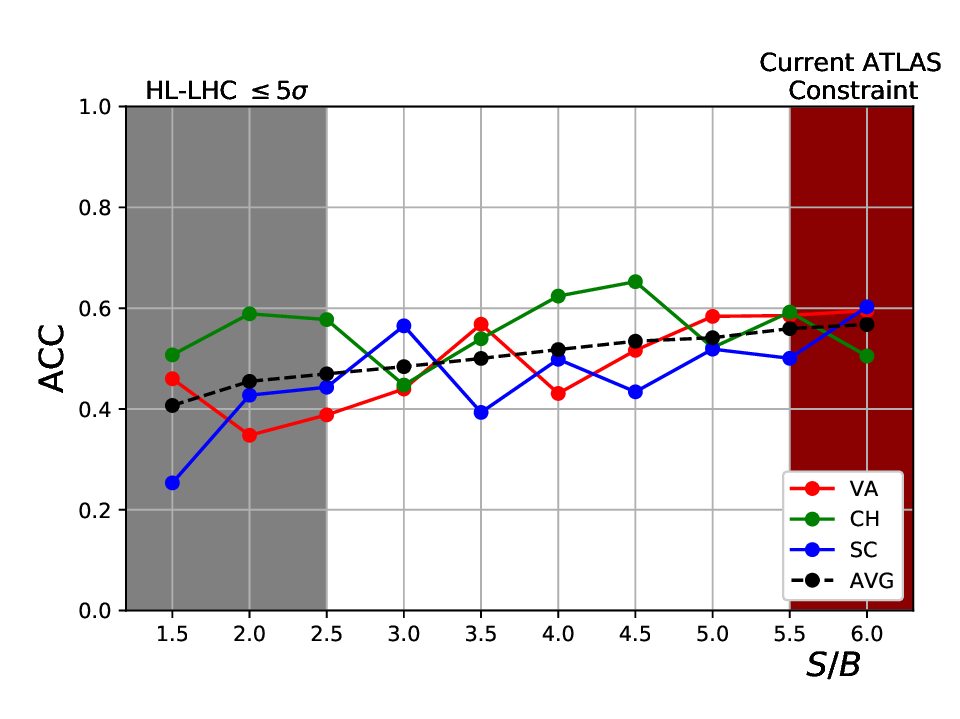}
	\vspace{-1.2cm}
	\caption{}
\end{subfigure}
\begin{subfigure}[t!]{0.45\textwidth}
	\includegraphics[width=\textwidth]{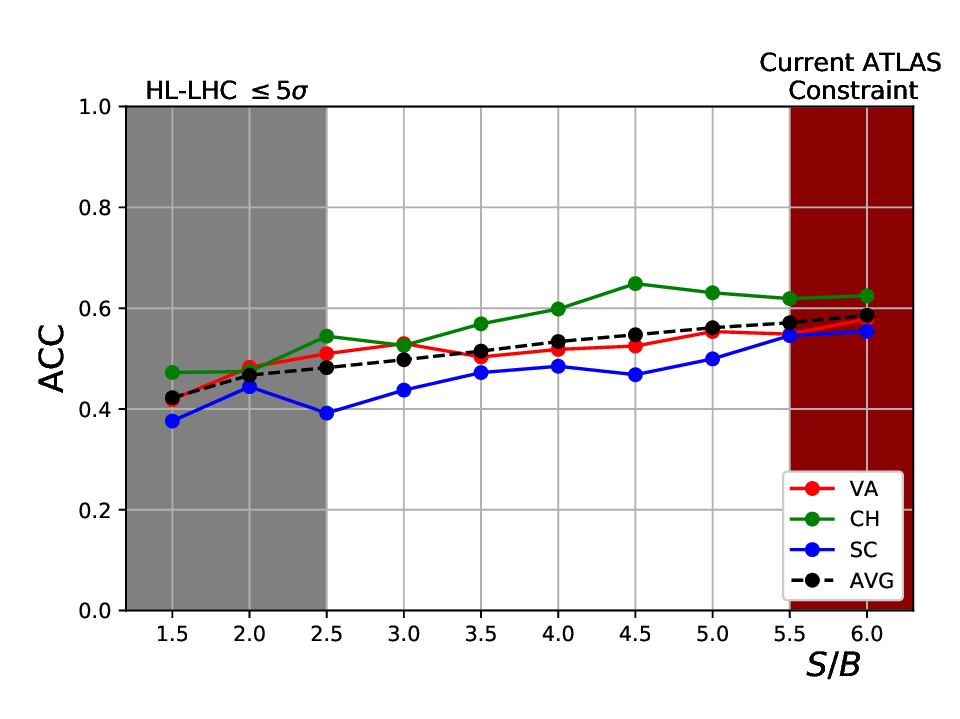}
	\vspace{-1.2cm}
	\caption{}
\end{subfigure}
\caption{0-jet ACCs for samples of (a) $4.5$-TeV and (c) $6$-TeV resonances using FNNh without CV, and (b) $4.5$-TeV and (d) $6$-TeV resonances with 10-fold CV applied.  Same color scheme as Fig.~\ref{LO:Light}.}
\label{LO:ACC}
\end{figure}

\begin{figure}[H]
\centering
\begin{subfigure}[t!]{0.49\textwidth}
	\includegraphics[width=\textwidth]{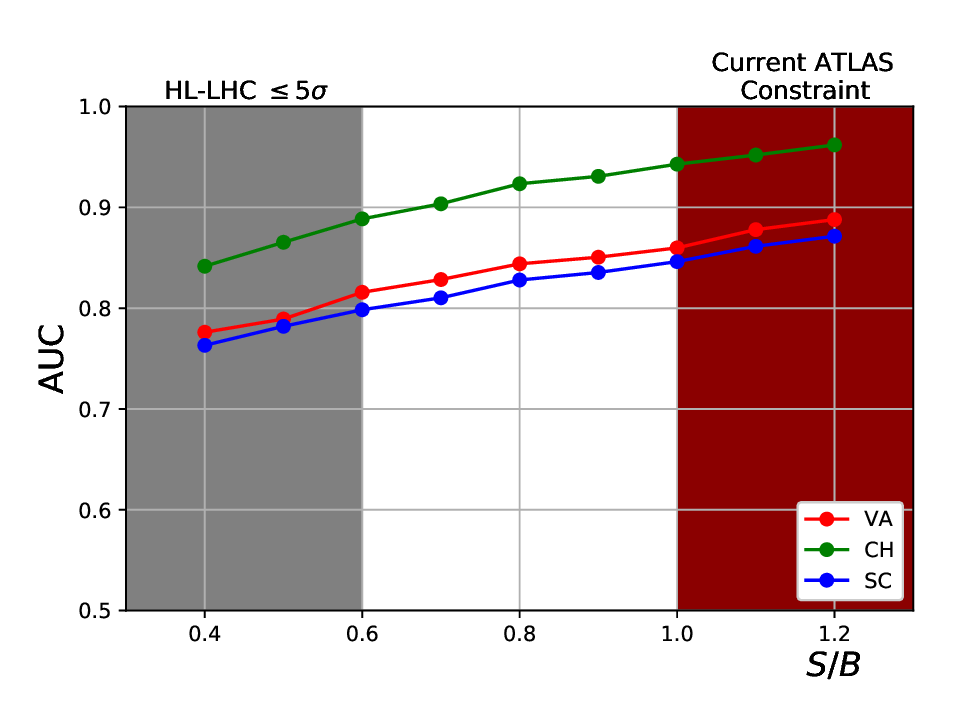}
	\vspace{-1.2cm}
	\caption{}
\end{subfigure}
\begin{subfigure}[t!]{0.49\textwidth}
	\includegraphics[width=\textwidth]{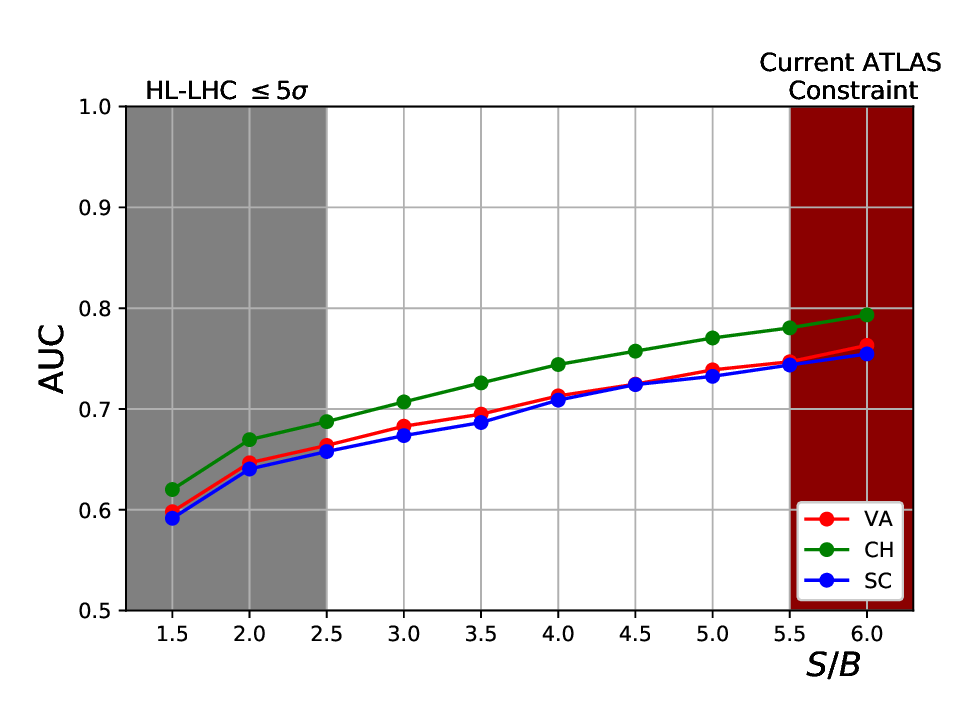}
	\vspace{-1.2cm}
	\caption{}
\end{subfigure}

\caption{0-jet AUCs for samples of (a) $4.5$-TeV and (b) $6$-TeV resonances using FNNh with 10-fold CV applied.  Same color scheme as Fig.~\ref{LO:Light}.}
\label{LO:AUC:CV}
\end{figure}

To give a more interpretable metric, we now present the ``accuracies'' (ACCs) of our FNNh. The ACC here (and the 1-jet case below) is to be understood as the class-wise true positive rate.  For this, we associate each testing histogram to the class for which it gets the highest score, and then calculate the true positive rate for each class.  We also calculate the average ACC curves, defined as the global true positive rate. Notice that although the average ACC curves, as shown in FIG.~\ref{LO:ACC}(a) and \ref{LO:ACC}(c), are stably improving, the class-wise ACCs are rather unstable. This is mainly due to model biases.  When evaluating the ACCs, we only pick the best class score of each event, and thus the relation between different class scores is in some sense ignored.  Unlike the AUCs which are evaluated using a sliding threshold, the ACCs are therefore more sensitive to model biases.  Thus, to improve the stability, we further apply a 10-fold cross validation (CV) to better address this issue, with the results shown in FIG.~\ref{LO:ACC}(b) and \ref{LO:ACC}(d).  As expected, CV helps stabilizing the class-wise accuracies and does not significantly alter the average.  For the sake of comparison, we also show in FIG.~\ref{LO:AUC:CV} the AUCs after applying 10-fold CV. Notice that the resulted AUCs are only at most $2\%$ and the ACCs at most $3\%$ better than those without applying CV, meaning that it does not matter much in the 0-jet case. However, CV does a nice job at stabilizing both $4.5$ and $6$-TeV testing performance.

Focusing once more on the average ACCs with CV applied, we find that the ACCs for $4.5$-TeV resonances are all above $0.6$, and can reach  $0.75$ at $S/B=1.2$; on the other hand, the ACCs for $6$-TeV resonances are around $0.5$ for $S/B\lesssim2.5$, and can reach almost $0.6$ at $S/B=6.0$.  All of these numbers improve significantly compared to random guess with ACC=$0.33$.  Even though the ACC metric is more interpretable, we will continue to focus on AUC as a more conventional metric to compare performance of our classifiers.  

\begin{figure}[H]
\centering
\begin{subfigure}[t!]{0.49\textwidth}
	\includegraphics[width=\textwidth]{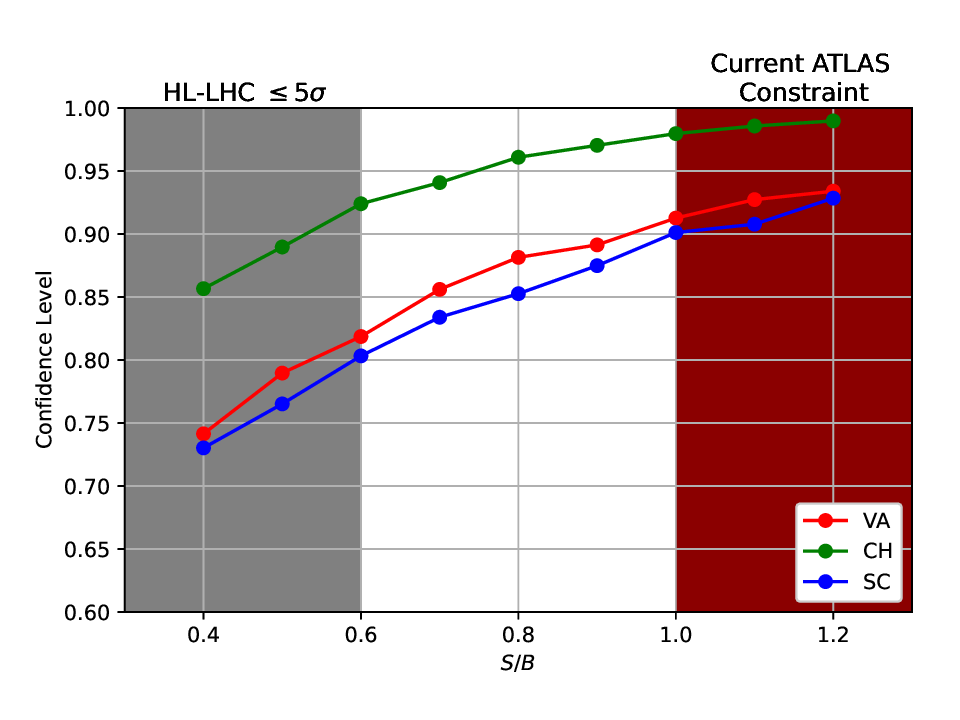}
	\vspace{-1.2cm}
	\caption{}
\end{subfigure}
\begin{subfigure}[t!]{0.49\textwidth}
	\includegraphics[width=\textwidth]{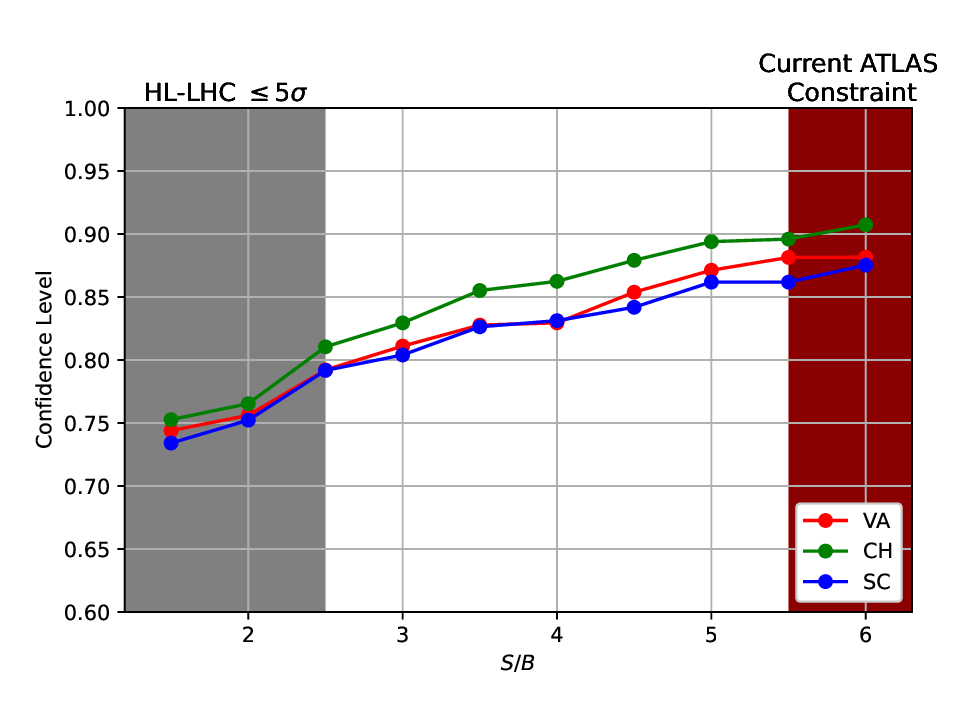}
	\vspace{-1.2cm}
	\caption{}
\end{subfigure}
\caption{Median 0-jet confidence levels at which the non-VA (red), non-CH (green), and non-SC (blue) hypotheses are excluded by the trained FNNh for samples of (a) $4.5$-TeV and (b) $6$-TeV resonances when assuming the VA, CH, and SC hypotheses are true, respectively.}
\label{CL:LO}
\end{figure}

Finally for FNNh, we analyze the confidence level at which it can rule out alternative hypotheses.  For this, we split the ternary scores and analyze the following three cases separately: VA vs. non-VA, CH vs. non-CH, and SC vs. non-SC. Note that these mirror the one-against-all strategy, allowing comparisons with the earlier AUC/ACC results.  For the VA vs. non-VA case at a fixed $S/B$, we assume that the VA hypothesis is true and use the VA score as the test statistic to constrain the non-VA hypothesis. We take the median value for the VA hypothesis and use it to determine the median expected $p$-value for the non-VA hypothesis, $p_{med}$, which then gives a median expected exclusion for the alternative hypothesis at a confidence level of CL$= 1-p_{med}$.  The modification for the other two cases requires swapping the assumed true and alternative hypotheses. We plot these CLs against the $S/B$ values for both $4.5$- and $6$-TeV resonances in Fig.~\ref{CL:LO}.  These CLs are correlated but not directly related to our AUC and ACC metrics, since the latter are derived with varying thresholds. For example, one can see that the CLs are higher (lower) than the AUCs for 4.5 (6)~TeV mass and that a CL value may correspond to very different corresponding AUC values.  For both $4.5$- and $6$-TeV resonances, all the alternative classes can be excluded at a CL $> 80\%$ in the $S/B$ region of our interest, with the CH class always surpassing the other two, as expected from the previous AUC/ACC results.  In particular, only the non-CH class can be excluded at $>95\%$ CL in the allowed $S/B$ range.

\subsection{1-jet results}\label{subsec:5:2}

\begin{figure}[H]
\hspace{3cm} 4.5~TeV \hspace{6.1cm} 6~TeV
\\
\centering
CNN
\begin{subfigure}[t!]{0.45\textwidth}
	\includegraphics[width=\textwidth]{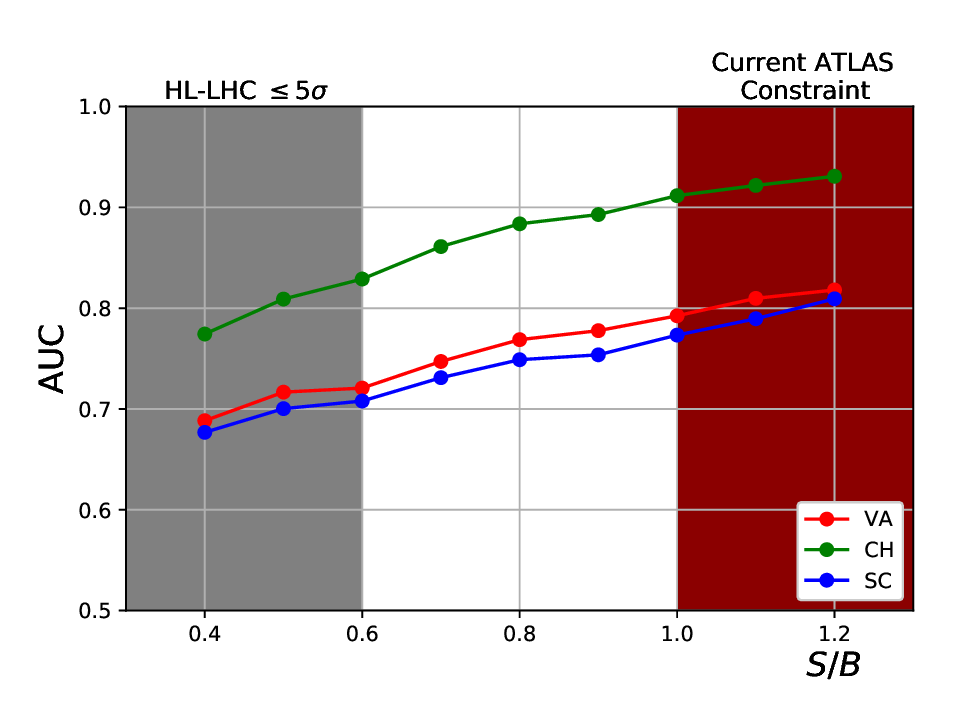}
	\vspace{-1.2cm}
	\caption{}
\end{subfigure}
\begin{subfigure}[t!]{0.45\textwidth}
	\includegraphics[width=\textwidth]{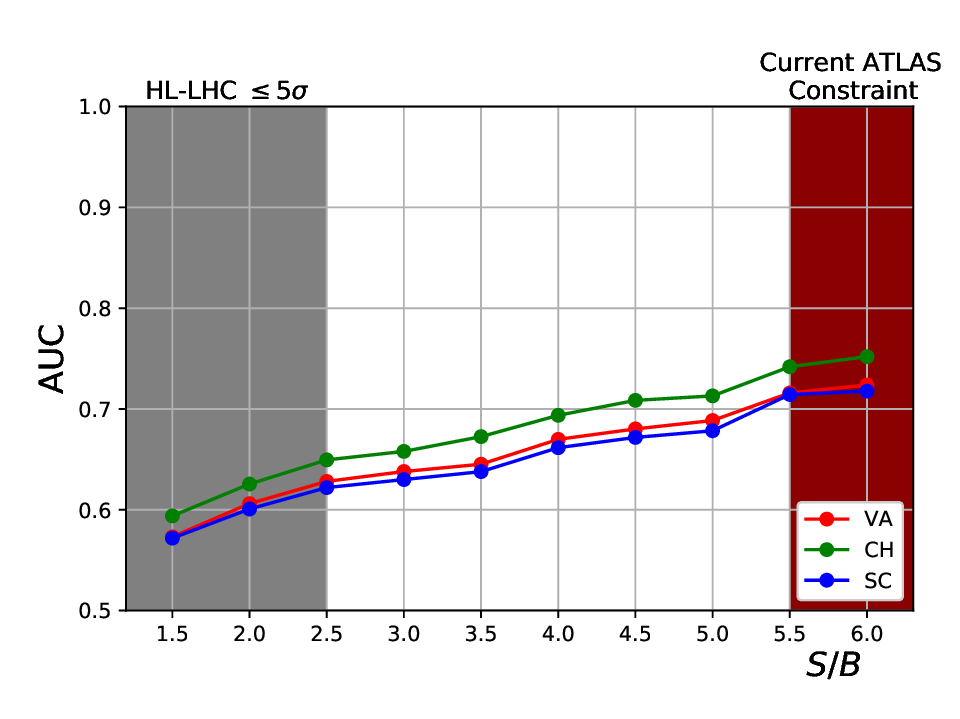}
	\vspace{-1.2cm}
	\caption{}
\end{subfigure}
\\
FNNh
\begin{subfigure}[t!]{0.45\textwidth}
	\includegraphics[width=\textwidth]{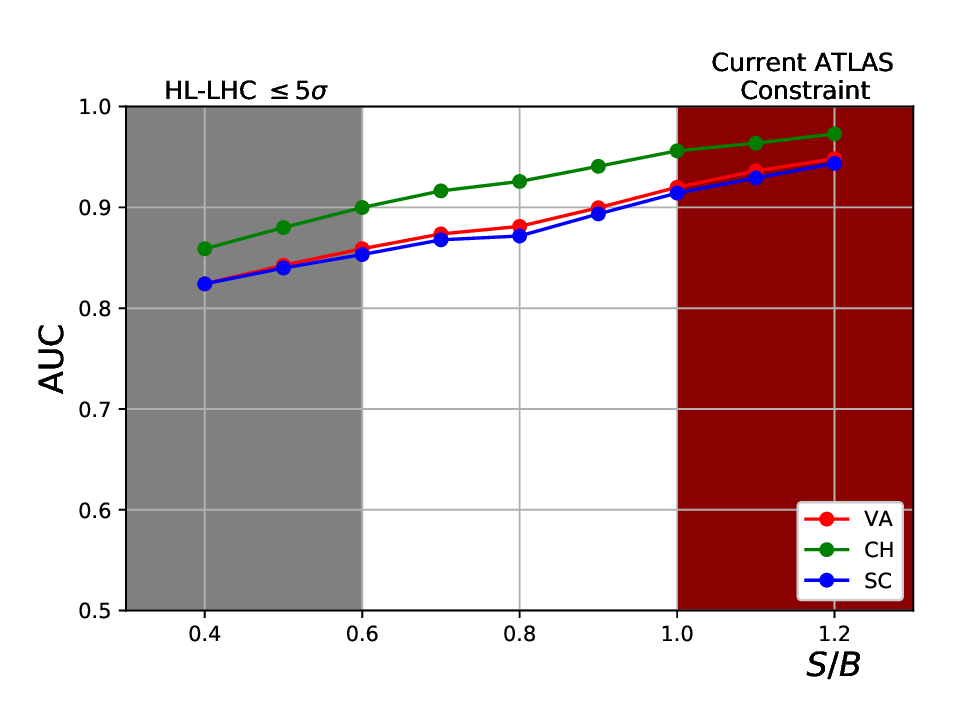}
	\vspace{-1.2cm}
	\caption{}
\end{subfigure}
\begin{subfigure}[t!]{0.45\textwidth}
	\includegraphics[width=\textwidth]{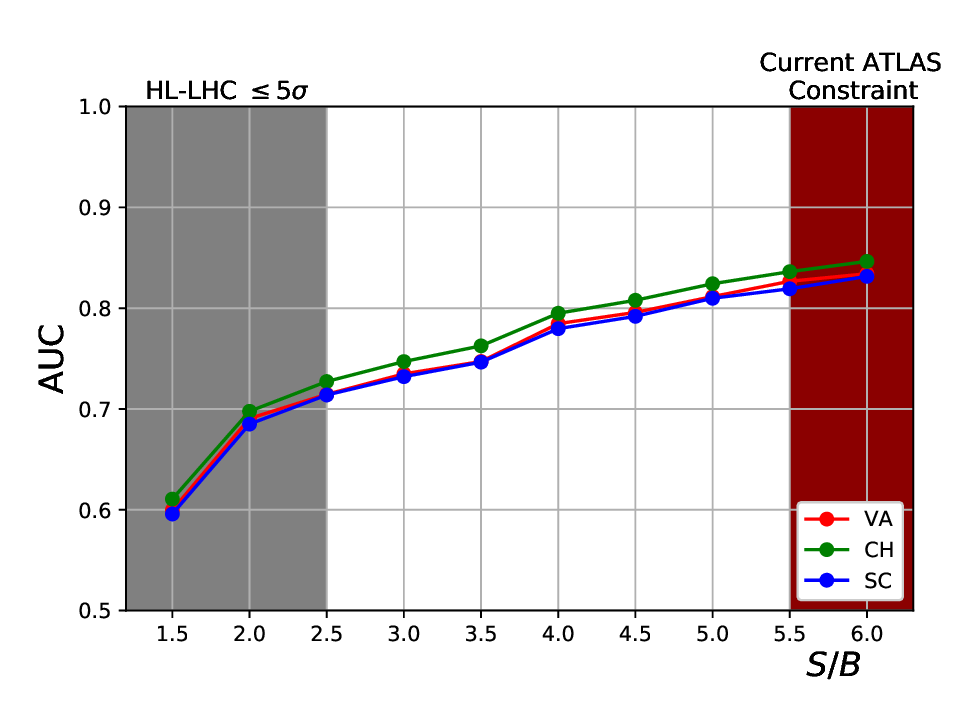}
	\vspace{-1.2cm}
	\caption{}
\end{subfigure}
\\
FNNi
\begin{subfigure}[t!]{0.45\textwidth}
	\includegraphics[width=\textwidth]{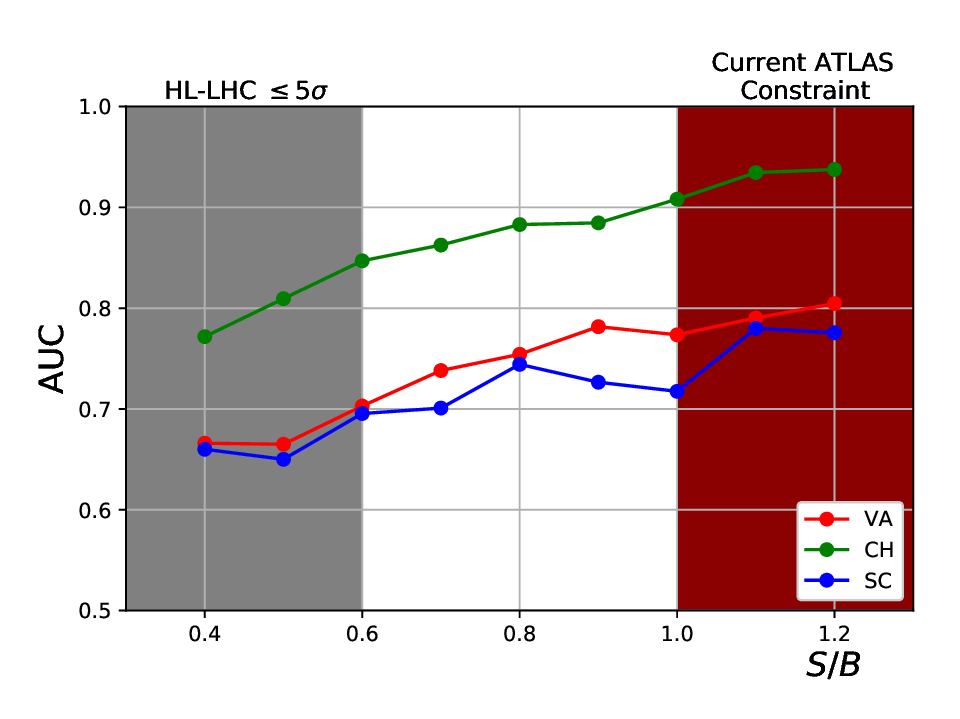}
	\vspace{-1.2cm}
	\caption{}
\end{subfigure}
\begin{subfigure}[t!]{0.45\textwidth}
	\includegraphics[width=\textwidth]{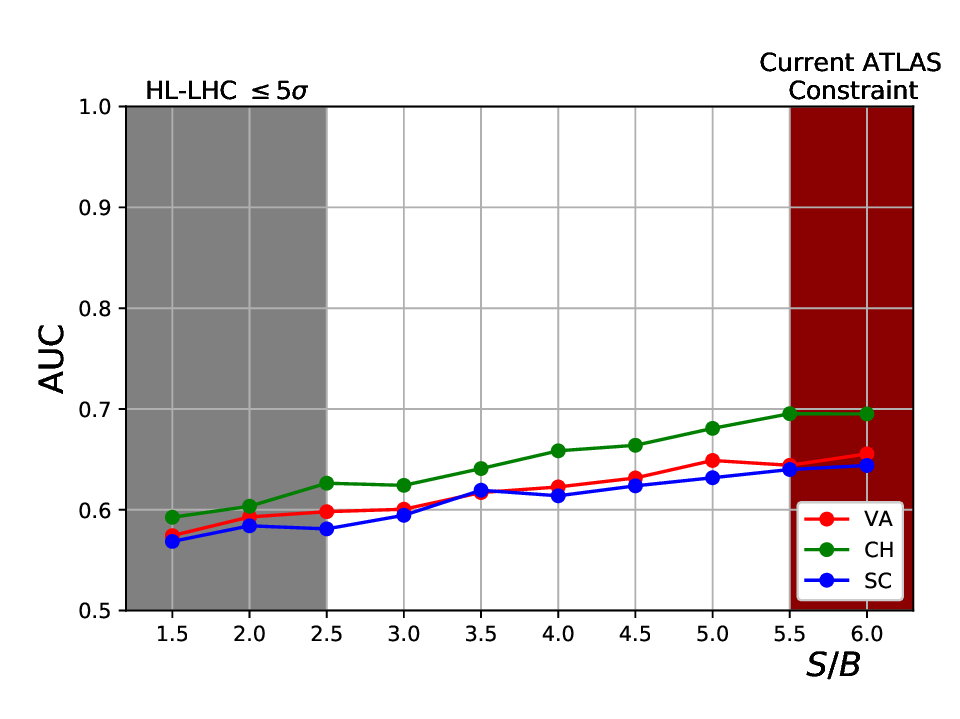}
	\vspace{-1.2cm}
	\caption{}
\end{subfigure}
\caption{Training outcomes of $4.5$-TeV resonances using 1-jet samples in Scheme 3 for (a) CNN and (c) FNNh, and (e) FNNi, and training outcomes of $6$-TeV resonances using 1-jet samples in Scheme 3 for (b) CNN and (d) FNNh, and (f) FNNi.  Same color scheme as Fig.~\ref{LO:Light}.}
\label{NLO:Light}
\end{figure}

In this section, we include the information of the leading jet in addition to the visible lepton and show how such additional information helps compensate for the lower event statistics. We will only present the result using Scheme 3 for $4.5$-TeV resonances with $\Gamma_{\rm NP}\approx200$~GeV and $6$-TeV resonances with $\Gamma_{\rm NP}\approx300$~GeV for reasons stated before.  We show the CNN, FNNh, and FNNi training outcomes in FIG.~\ref{NLO:Light}.  First of all, we see again that for $4.5$-TeV resonances FNNh outperforms the other two.  There is an intriguing trend in the $6$-TeV results: as CNN is consistently better than FNNi, FNNh is only slightly better than both of them at $S/B=1.5$.  As soon as $S/B$ reaches $2.0$, FNNh makes a sudden jump and significantly outperforms the other two henceforth. 

Comparing FIG.~\ref{NLO:Light}(c) with FIG.~\ref{LO:Light}(c), we see that the 1-jet FNNh performance for $4.5$-TeV resonances is much better than that of the 0-jet in terms of the VA and SC classes, both of which can reach AUCs of $0.8$ even at $S/B=0.4$, while CH seems to be only slightly better.  A comparison between FIG.~\ref{NLO:Light}(d) and FIG.~\ref{LO:Light}(d) shows an even more interesting trend for $6$-TeV resonances: all three classes can be better classified using the 1-jet strategy except for $S/B=1.5$, and can even reach AUCs of $0.8$ for $S/B \agt 5.0$.  This shows that even with the drop in statistics by going to 1-jet events, there is improved discriminating power over the 0-jet analysis.  Thus, this proves that this technique is promising for higher-dimensional histograms, thus broadening the range of viable channels to be studied and even potentially granting better distinguishing power. 

\begin{figure}[H]
\hspace{3.3cm} w/o CV \hspace{6.1cm} with CV
\\
\centering
4.5~TeV
\begin{subfigure}[t!]{0.45\textwidth}
	\includegraphics[width=\textwidth]{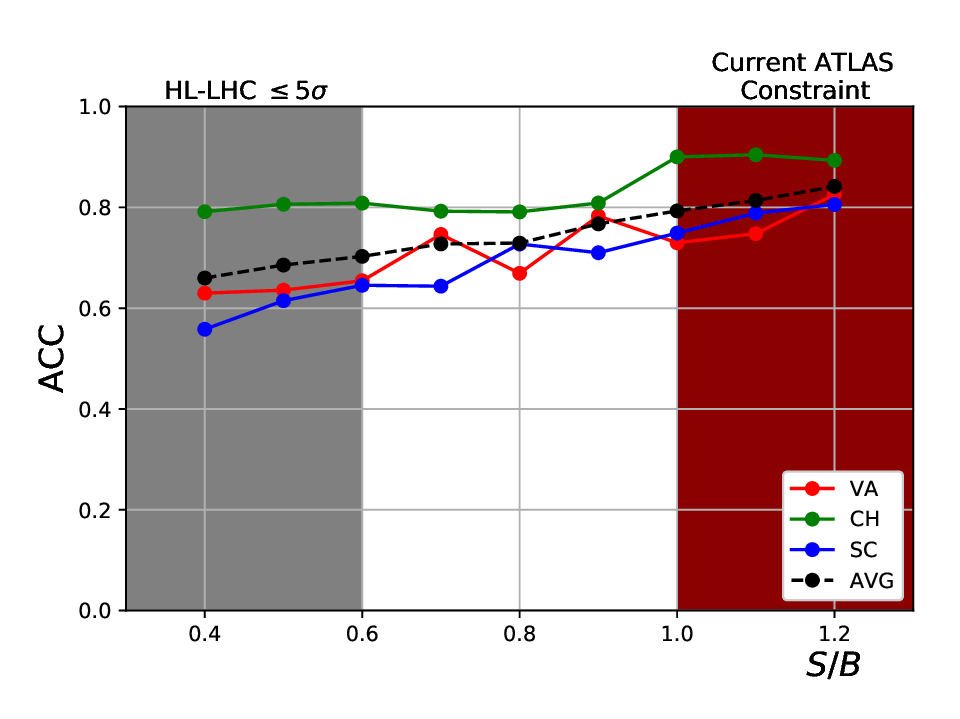}
	\vspace{-1.2cm}
	\caption{}	
\end{subfigure}
\begin{subfigure}[t!]{0.45\textwidth}
	\includegraphics[width=\textwidth]{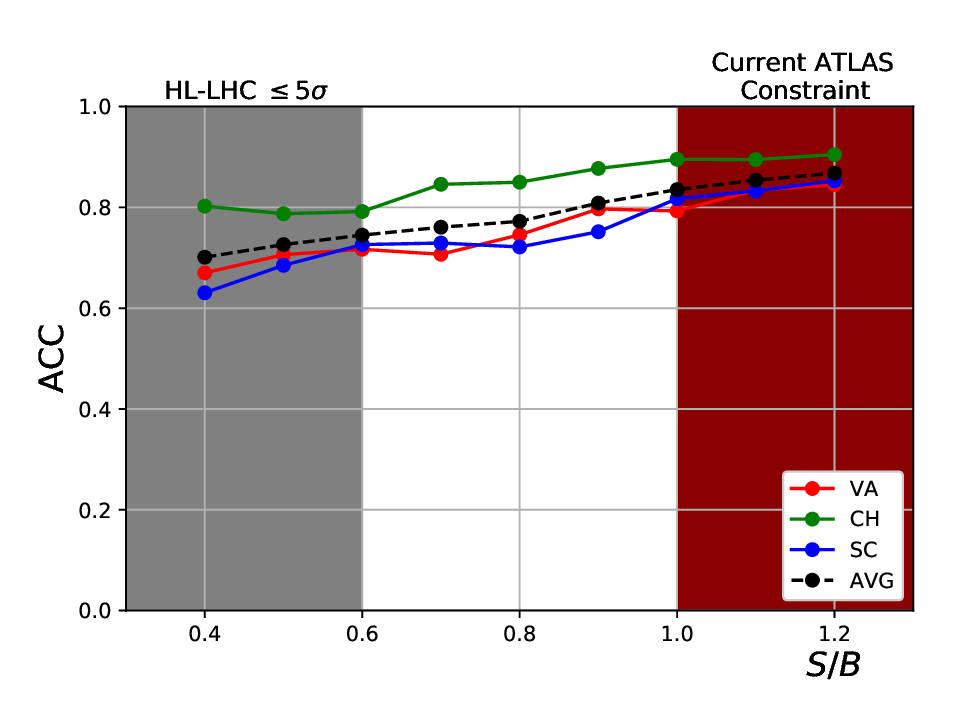}
	\vspace{-1.2cm}
	\caption{}
\end{subfigure}
\\
6~TeV
\begin{subfigure}[t!]{0.45\textwidth}
	\includegraphics[width=\textwidth]{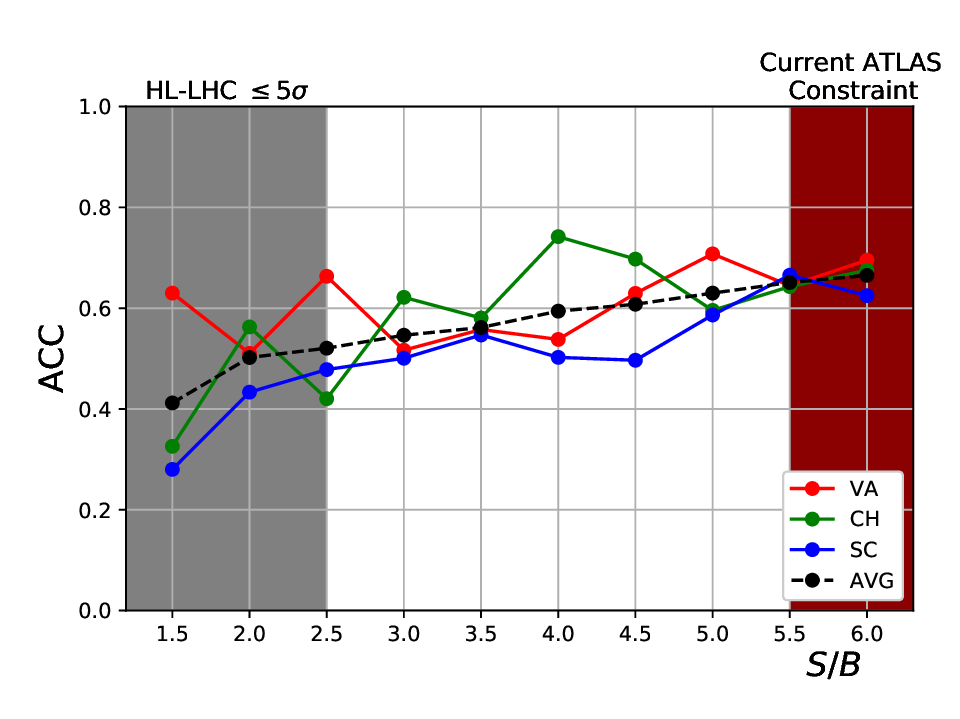}
	\vspace{-1.2cm}
	\caption{}
\end{subfigure}
\begin{subfigure}[t!]{0.45\textwidth}
	\includegraphics[width=\textwidth]{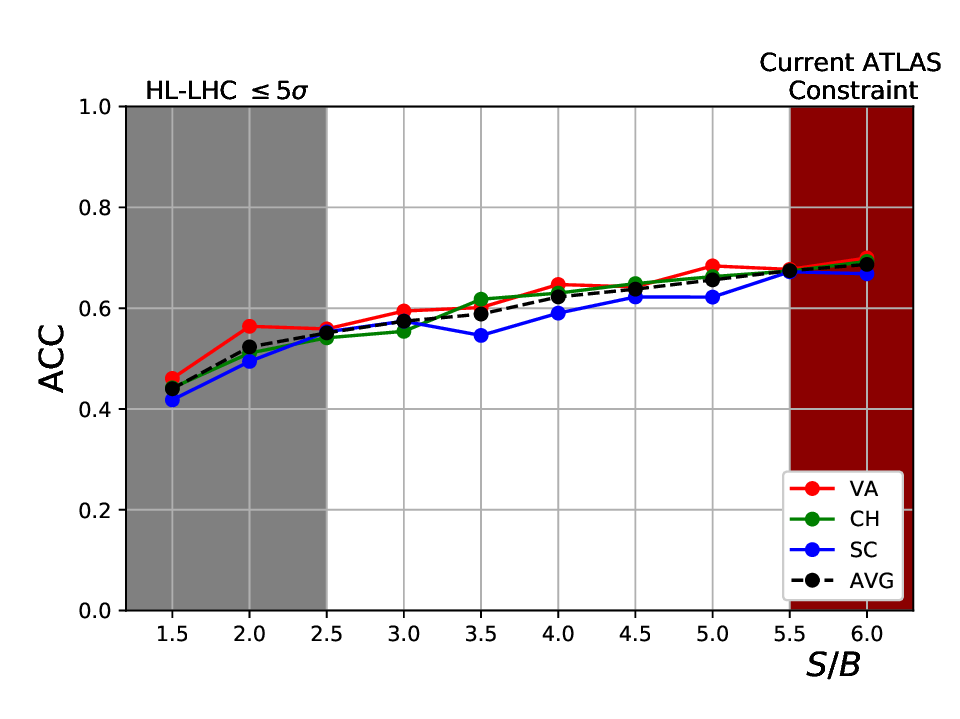}
	\vspace{-1.2cm}
	\caption{}
\end{subfigure}
\caption{1-jet (a) $4.5$ and (c) $6$-TeV resonance FNNh Scheme 3 ACCs without CV, and (b) $4.5$ and (d) $6$-TeV resonance with 10-fold CV applied.  Same color scheme as Fig.~\ref{LO:Light}.}
\label{NLO:ACC}
\end{figure}

\begin{figure}[H]
\centering
\begin{subfigure}[t!]{0.49\textwidth}
	\includegraphics[width=\textwidth]{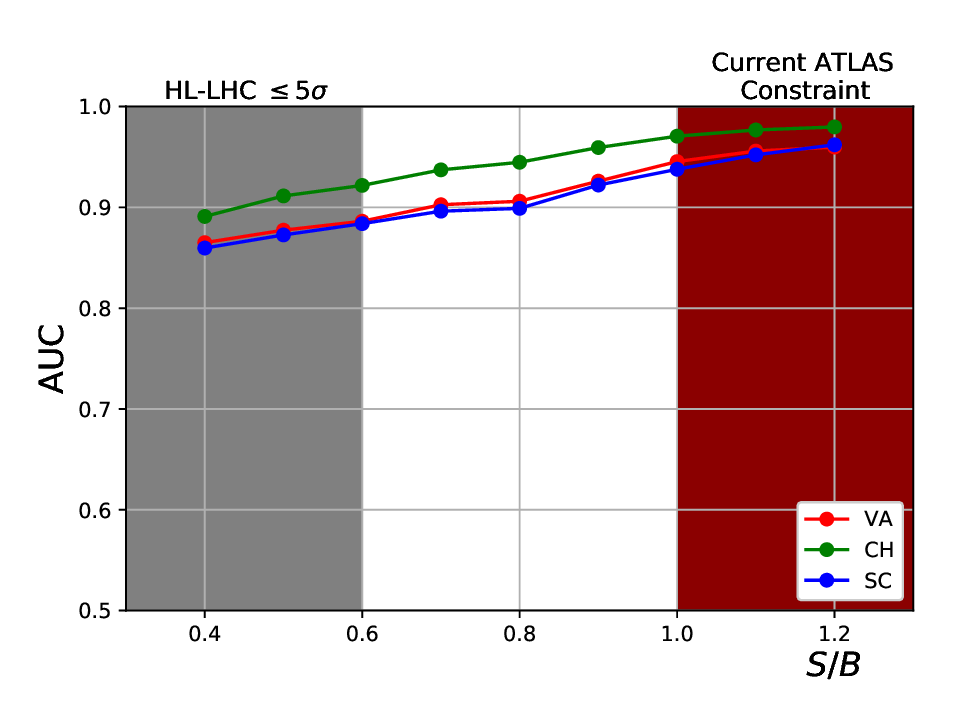}
	\vspace{-1.2cm}
	\caption{}
\end{subfigure}
\begin{subfigure}[t!]{0.49\textwidth}
	\includegraphics[width=\textwidth]{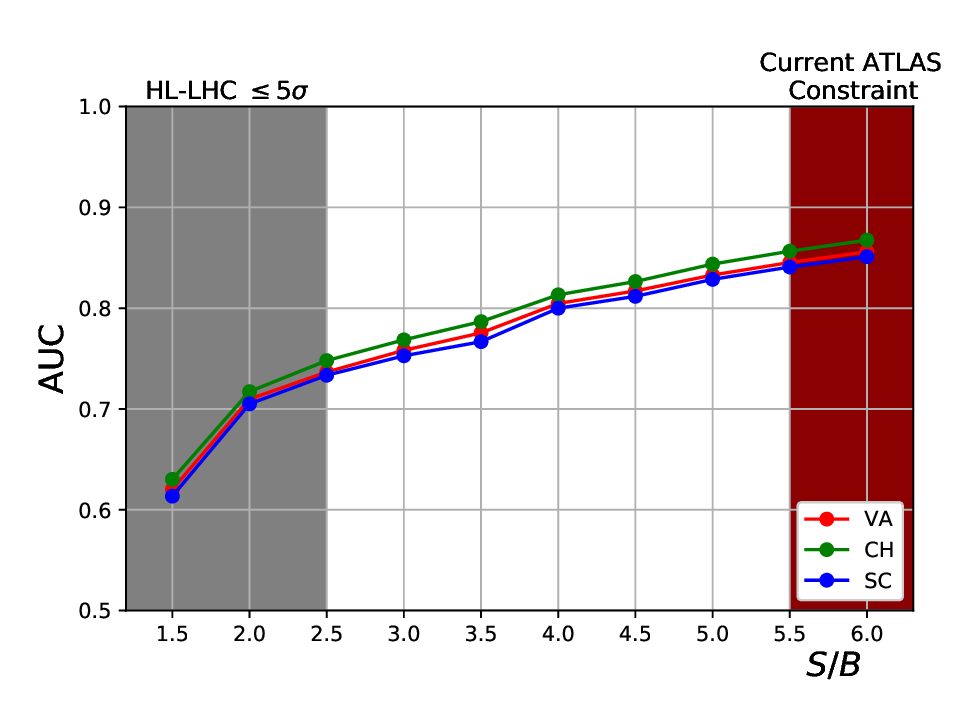}
	\vspace{-1.2cm}
	\caption{}
\end{subfigure}
\caption{1-jet AUCs for samples of (a) $4.5$ and (b) $6$-TeV resonances using FNNh in Scheme 3 with 10-fold CV applied.  Same color scheme as Fig.~\ref{LO:Light}.}
\label{NLO:AUC:CV}
\end{figure}

We present in FIG.~\ref{NLO:ACC} the FNNh ACCs without (left column) and with (right column) CV for 1-jet processes, and in FIG.~\ref{NLO:AUC:CV} the FNNh AUCs with CV applied. Compared to the 0-jet results, CV does an even better job stabilizing the 1-jet results. For $4.5$-TeV resonances, the AUCs are on the average $4-5\%$ better and the average ACCs $6-7\%$ better than those without CV.  On the other hand, for the $6$-TeV resonances the AUCs are boosted by $1-2\%$ and the average ACCs by $2-3\%$. Comparing the ACCs to those of the 0-jet study, we can see that they are much better except for the $6$-TeV case when $S/B=1.5$, indicating that the 1-jet strategy is more powerful in distinguishing different interaction hypotheses.

\begin{figure}[H]
\centering
\begin{subfigure}[t!]{0.49\textwidth}
	\includegraphics[width=\textwidth]{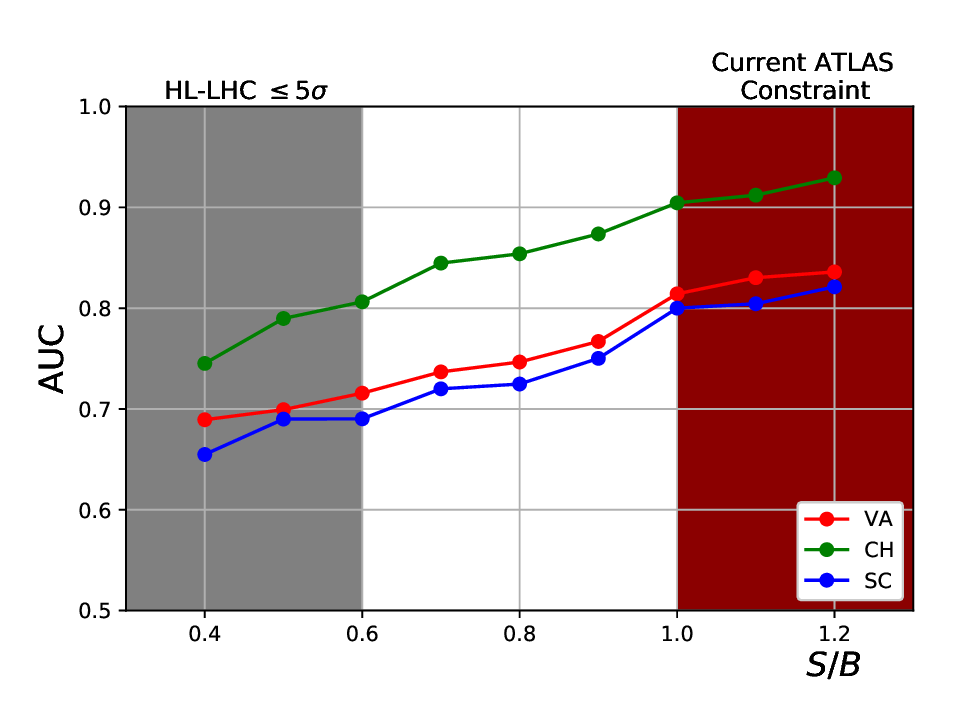}
	\vspace{-1.2cm}
	\caption{}
\end{subfigure}
\begin{subfigure}[t!]{0.49\textwidth}
	\includegraphics[width=\textwidth]{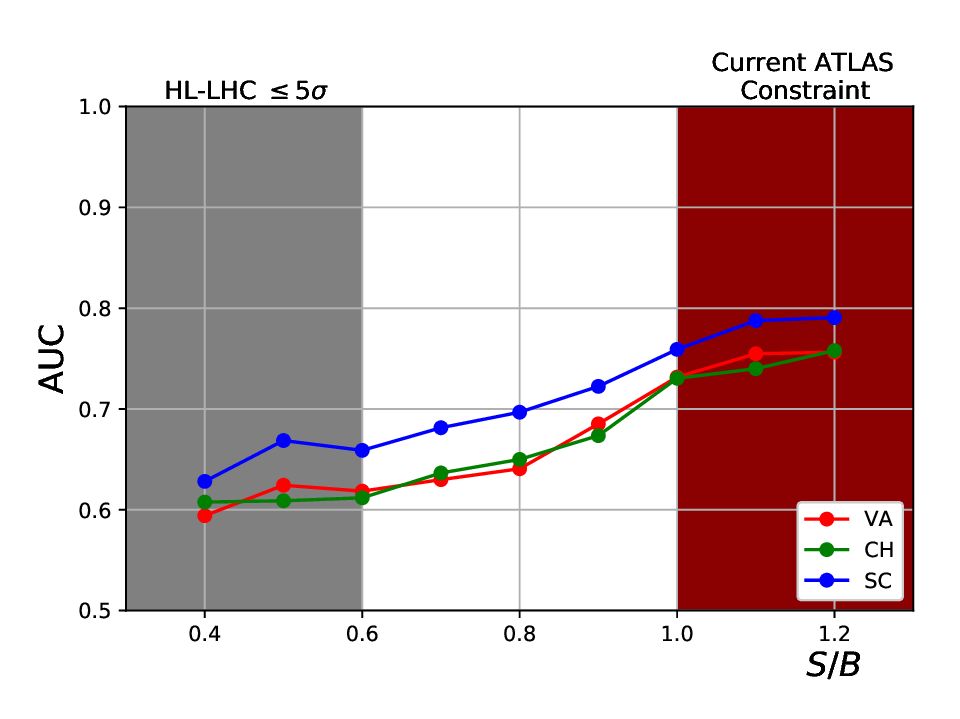}
	\vspace{-1.2cm}
	\caption{}
\end{subfigure}
\begin{subfigure}[t!]{0.49\textwidth}
	\includegraphics[width=\textwidth]{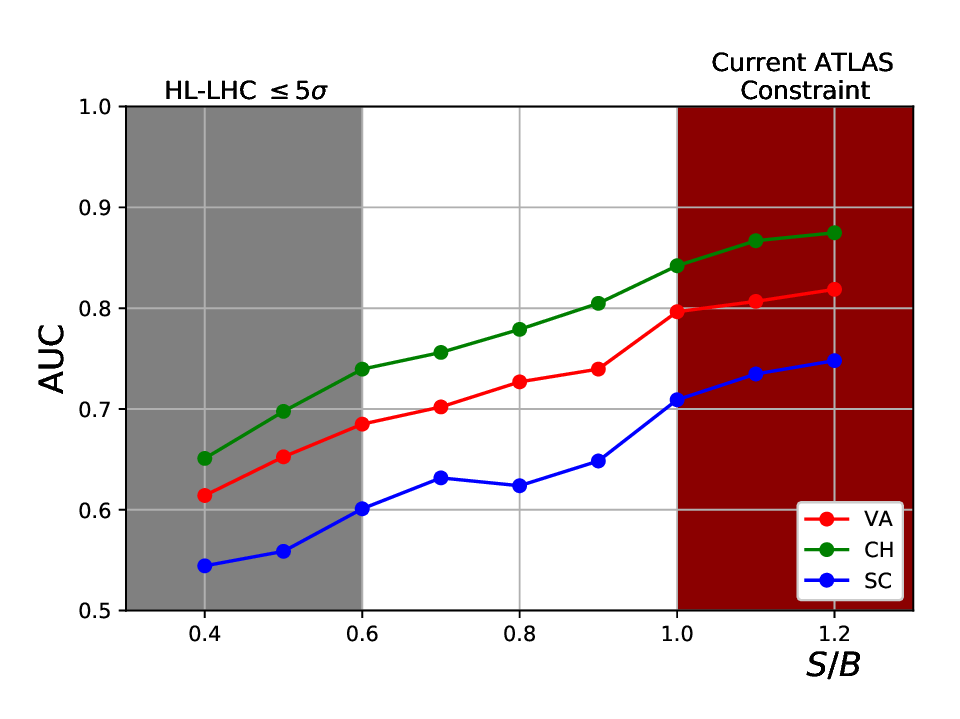}
	\vspace{-1.2cm}
	\caption{}
\end{subfigure}
\begin{subfigure}[t!]{0.49\textwidth}
	\includegraphics[width=\textwidth]{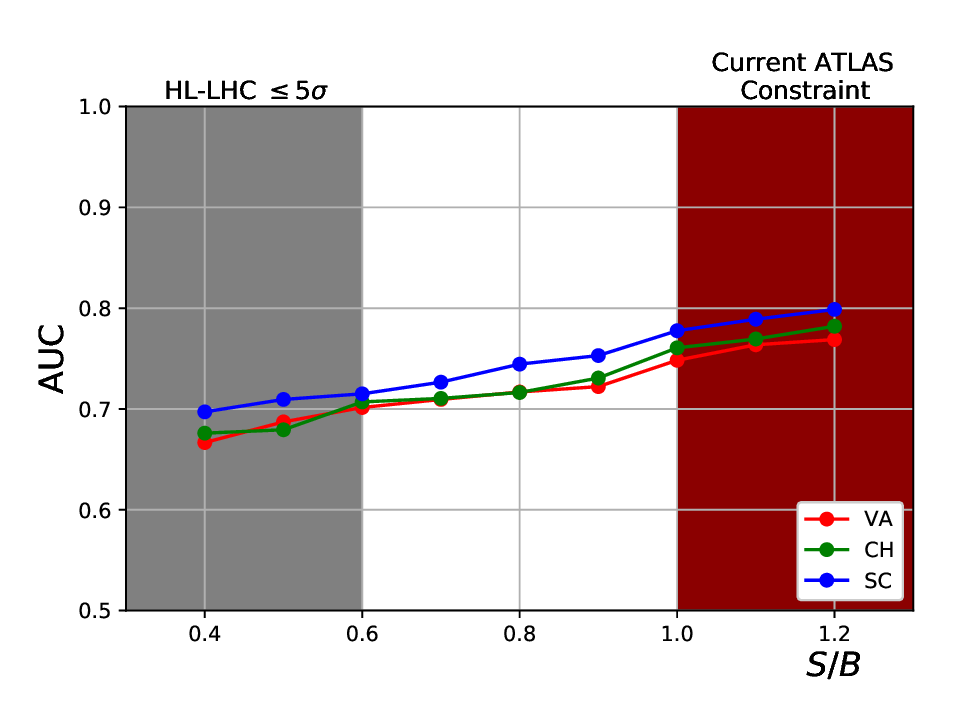}
	\vspace{-1.2cm}
	\caption{}
\end{subfigure}
\caption{1-jet FNNh training outcomes for $4.5$-TeV resonances using individual channels: (a) $p_T^e$ vs. $\eta^e$, (b) $p_T^e$ vs. $\cancel{E}_T$, (c) $\eta^e$ vs. $\eta^{j}$, and (d) $p_T^e$ vs. $\Delta\phi_{ej}$.  Same color scheme as Fig.~\ref{LO:Light}.}
\label{NLO:individual}
\end{figure}

To understand the importance of each individual variable pair in the 1-jet FNNh training, we have also trained the FNNh on single pair histograms for $4.5$-TeV resonances.  The FNNh training outcomes for the most powerful individual histograms mentioned in all four 1-jet schemes are shown in FIG.~\ref{NLO:individual}. We dropped the results of $p_T^j$~vs.~$\eta^j$, $\Delta\phi_{ej}$~vs.~$\Delta\phi_{e\cancel{E}_T}$, and $\Delta\phi_{e\cancel{E}_T}$~vs.~$\Delta\phi_{j\cancel{E}_T}$ here as they barely have any distinguishing power.  Clearly, $p_T^e$ vs.~$\eta^e$ plays the most important role in the class discrimination.  This is physically understandable as we expect the angular and coupling information of the leptonic decay to be preserved mostly in the charged lepton, which is a direct decay product of the new charged bosons, rather than in $j$.  Following $p_T^e$ vs.~$\eta^e$ are $\eta^e$ vs.~$\eta^j$, $p_T^e$ vs.~$\cancel{E}_T$, and $p_T^e$ vs.~$\Delta\phi_{ej}$, with the first two best at identifying the CH class and the latter two identifying the SC class. Compared to FIG.~\ref{NLO:Light}, we see that combining different channels does lead to a better overall performance, thus demonstrating that the multi-dimensional FNNh can successfully utilize the additional information in these channels.

\begin{figure}[H]
\centering
\begin{subfigure}[t!]{0.45\textwidth}
	\includegraphics[width=\textwidth]{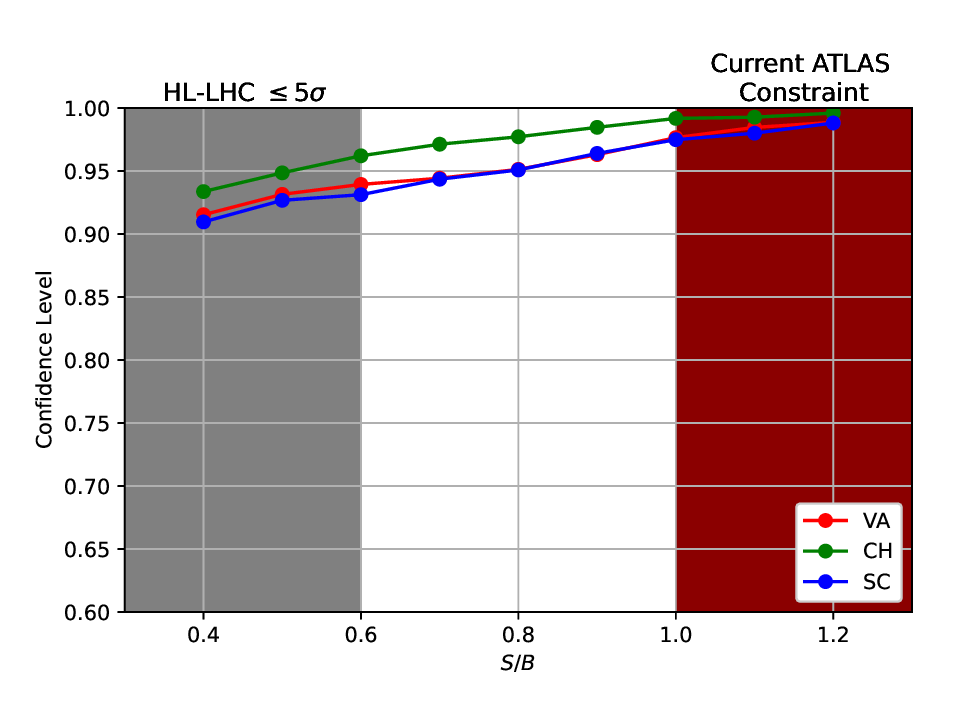}
	\vspace{-1.2cm}
	\caption{}
\end{subfigure}
\begin{subfigure}[t!]{0.45\textwidth}
	\includegraphics[width=\textwidth]{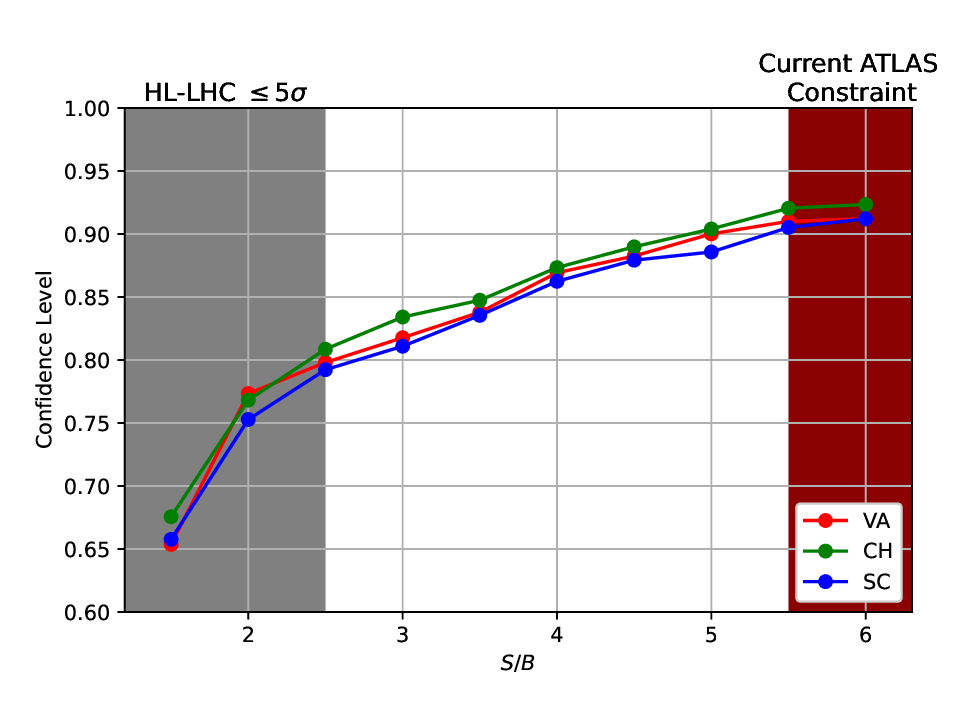}
	\vspace{-1.2cm}
	\caption{}
\end{subfigure}
\caption{Median 1-jet confidence levels at which the non-VA (red), non-CH (green), and non-SC (blue) hypotheses are excluded by the trained FNNh for samples of (a) $4.5$-TeV and (b) $6$-TeV resonances when assuming the VA, CH, and SC hypotheses are true, respectively.}
\label{CL:NLO}
\end{figure}

Finally, we also analyze the CL at which the FNNh can rule out alternative hypotheses.  We plot these CLs against the $S/B$ values for both $4.5$- and $6$-TeV resonances in Fig.~\ref{CL:NLO}.  For $6$-TeV resonances, all the alternative classes can again be excluded at a CL $\gtrsim 80\%$ in the $S/B$ region of our interest, while for $4$-TeV resonances, they can reach CLs of over $95\%$. Both of these are better than their 0-jet counterparts, providing us yet another metric to highlight the improvement.

\section{Comparison with Bayesian hypothesis test}\label{sec:6}

Finally, to give context for our NN approach, we compare the 0- and 1-jet FNNh $4.5$-TeV-resonance results with a standard hypothesis test, the Bayesian hypothesis (BH) test.\footnote{We have also tried to compare with the $\chi^2$ test.  However, it suffers from serious issues in the presence of bins with small or zero expected events, which has to be resolved through coarser binning and decomposing the ternary test to multiple binary tests.}
%
%
In the Bayesian approach, for a specific observed dataset $D$, the probability for it to suggest a specific hypothesis $H^k$ is given by
\begin{equation}
	P(H^k\vert D)=\frac{P(D\vert H^k)\times P(H^k)}{\sum_kP(D\vert H^k)\times P(H^k)}
	~,
\end{equation}
where $P(H^k)$ denotes the prior that the hypothesis $H^k$ is correct, and $P(D\vert H^k)$ gives the conditional probability to obtain dataset $D$ given the fact that $H^k$ is correct.
In our study, we assume that it is equally likely for all the hypotheses ($k =$ VA, CH, and SC) to be correct and hence $P(H^k)=1/3$. We assume Poisson distributions for all individual bin counts, and the conditional probabilities $P(D\vert H^k)$ are then given by
\begin{equation}
	P(D\vert H^k)=\prod_{m,n}f(h^D_{mn},H^k_{mn})
	~,
	\label{BH:Poisson}
\end{equation}
where $f(h^D_{mn},H^k_{mn})$ denotes the Poisson probability for an observed number of counts $h^D_{mn}$ at the pixel $(m,n)$ in the 2D histogram, assuming an expectation value of $H^k_{mn}$.  From the definition of Eq.~(\ref{BH:Poisson}), one can see that there would be a problem if any $H^k_{mn}=0$ because an observed count in this pixel would have an extremely high weight in determining the hypothesis. An even worse case is that suppose $H^k_{mn}=0$ while $H^{k'}_{mn}\neq0$ for $k' \neq k$, any sample histogram with $h^D_{mn}\neq0$ will definitely have zero probability to be identified as class $k$. This results from the fact that this approach does not take into account possible systematic or statistical errors, and hence it cannot be trusted around these low-statistics regions.  To overcome this problem, we first symmetrize $H^k$ with respect to the $\eta^e = 0$ axis and then exclude the problematic bins (those where $H^k_{mn}=0$ while $H^{k'}_{mn}\neq0$). To fairly compare the BH test with the FNNh, we apply the same binning configuration to the training and testing samples with 10-fold CV.

\begin{figure}[H]
\centering
0-jet
\begin{subfigure}[t!]{0.45\textwidth}
	\includegraphics[width=\textwidth]{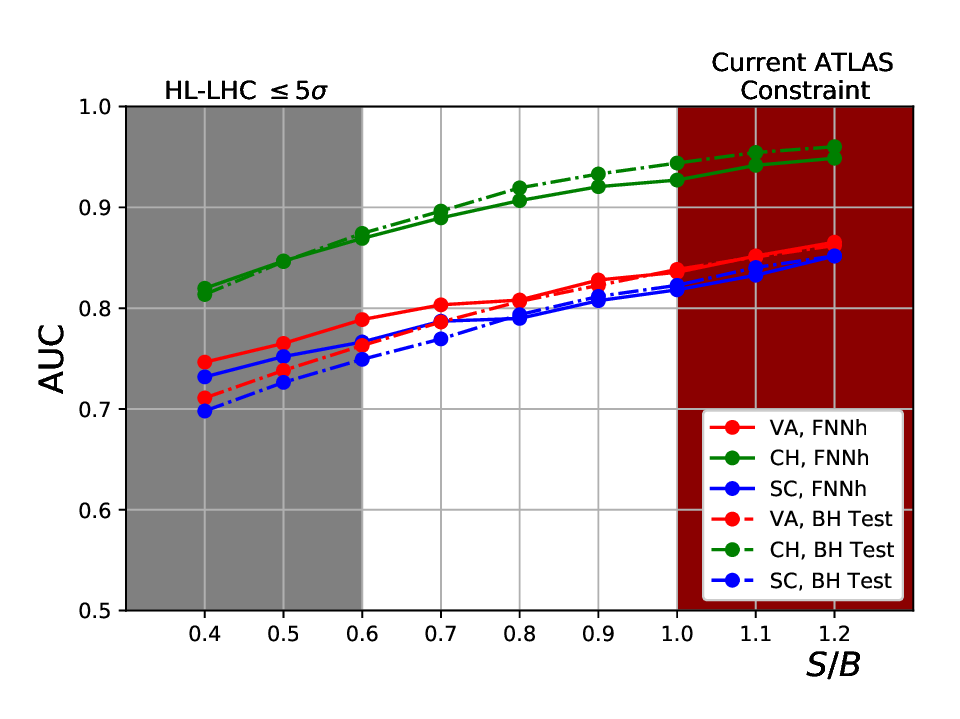}
\end{subfigure}
\begin{subfigure}[t!]{0.45\textwidth}
	\includegraphics[width=\textwidth]{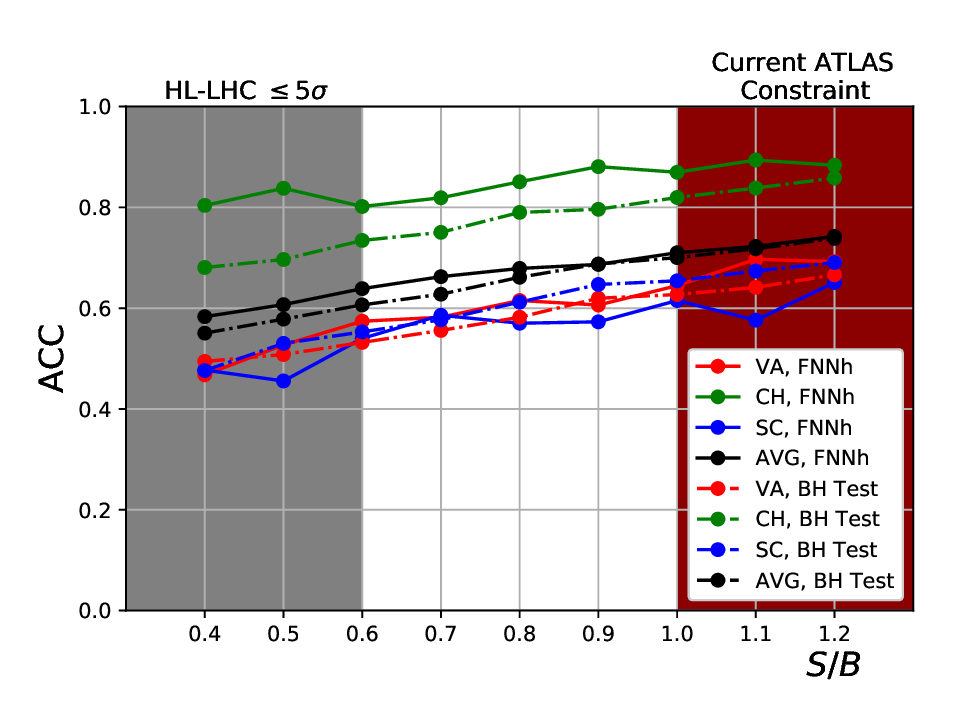}
\end{subfigure}
\\
1-jet
\begin{subfigure}[t!]{0.45\textwidth}
	\includegraphics[width=\textwidth]{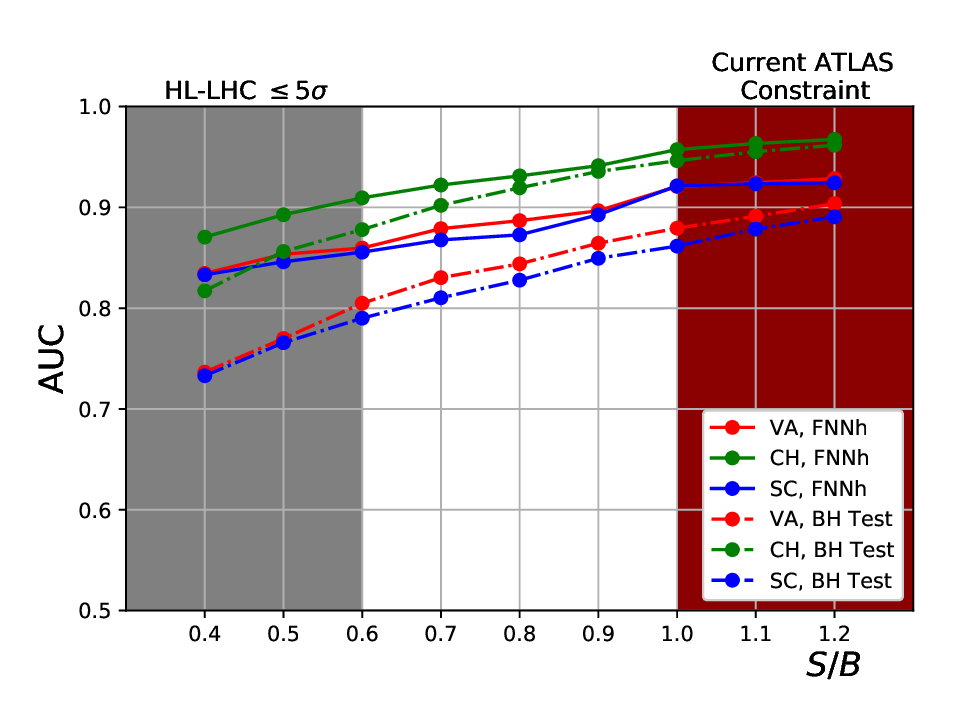}
\end{subfigure}
\begin{subfigure}[t!]{0.45\textwidth}
	\includegraphics[width=\textwidth]{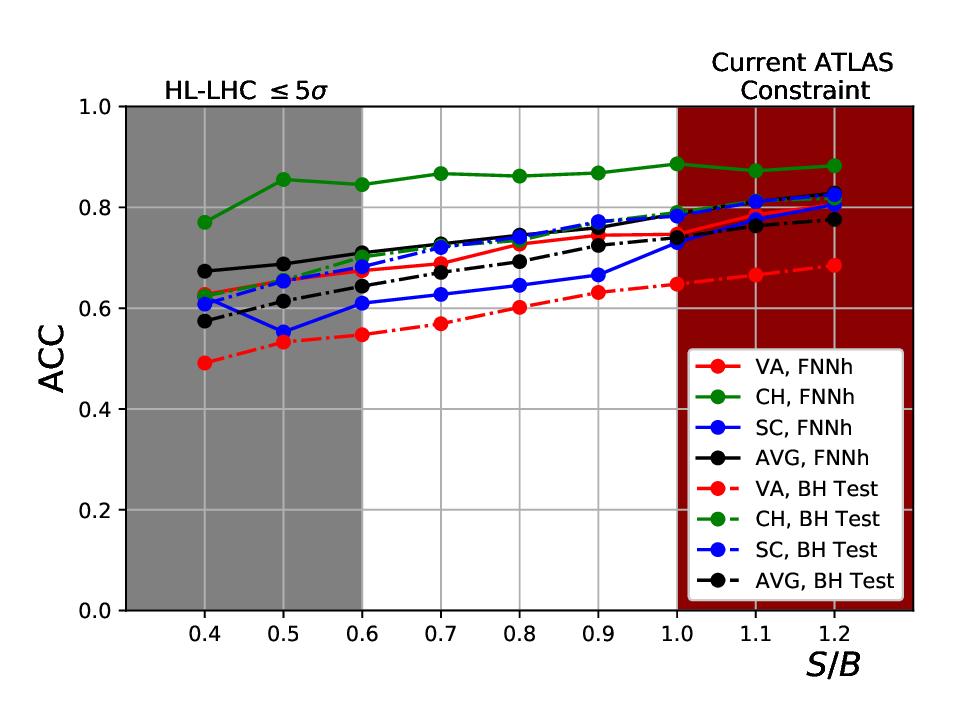}
\end{subfigure}
\caption{(a) 0-jet AUCs, (b) 0-jet ACCs, (c) 1-jet Scheme 3 AUCs, and (d) 1-jet Scheme 3 ACCs for $4.5$-TeV resonances, obtained using FNNh with 10-fold CV and the BH test.  The histogram dimension is set to $60\times60$ and the binning configuration mentioned in the main text is applied.  Same color scheme as Fig.~\ref{LO:Light}.}
\label{BH}
\end{figure}

The 0-jet AUCs and ACCs for both FNNh with 10-fold CV (solid lines) and BH tests (dashed lines) are shown in FIG.~\ref{BH}~(a) and (b) respectively. The VA, SC AUCs as well as the VA, CH, and average ACCs indicate that the FNNh approach is able to produce the same level of performance as the BH test for $S/B\gtrsim0.8$, and even performs better when $S/B\lesssim0.8$. One exception happens for the CH AUCs when $S/B\gtrsim0.6$ although the difference is less than $1.5\%$.  Another exception shows up in the SC ACCs, where the BH test is better than FNNh. 

This kind of binning strategy used in the BH test could be more difficult to implement in other cases.  For example, for a higher mass resonance the $p_T$ range to be studied would be wider, with more chances to get empty bins.  Therefore, either more events need to be generated, or more bins need to be excluded, or the bins should be made coarser; otherwise, the BH test cannot be applied properly.  Another complication could occur if more kinematic variables are needed.  As the dimension of the phase space to be studied increases, proper binning will become more challenging.
In fact, we have encountered such an issue when we turn to 1-jet samples.  To compare with the 1-jet FNNh in  Scheme 3, we performed a BH test using
\begin{equation}
	P(D\vert H^k)=\prod_{a=1}^3P(D_a\vert H^k_a)
	~,
\end{equation}
where $a$ denotes the three input channels.\footnote{In principle, a 4D version of the BH test could be done with the full knowledge of the probability density function in the 4D phase space of $(p_T^e,\cancel{E}_T,\eta^e,\Delta\phi_{ej})$.  This is computationally challenging.  But it would be interesting to compare with either a 4D CNN or a 6-color 2D CNN taking in all the variables.}  The resulted AUCs and ACCs are shown in FIG.~\ref{BH}(c) and (d). We can see that for the 1-jet case, FNNh is consistently better than the BH test in terms of all metrics but the SC ACC. Thus, we can conclude that FNNh generally has a better adapting ability against the binning issue, as well as a stronger power to combine information from different channels, except when it comes to the identification of SC signals. However, as the previous results have shown, FNNh does not suffer from the binning issue and thus can in general improve.

As these two comparisons illustrate, the FNNh compares favorably in performance with the standard BH test in most cases.  To summarize the pros and cons compared to the BH test, the FNNh has the advantages that it automatically takes care of the binning issue, does not require a large sample to approximate the probability density functions, and easily generalizes to higher dimensions while it has the usual neural network disadvantages of proper training and validation.

\section{Conclusions}\label{sec:7}

In this paper, we have investigated the ability of using deep neural networks to distinguish different resonances in the $pp\rightarrow W'/H\rightarrow \ell\nu_\ell$ process at the HL-LHC.  We showed that the original event-by-event ambiguities in the coupling differentiation problem could be tackled by classifiers with a neural network (NN) architecture that takes binned histograms as the input.  The predicted $p_T^e$ distributions allow a discrimination between $H$ and $W'$, and because of the boosted parton collision frame, $W'$ bosons with different couplings further manifest different $\eta$ distributions.

Extending previous signal-only analyses~\cite{Khosa:2019kxd}, we demonstrated that simple NNs could start distinguishing the signals even with low signal-to-background ratio, $S/B$.  Of the three NN approaches we studied, the best was a fully connected neural network whose inputs were flattened histograms of kinematic variables (FNNh), though FNNi could potentially compete with it if more careful fine-tunings were applied as pointed out in Ref.~\cite{Nachman:2021yvi}.  We found the FNNh could achieve AUCs over $0.80$ when $S/B\gtrsim 0.8$ for $4.5$-TeV resonances, and over $0.60$ when $S/B\gtrsim3.0$ for $6$-TeV resonances.  As our 1-jet schemes showed, the 2D approach of Ref.~\cite{Khosa:2019kxd} could also be generalized to higher dimensions, where we took into account the extra information of the jet by using the ``RGB'' channels to represent different kinematic variable pairs. This additional jet information compensated for the drop in the event statistics, generally leading to better performance for both $4.5$- and $6$-TeV resonances. We also investigated the usefulness of cross validation, and discovered that it helped less for the 0-jet study, but boosted the performance and stability much for the 1-jet study. Performance differences resulting from different boson widths and the four pairing schemes were also investigated, and it was concluded that there was no major difference among the training results.

Finally, we studied the importance of each individual variable pair in the 1-jet FNNhs, and found that they had different discriminating power for the three signal classes, with some variable pairs being more suited to picking out certain classes.  Out of all the variable pairs, the FNNh still relied mostly on the information of the charged lepton, although our results showed that the RGB color scheme successfully combined multiple channels to produce a better overall performance.  As a final comparison, we also showed that this technique was as good or better than the conventional Bayesian hypothesis testing procedure, without having to worry about binning issues or how to generalize to higher dimensions.

Even though this study is based upon the specific choice of $4.5$ and $6$-TeV mass for the new charged resonance, it can be readily extended to other mass ranges at future colliders, in which case sufficient event statistics apparently is a critical factor for the success of the NN technique. Moreover, more general studies can also be considered, such as modifying the hypotheses (e.g. spin and couplings), analyzing channels other than the 0- and 1-jet processes presented here, or constructing NN's with inputs of more than three channels and higher-dimensional ``super-images.''

\begin{acknowledgements}
The work of SC was supported in part by the U.S. Department of Energy under Grant Number DE-SC0011640.  The works of TKC and CWC were supported in part by the Ministry of Science and Technology (MOST) of Taiwan under Grant Number MOST-108-2112-M-002-005-MY3.
We appreciate the support of NVIDIA Corporation with the donation of a Titan Xp GPU used in this study.  We also thank Kai-Feng Chen for the suggestions about hypothesis tests and Yu-Chen Janice Chen for support and discussion about NNs.  SC thanks the hospitality of the Physics Department of National Taiwan University and the sabbatical support of the MOST of Taiwan when this project was initiated.
\end{acknowledgements}

\appendix
\section{Technical studies}\label{sec:a}

To better understand the technical details of our method, we investigate the dependence of the 0-jet FNNh on variables such as the resolution and kinematic window.  We also confirm the consistency between 0- and 1-jet binary and ternary classifiers by introducing a projection of scores in the latter case, which has also been studied in Ref.~\cite{Chen:2019uar}.  Finally, we demonstrate how robust the performance of the FNNh is even when applied to testing samples from a distribution which it is not trained on with different $S/B$ ratios and decay widths. All the following studies are based upon $4.5$-TeV resonances.

\subsection{Kinematic window and resolution}

We expect the performances of the NNs to be better if we extend the phase space from $p_{T}^{e}>1350$~GeV to a lower $p_{T}^{e}$ minimum as it would include more information about the signal.  However, there are two problems associated with an unchecked extension of this lower bound:
\begin{itemize}
\item First, when $p_T^{e}$ gets closer to $m_W/2$, the number of NP signals will be overwhelmed by the number of SM signals around the $W$ boson Jacobian peak.  Therefore, including information from this region would contribute little to none.  What is even worse is that the excess of SM signals may confuse the NNs and reduce its efficiency.
 
\item Second, if one were to maintain the same $p_T$ resolution for the histogram bins, the required NN complexity and computational resources for training would increase rapidly as $p_{T,min}$ lowers.  Yet if one wants to maintain the same level of input bins for the NNs, the resolution in $p_T^e$ would be compromised.
\end{itemize}

As a result, we expect a ``sweet window'' that balances among these issues.  We base our study upon samples of $S/B=0.4$ and $1.2$ under the cuts given in TABLE~\ref{Sample:cut}. To extend $p_T^{e}$ to lower regions, we first define the following parameters:
\begin{center}
\centering
	$B$, $B'(k)$: numbers of SM events for $p_T^e\geq1350$~GeV, $p_{T,min}$, respectively. \\
	$S_{\text{c}}$, $S'_{\text{c}}(k)$: numbers of class c events for $p_T^e\geq1350$~GeV, $p_{T,min}$, respectively.
\end{center}

We generate another set of samples based upon the same settings as before, but change the selection cut from $p_T^e>1350$~GeV to $p_T^e>750$~GeV as we are setting $p_{T,min}$ to $750$. Introducing the ratios $r_B \equiv B' / B$ and $r_{S_c} \equiv S'_{\text{c}} / S_{\text{c}}$, the mixing ratio between the NP and SM events should then be modified to
\begin{equation}
\label{new_ratio}
	\frac{S'_{\text{c}}}{B'}=\frac{r_{S_{\text{c}}}}{r_B}\frac{S_{\text{c}}}{B}
	~.
\end{equation}

In general, $r_{S_{\text{VA}}}$, $r_{S_{\text{CH}}}$, and $r_{S_{\text{SC}}}$ are all different. This would lead to histograms with different number of events.  Instead, we define $r_S \equiv \sum_c r_{S_{\text{c}}} / 3$ and mix the new samples of all three classes according to:
\begin{equation}
 \frac{S'}{B'}=\frac{r_S}{r_B}\frac{S}{B}
 ~,
\end{equation}
This procedure is carried out for $p_{T,min} = 750,950,\cdots,2150$~GeV, respectively, and the corresponding histograms of dimension $60\times60$ are then made from the mixed samples.  Note that in these studies, we fix the bin size of $\eta^e$.

\begin{figure}[H]
\centering
\begin{subfigure}[t!]{0.49\textwidth}
	\includegraphics[width=\textwidth]{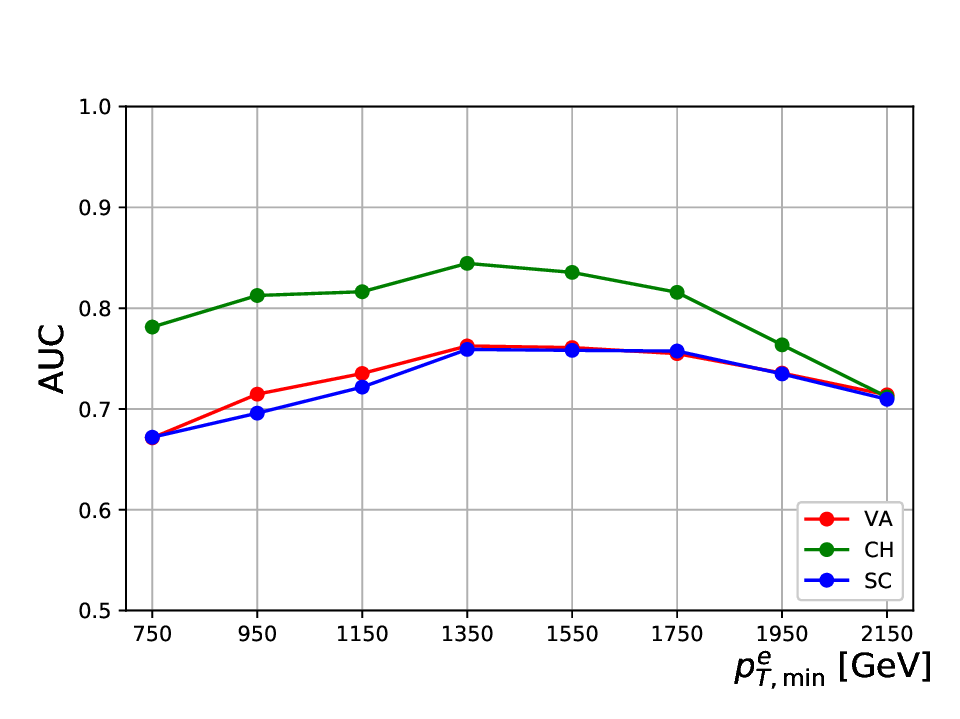}
	\vspace{-1.2cm}
	\caption{}
\end{subfigure}
\begin{subfigure}[t!]{0.49\textwidth}
	\includegraphics[width=\textwidth]{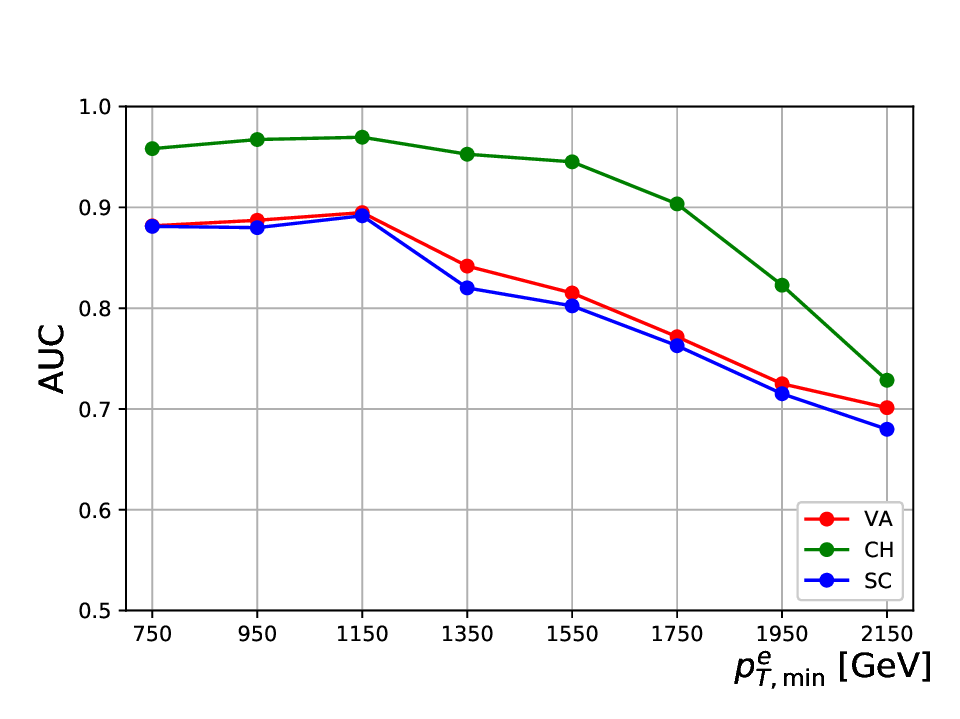}
	\vspace{-1.2cm}
	\caption{}
\end{subfigure}
\caption{0-jet AUCs of training upon histograms of different $p_{T,min}$, with their dimensions fixed to $60\times60$ and covering the entire $p_T$ range. The histograms are  made from samples of $S/B=$ (a) $0.4$ and (b) $1.2$. The $\eta^e$ resolution remains the same as the value implemented in previous training.}
\label{Low_Bound}
\end{figure}

The AUCs of FNNh trained upon 0-jet histograms of different $p_{T,min}$ are plotted in FIG.~\ref{Low_Bound}. It is clear that there exists a ``sweet window'' for the cut at $1350$~GeV for $S/B=0.4$ and within $[950,1150]$ for $S/B=1.2$. The performance deteriorates for $p_{T,min}^e$ either lower or higher than the window boundaries. The windows for the two different $S/B$ ratios are different because with higher $S/B$, it is more likely to get ``useful'' signals as $p_{T,min}$ lowers, and hence the optimal $p_T$ cut which balances the previous issues should naturally lie somewhat lower.

\begin{figure}[H]
\centering
\begin{subfigure}[t!]{0.49\textwidth}
	\includegraphics[width=\textwidth]{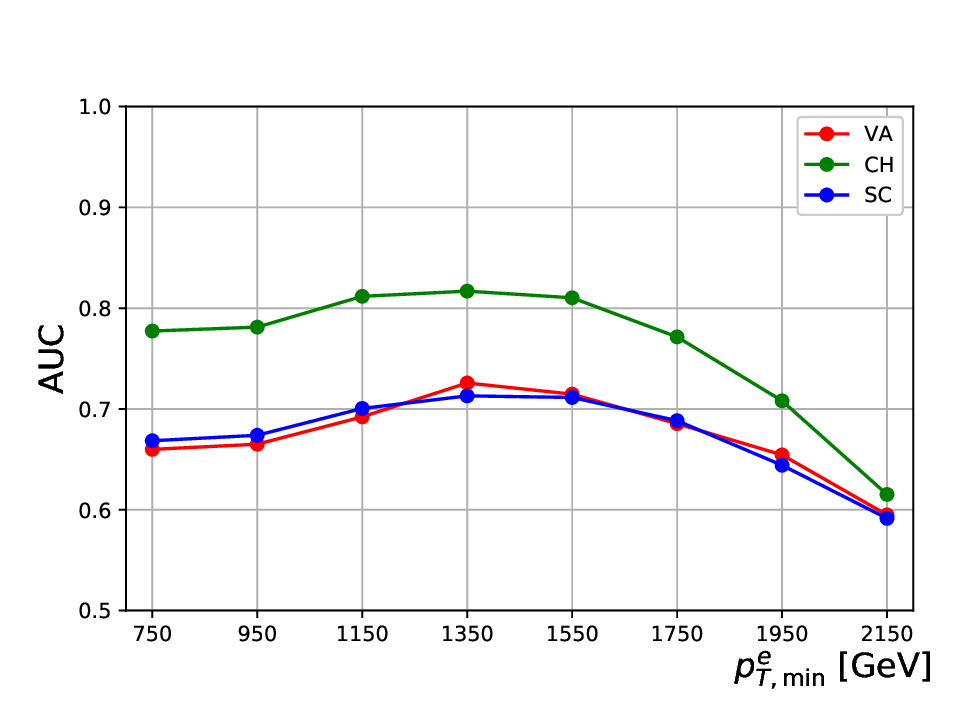}
	\vspace{-1.2cm}
	\caption{}
\end{subfigure}
\begin{subfigure}[t!]{0.49\textwidth}
	\includegraphics[width=\textwidth]{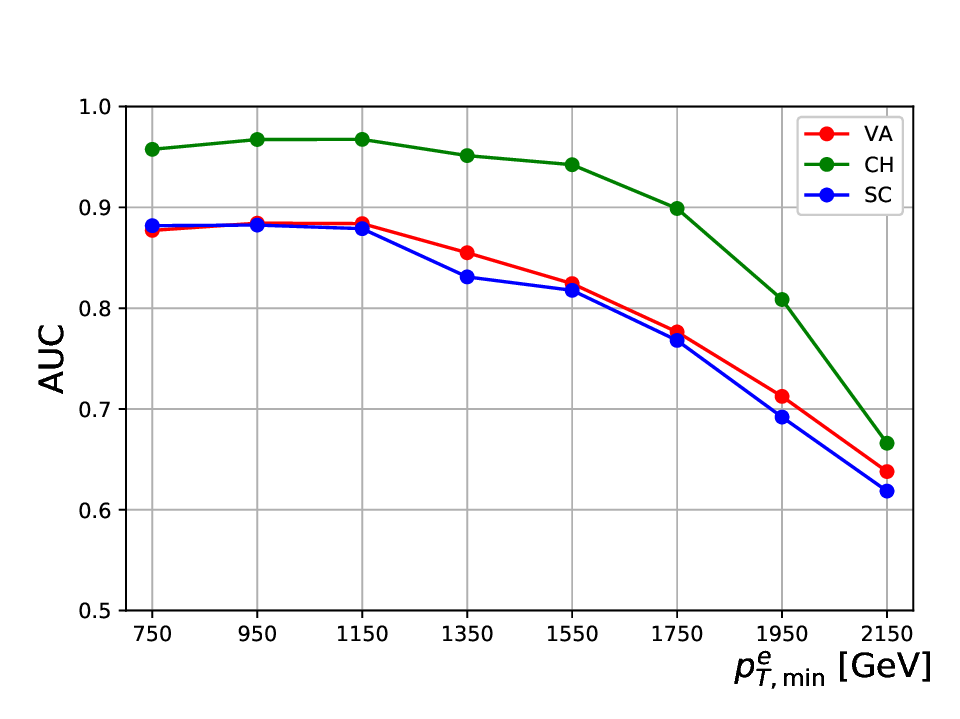}
	\vspace{-1.2cm}
	\caption{}
\end{subfigure}
\caption{0-jet AUCs of training upon histograms of different $p_{T,min}$, with their dimensions fixed to $60\times60$, $p_T$ bin size to $30$~GeV, and the uncovered bins left empty.  The histograms are made from samples of $S/B=$ (a) $0.4$ and (b) $1.2$. The $\eta^e$ bin number is fixed at $60$.}
\label{Range}
\end{figure}

To pin down whether the effect of $p_{T,min}$ is due to resolution, we also fix the $p_T$ bin size to $30$~GeV. The bins outside the $p_T$ cut are then filled with zeros so as to retain a uniform structure for our NNs.  The results are given in FIG.~\ref{Range}.  The overall trend suggests that the reduced performance of a lower $p_{T,min}$ is mainly due to incomplete training rather than the resolution, which appears to be more important for the $S/B=0.4$ scenario due to lower event numbers.  

\begin{figure}[H]
\centering
\begin{subfigure}[t!]{0.49\textwidth}
	\includegraphics[width=\textwidth]{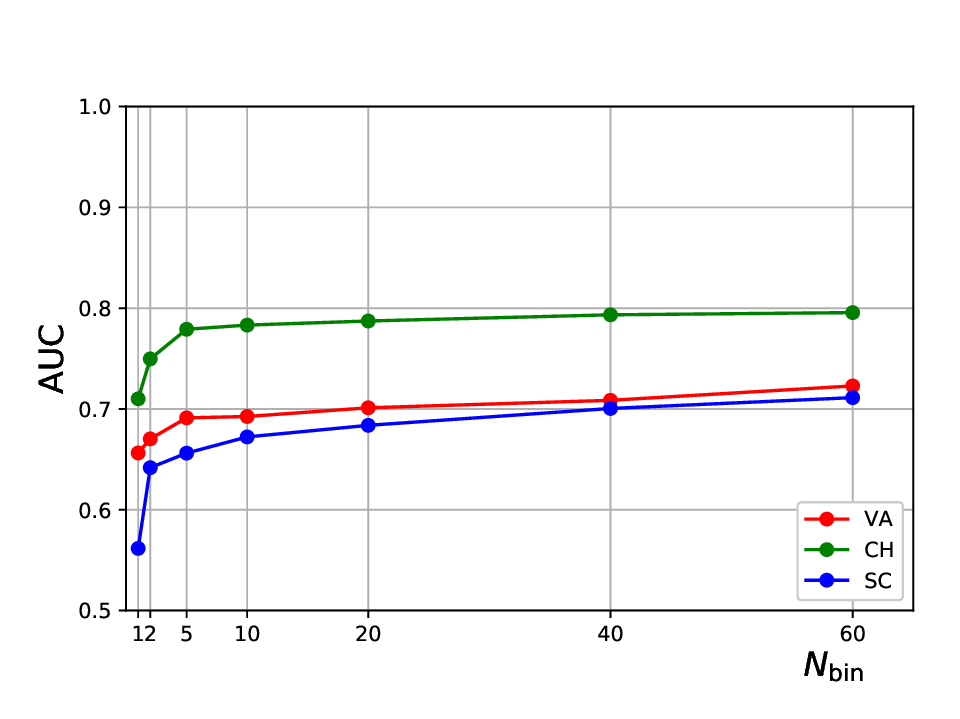}
	\vspace{-1.2cm}
	\caption{}
\end{subfigure}
\begin{subfigure}[t!]{0.49\textwidth}
	\includegraphics[width=\textwidth]{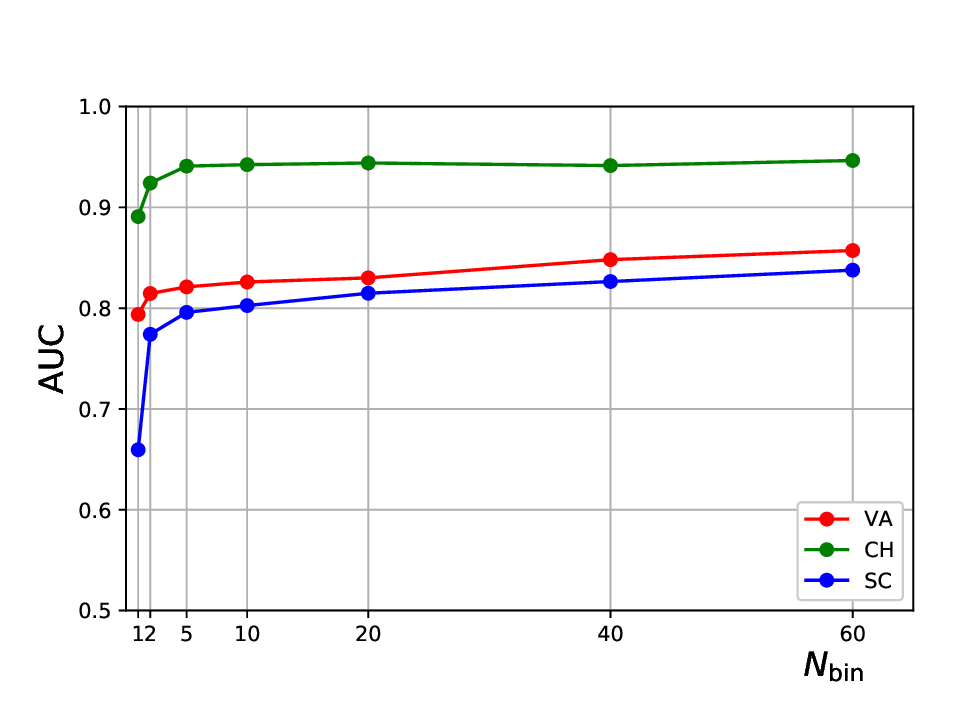}
	\vspace{-1.2cm}
	\caption{}
\end{subfigure}
\caption{0-jet AUCs of trainings upon histograms in which the $p_T^{e}$ range $[1350,2550]$~GeV is binned into $1,2,3,4,5,10,20,40,60$ bins with uncovered bins left empty.  Samples of $S/B=$ (a) $0.4$ and (b) $1.2$ are used.}
\label{LO:Reso}
\end{figure}

We further study the effect of $p_T^e$ resolution in the following way: we only use events with $p_T^e\in[1350,2550]$~GeV and partition them into $1,2,3,4,5,10,20,40,60$ bins, respectively.  Samples of $S/B=0.4$ and $1.2$ are again used.  To retain the same NN structure, we fill in null bins so that the histograms are still of dimension $60\times60$.  The training outcomes are shown in FIG.~\ref{LO:Reso}. The AUCs apparently drop as the bin number decreases, but only when there are five or fewer bins, confirming that the $p_T$ resolution does play a role in the NN performance but only when the binning is extremely coarse.  As one increases the number of bins, the AUCs nearly saturate their maximum values way before $N_{bin} = 60$.  Consequently, we can infer that as long as the $\eta^e$ resolution remains sufficiently high, the $p_T^e$ resolution does not need to be maximized to obtain the optimal NN performance.

\subsection{Consistency between binary and ternary classifiers}

Even though we are dealing with a three-class problem, one alternative other than training a ternary classifier to tag a specific sample set is to test it with multiple binary classifiers. If the NNs are all properly trained, we should expect a consistency in their performances. Therefore, we compare the two methods in the following way.

After each individual testing sample is tested by a trained ternary NN classifier, it will be assigned with a three-component score array, $(P_1,P_2,P_3)$, denoting its ``probabilities'' of belonging to one of the three classes. Suppose we are trying to compare a ternary NN's performance with that of a binary NN concerning the discrimination between class $i$ and class $j$, we \textit{project} the score components of the ternary by defining
\begin{equation}
\label{rescaling}
	P'_{k}=\frac{P_{k}}{P_i+P_j},\quad k=i,j~.
\end{equation}

\begin{figure}[H]
\centering
\begin{subfigure}[t!]{0.49\textwidth}
	\includegraphics[width=\textwidth]{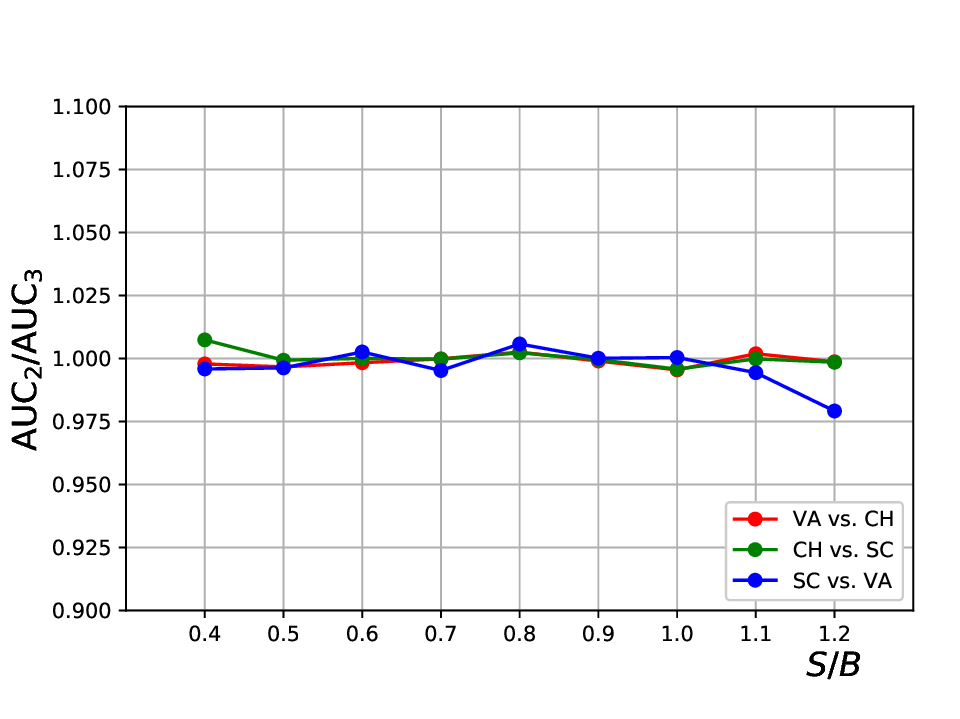}
	\vspace{-1.2cm}
	\caption{}
\end{subfigure}
\begin{subfigure}[t!]{0.49\textwidth}
	\includegraphics[width=\textwidth]{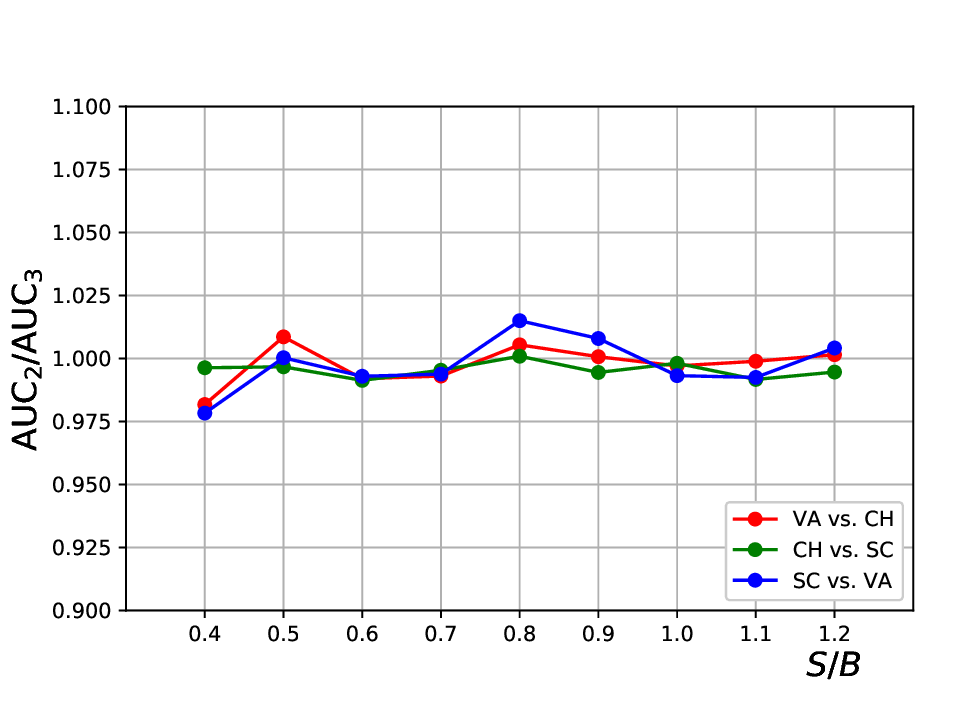}
	\vspace{-1.2cm}
	\caption{}
\end{subfigure}
\caption{AUCs ratios of the binary to the projected ternary of (a) 0-jet and (b) 1-jet dedicated to VA vs. CH, CH vs. SC, and SC vs. VA for different $S/B$ ratios. The AUCs for VA vs. CH are depicted in red, CH vs. SC in green, and SC vs. VA in blue.}
\label{ternary:binary}
\end{figure}

We then go on to compare the projected  0- and 1-jet AUCs (AUC$_3$) with the AUCs given by the true binary classifier (AUC$_2$) dedicated to classes $i$ and $j$ in terms of the ratio ${\rm AUC}_2/{\rm AUC}_3$. FIG.~\ref{ternary:binary} shows these ratios dedicated to VA vs. CH, CH vs. SC, and SC vs. VA for different $S/B$ ratios.  The plots show that the projected ternary AUCs are consistent with the binary AUCs, implying that the ternary classifier gives the same level of performance in binary classifications as the dedicated binary classifiers.  

\subsection{Applying the wrong models}

Another interesting question is what would happen if the \textit{wrong} models are applied to a set of testing samples. There are two variables to test this in our analysis: wrong significance and wrong decay widths.  In the following, we show the two corresponding tests.

\begin{figure}[H]
\centering
\begin{subfigure}[t!]{0.49\textwidth}
	\includegraphics[width=\textwidth]{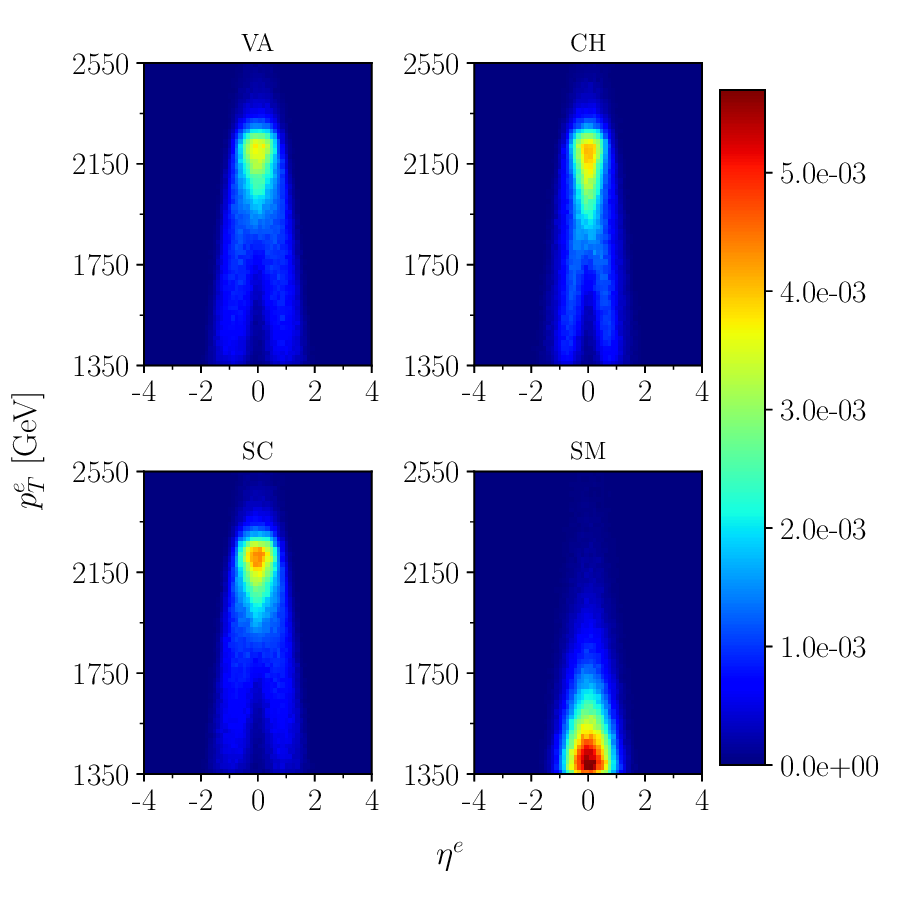}
	\vspace{-1.2cm}
	\caption{}
\end{subfigure}
\begin{subfigure}[t!]{0.49\textwidth}
	\includegraphics[width=\textwidth]{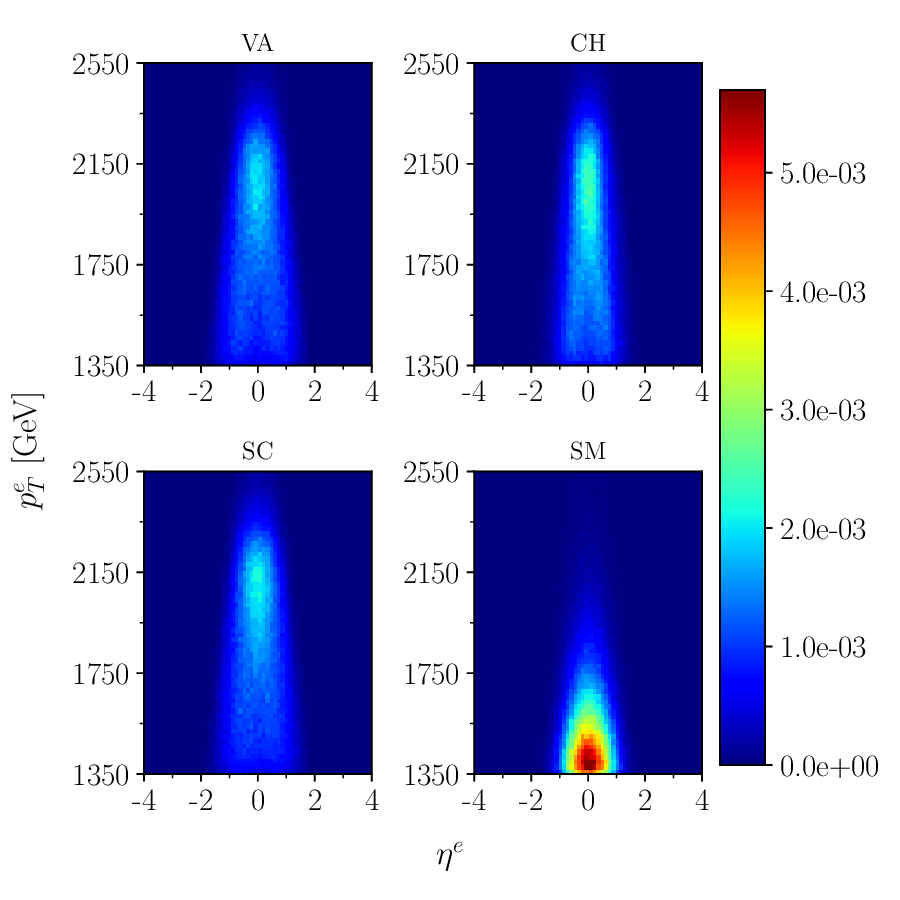}
	\vspace{-1.2cm}
	\caption{}
\end{subfigure}
\caption{$p_T^e$ vs. $\eta^e$ distributions for $4.5$-TeV resonances with $\Gamma_{\rm NP}\approx$ (a) $50$ and (b) $500$~GeV.}
\label{LO:2D}
\end{figure}

The first is to use the models trained upon 0-jet samples of $4.5$-TeV resonances with $\Gamma_{\rm NP}\approx500$~GeV to test the samples of $\Gamma_{\rm NP}\approx200$~GeV at a fixed $S/B$ and vice versa, as well as between samples of $\Gamma_{\rm NP}\approx200$~GeV and $\Gamma_{\rm NP}\approx50$~GeV, and samples of $\Gamma_{\rm NP}\approx500$~GeV and $\Gamma_{\rm NP}\approx50$~GeV. We then calculate the ratios of the ``wrong AUCs" (AUC) to the ``correct AUCs" (AUC$_0$) with respect to different significances.  To compare with FIG.~\ref{LO:Kinematics}(e), we show the $p_T^e$ vs. $\eta^e$ distributions for $\Gamma_{\rm NP}\approx500,50$~GeV in FIG.~\ref{LO:2D}. The training results are shown in FIG.~\ref{LO:Wrong_Width}. We can see that applying models of the wrong widths still has some discriminating power, yet they are consistently worse than applying the correct models.  This indicates the importance of getting the right order of magnitude for $\Gamma_{\rm NP}$ before setting up the trainings, and shows that even an incorrectly trained NN still has an AUC within $\sim 10-25\%$ of the correctly trained model.

\begin{figure}[H]
\centering
\begin{subfigure}[t!]{0.49\textwidth}
	\includegraphics[width=\textwidth]{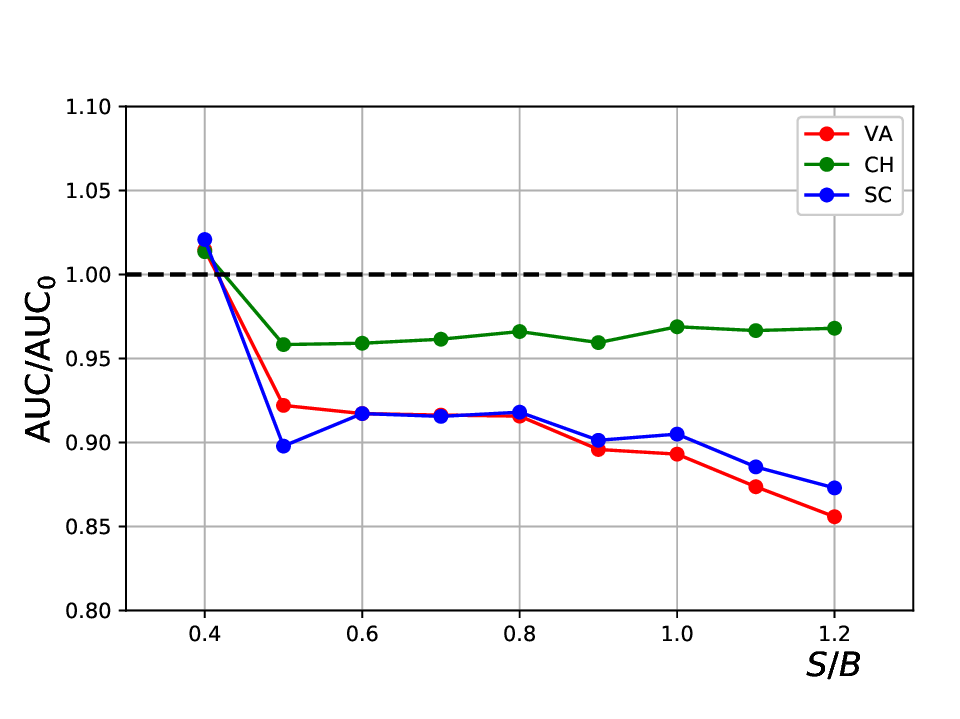}
	\vspace{-1.2cm}
	\caption{}
\end{subfigure}
\begin{subfigure}[t!]{0.49\textwidth}
	\includegraphics[width=\textwidth]{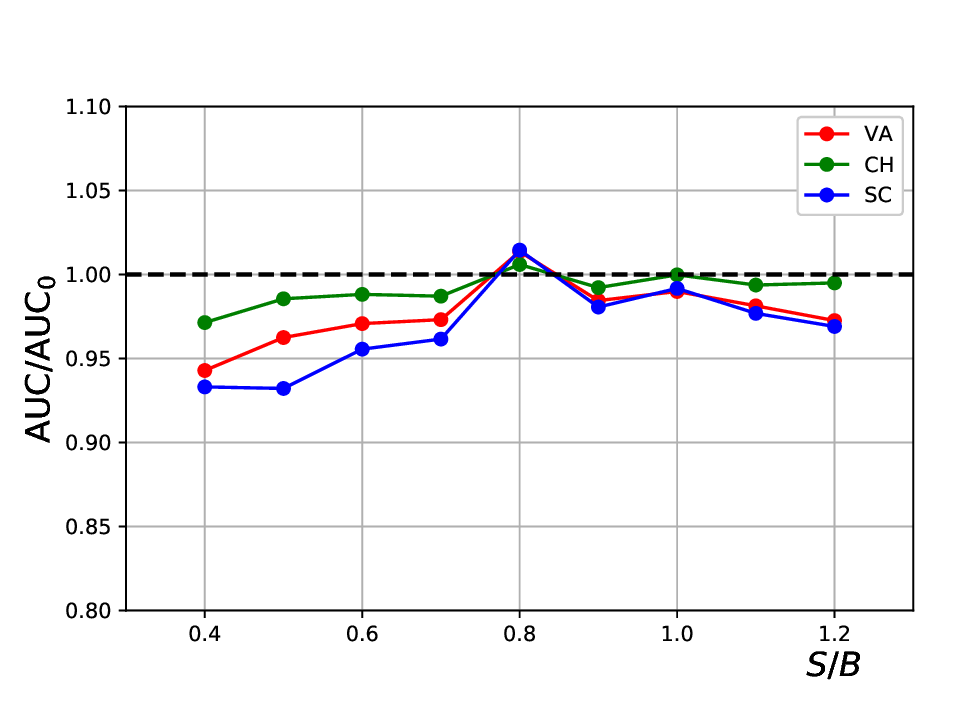}
	\vspace{-1.2cm}
	\caption{}
\end{subfigure}
\begin{subfigure}[t!]{0.49\textwidth}
	\includegraphics[width=\textwidth]{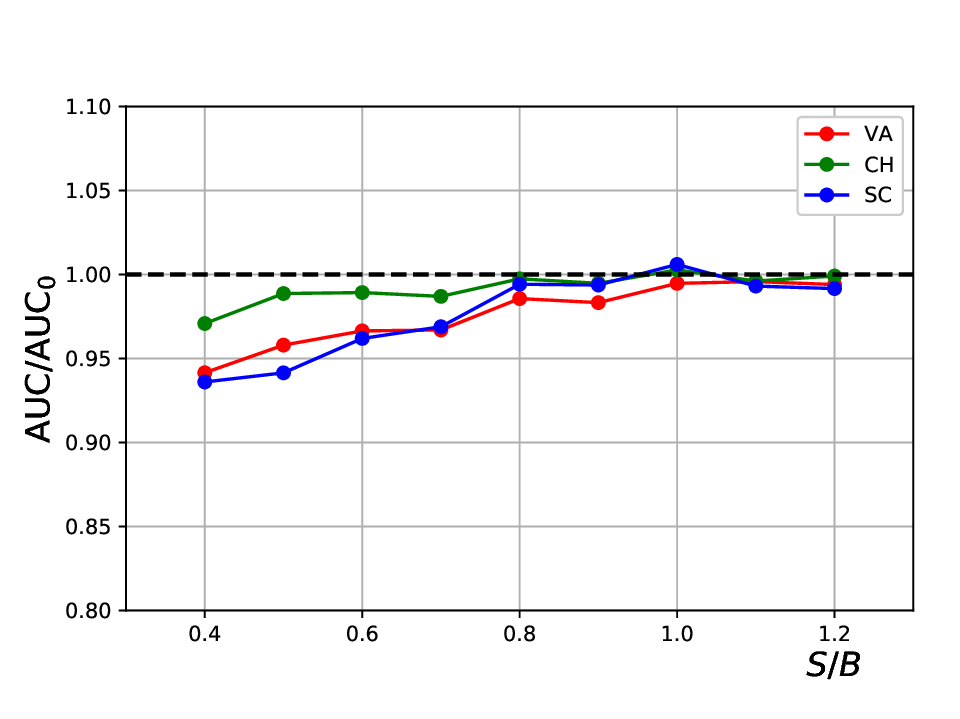}
	\vspace{-1.2cm}
	\caption{}
\end{subfigure}
\caption{Ratios of the 0-jet AUCs from the tests with (a) $S/B=0.4$, (b) $S/B=0.8$, and (c) $S/B=0.12$ models applied to samples of different $S/B$ ratios to the correct AUCs, using $4.5$-TeV resonance samples with $\Gamma_{\rm NP}\approx200$~GeV.}
\label{LO:Wrong_Sig}
\end{figure}

The second is to use the models trained upon 0-jet samples of $S/B=0.4,0.8,0.12$ to test the samples of $S/B\in[0.4,1.2]$ for $4.5$-TeV resonances with fixed $\Gamma_{\rm NP}\approx200$~GeV. We also calculate the ratios of the ``wrong AUCs'' (AUC) to the ``correct AUCs'' (AUC$_0$) for different significances and show them in FIG.~\ref{LO:Wrong_Sig}. The plots show that the wrong models are still able to yield reasonable results in the vicinity of the trained significance level. This result shows that some deviation from the correct significance is all right if one is satisfied with performance within 10\%.

\begin{figure}[H]
\centering
\begin{subfigure}[t!]{0.49\textwidth}
	\includegraphics[width=\textwidth]{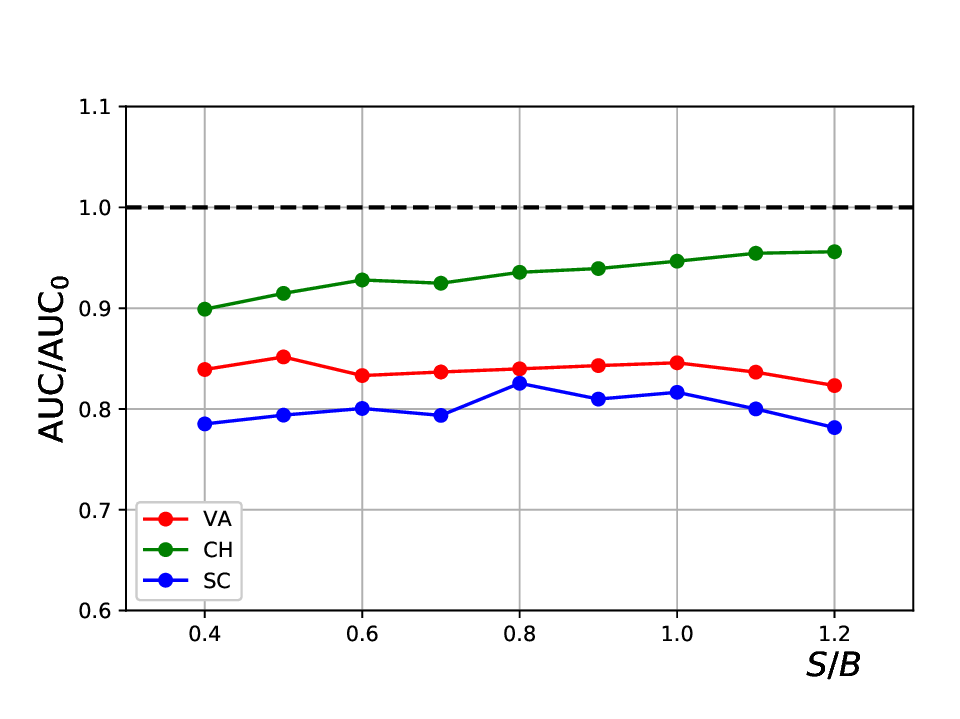}
	\vspace{-1.2cm}
	\caption{}
\end{subfigure}
\begin{subfigure}[t!]{0.49\textwidth}
	\includegraphics[width=\textwidth]{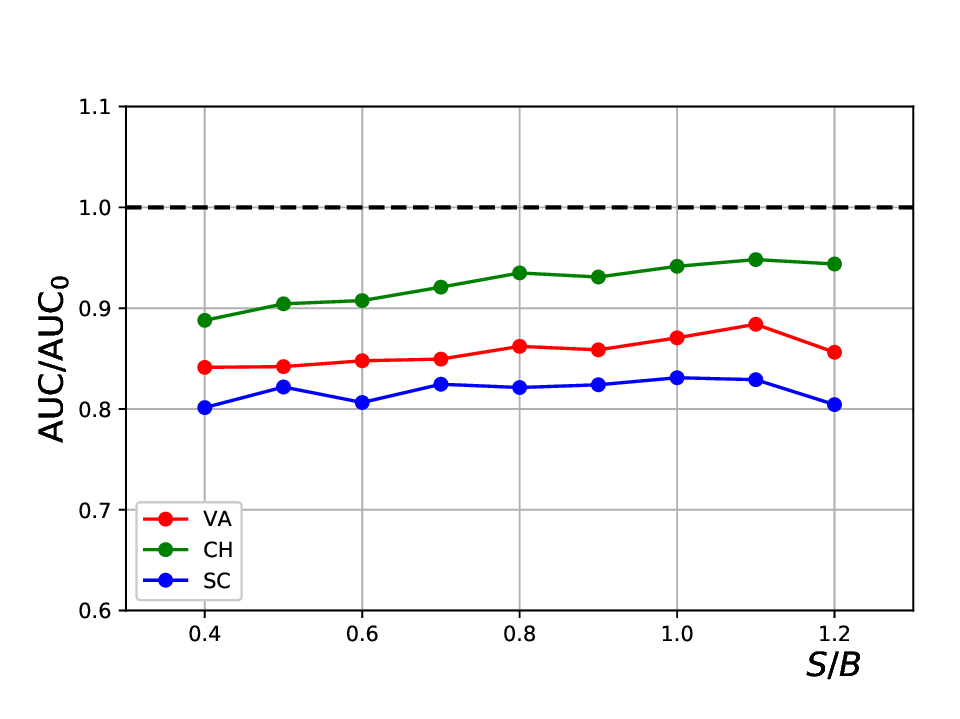}
	\vspace{-1.2cm}
	\caption{}
\end{subfigure}
\end{figure}
\begin{figure}[H]\ContinuedFloat
\centering
\begin{subfigure}[t!]{0.49\textwidth}
	\includegraphics[width=\textwidth]{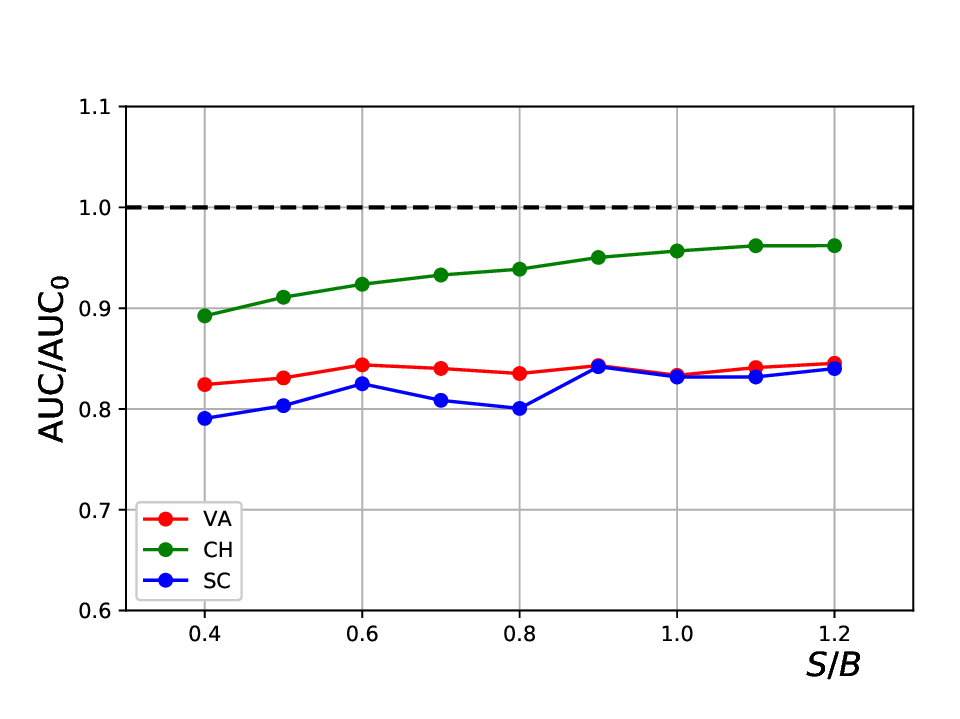}
	\vspace{-1.2cm}
	\caption{}
\end{subfigure}
\begin{subfigure}[t!]{0.49\textwidth}
	\includegraphics[width=\textwidth]{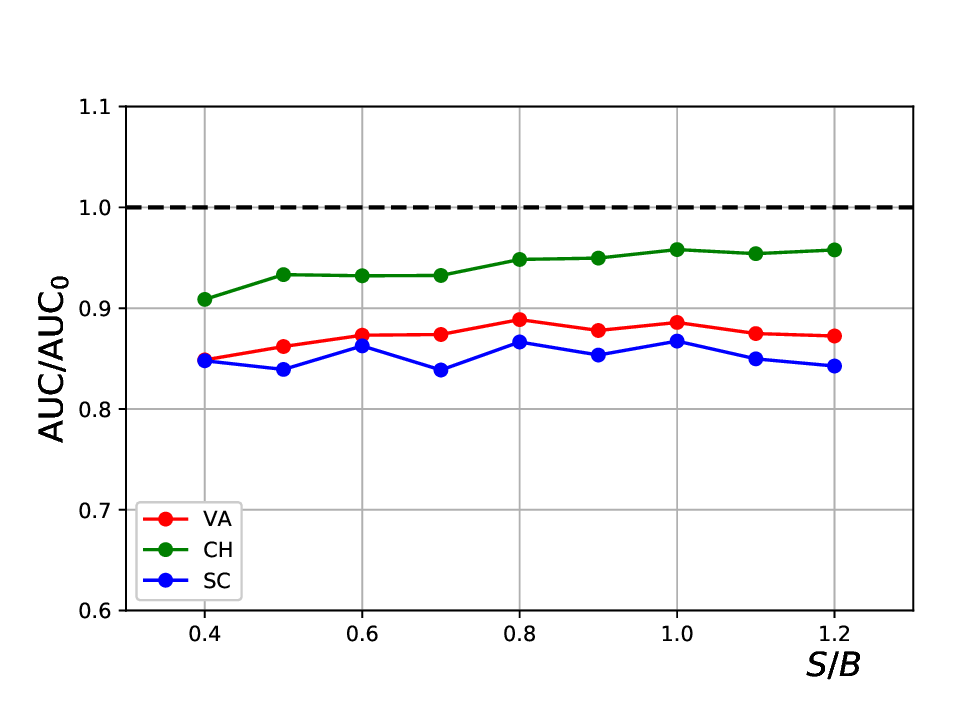}
	\vspace{-1.2cm}
	\caption{}
\end{subfigure}
\begin{subfigure}[t!]{0.49\textwidth}
	\includegraphics[width=\textwidth]{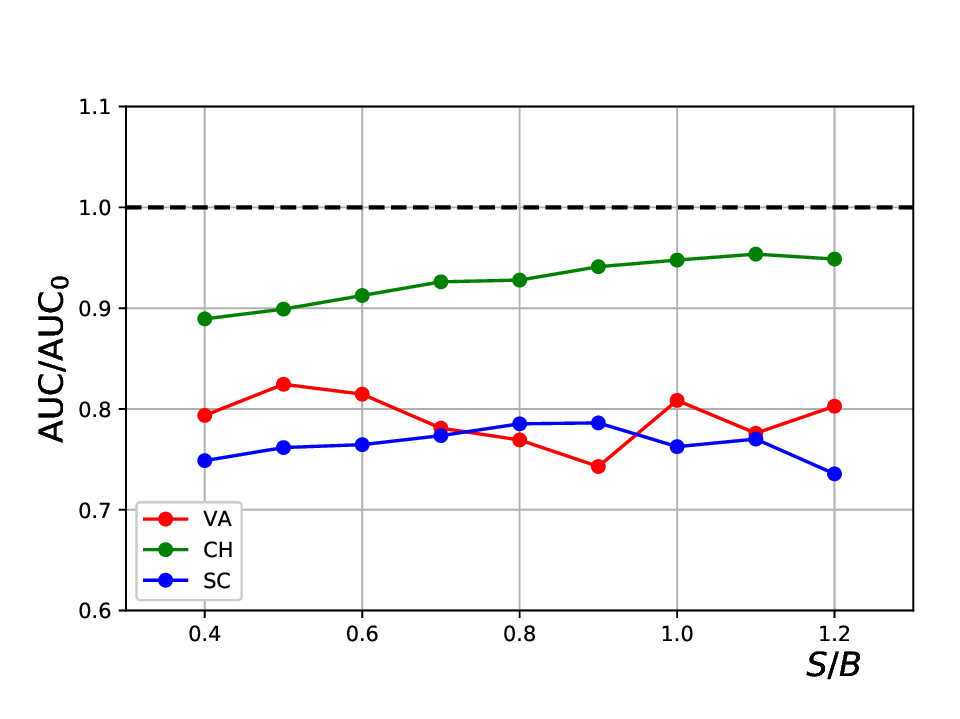}
	\vspace{-1.2cm}
	\caption{}
\end{subfigure}
\begin{subfigure}[t!]{0.49\textwidth}
	\includegraphics[width=\textwidth]{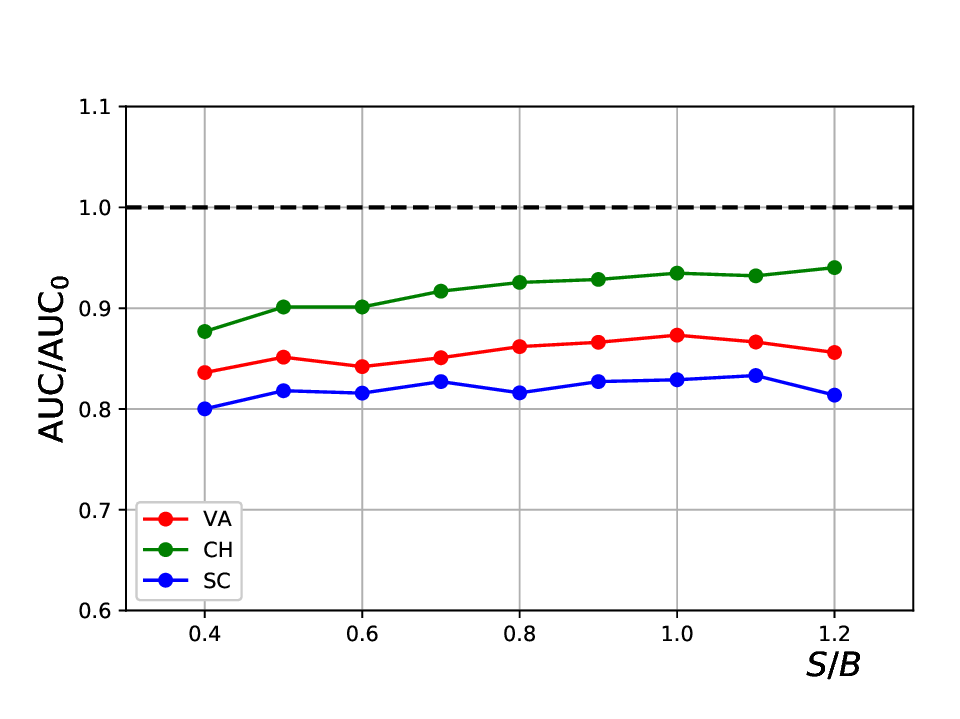}
	\vspace{-1.2cm}
	\caption{}
\end{subfigure}
\caption{Ratios of the 0-jet AUCs from the tests with (a) $\Gamma_{\rm NP}\approx500$~GeV models applied to $\Gamma_{\rm NP}\approx200$~GeV samples, (b) $\Gamma_{\rm NP}\approx200$~GeV models applied to $\Gamma_{\rm NP}\approx500$~GeV samples, (c) $\Gamma_{\rm NP}\approx200$~GeV models applied to $\Gamma_{\rm NP}\approx50$~GeV samples, (d) $\Gamma_{\rm NP}\approx50$~GeV models applied to $\Gamma_{\rm NP}\approx200$~GeV samples, (e) $\Gamma_{\rm NP}\approx500$~GeV models applied to $\Gamma_{\rm NP}\approx50$~GeV samples, and (f) $\Gamma_{\rm NP}\approx50$~GeV models applied to $\Gamma_{\rm NP}\approx500$~GeV samples, to the AUCs using the correct models in the low-significance scenarios.}
\label{LO:Wrong_Width}
\end{figure}

These two comparisons indicate that when applying our analysis to the parameter space of the signal hypotheses, even a coarse set of FNNhs covering the allowed parameter space will still have reasonable performance for a model with a decay width or significance different than the ones used for the set of FNNhs, allowing a reduction of computing resources with a trade off of a small drop in performance.  

\bibliographystyle{apsrev4-1}
%

\end{document}